\documentclass[%
 reprint,
 amsmath,amssymb,
prl,
]{revtex4-2}

\usepackage{amsthm}
\usepackage{xcolor}
\usepackage{graphicx}
\usepackage{dcolumn}
\usepackage{bm}

\usepackage{natbib}

\global\long\def\argmin{\operatornamewithlimits{argmin}}

 \theoremstyle{plain}
  \newtheorem*{thm*}{\protect\theoremname}
  \providecommand{\theoremname}{Theorem}
  
\DeclareMathOperator{\Tr}{Tr}
\DeclareMathOperator\arctanh{arctanh}

\makeatletter
\newcommand*{\balancecolsandclearpage}{%
  \close@column@grid
  \clearpage
  \twocolumngrid
}
\makeatother

\begin{document}

\title{Programmable Quantum Annealers as Noisy Gibbs Samplers}

\author{Marc Vuffray$^{1,2}$, Carleton Coffrin$^{2}$, Yaroslav A. Kharkov$^{3}$, Andrey Y. Lokhov$^{1,2}$}
\email{lokhov@lanl.gov}
\affiliation{$^1$Theoretical Division, Los Alamos National Laboratory, Los Alamos, NM 87545}
\affiliation{$^2$Advanced Network Science Initiative, Los Alamos National Laboratory, Los Alamos, NM 87545}
\affiliation{$^3$Joint Center for Quantum Information and Computer Science, and Joint Quantum Institute, NIST/University of Maryland, College Park, Maryland 20742}

\begin{abstract}
\end{abstract}

\maketitle

{\bf Drawing independent samples from high-dimensional probability distributions represents the major computational bottleneck for modern algorithms, including powerful machine learning frameworks such as deep learning \cite{lecun2015deep}. The quest for discovering larger families of distributions for which sampling can be efficiently realized has inspired an exploration beyond established computing methods \cite{temme2011quantum, biamonte2017quantum} and turning to novel physical devices that leverage the principles of quantum computation \cite{ladd2010quantum}. Quantum annealing \cite{das2008colloquium} embodies a promising computational paradigm that is intimately related to the complexity of energy landscapes in Gibbs distributions, which relate the probabilities of system states to the energies of these states. Here, we study the sampling properties of physical realizations of quantum annealers which are implemented through programmable lattices of superconducting flux qubits \cite{johnson2011quantum}. Comprehensive statistical analysis of the data produced by these quantum machines shows that quantum annealers behave as samplers that generate independent configurations from low-temperature noisy Gibbs distributions. We show that the structure of the output distribution probes the intrinsic physical properties of the quantum device such as effective temperature of individual qubits and magnitude of local qubit noise, which result in a non-linear response function and spurious interactions that are absent in the hardware implementation. We anticipate that our methodology will find widespread use in characterization of future generations of quantum annealers and other emerging analog computing devices.}

Sampling -- the task of producing independent configurations of random variables from a given distribution -- is believed to be among the most challenging computational problems. In particular, many sampling tasks cannot be performed in polynomial time, unless strong and widely accepted conjectures in approximation theory are refuted \cite{sinclair1989approximate,jerrum1986random, jerrum1993polynomial}. The potential value of quickly generating high-quality samples is exemplified by the recent application of emerging analog computing devices, including those based on optical \cite{inagaki2016coherent} and quantum gate \cite{arute2019quantum} technologies, to sampling tasks. However, analog computers are inevitably impacted by hardware imperfections and environmental noise, which distort the computations that they are designed to perform. Assessment of the quality of samples produced by such devices provides a key diagnostic to understand the nature of interactions, biases, and noise inside analog computational machines. In this Report, we leverage state-of-the-art statistical learning methods to conduct a precise fidelity assessment of the sampling properties of analog quantum annealing devices, providing a key foundation for using these devices in high-value sampling tasks.

Adiabatic quantum computing \cite{farhi2001quantum} exemplifies a promising physical principle that may lead to an enhanced exploration of the potentially rough energy landscape due to quantum tunneling \cite{das2008colloquium}. State-of-the-art quantum annealing processors \cite{johnson2011quantum} were recently used to push the frontiers of quantum simulations \cite{king2018observation,harris2018phase}, optimization \cite{mott2017solving}, and machine learning \cite{amin2018quantum}. Comparably, the use of quantum annealers for sampling \cite{amin2015searching, benedetti2016estimation, benedetti2017quantum, li2020limitations} is not as well understood. This is partially due to the lack of methods for a rigorous characterization of empirical distributions produced by independent runs of quantum annealers. Additionally, as is the case with any sophisticated analog device, quantum annealing processors are inevitably affected by noise and biases of diverse nature that are difficult to characterize \cite{albash2019analog, pearson2019analog} and complicate the use of these devices as samplers.

In this study, we focus on the family of D-Wave quantum processing units (QPUs) \cite{johnson2011quantum}. An elementary unit of the D-Wave quantum annealer is a superconducting quantum qubit $i$ whose final state is specified by a binary spin variable $\sigma^{z}_i$ that takes value $+1$ or $-1$ during a read-out process in the computational basis denoted by $z$. Depending on the particular device, the total number of qubits can vary between $1152$ for D-Wave 2X to $2048$ for the D-Wave 2000Q machine. The qubits are interconnected through superconducting \emph{couplers} that form the so-called chimera graph $G=(V,E)$, where $V$ denotes the ensemble of qubits, and $E$ is the set of couplers defined by the connectivity of the chip, see Fig.~1a. The magnitude of currents circulating in the superconducting couplers define the strength of pairwise interactions between individual qubits that can also be biased towards a particular state through a local field.

A D-Wave QPU implements \cite{D-Wave_documentation} the following interpolating Hamiltonian, also known as \emph{energy function}, for $s \in [0,1]$:
\begin{equation}
    H(s) = A(s)\sum_{i \in V} \sigma^{x}_i + B(s) H_{\text{Ising}},
\end{equation}
where $H_{\text{Ising}} = \sum_{ij \in E} J^{\rm{in}}_{ij} \sigma^{z}_i \sigma^{z}_j + \sum_{i \in V}  h^{\rm{in}}_{i} \sigma^{z}_i$ is the target Hamiltonian of the Ising type, i.e. containing only pairwise interactions and local terms. This Ising energy function is specified through the user-defined \emph{input} parameters: Pairwise couplings $\mathbf{J}^{\rm{in}} \equiv \{J^{\rm{in}}_{ij}\}_{(ij) \in E}$ and local fields $\mathbf{h}^{\rm{in}} \equiv \{h^{\rm{in}}_{i}\}_{i \in V}$, where each $J^{\rm{in}}_{ij}$ can be set in the range $[-1,1]$ and each $h^{\rm{in}}_{i}$ in the range $[-2,2]$. The annealing schedule is controlled by two monotonic functions $A(s)$ and $B(s)$ satisfying $A(0) \gg B(0)$ and $B(1) \gg A(1)$. The value $\sigma^{z}_i$ (that we will refer to as ``spin'') for each qubit $i$ is read out in the end of the annealing procedure; In what follows, we will drop index $z$ when discussing these classical measurements of the qubit state. While QPU takes $\mathbf{J}^{\rm{in}}$ and $\mathbf{h}^{\rm{in}}$ as input values, the real values of couplings and magnetic fields implemented on the chip can significantly differ due to a combination of several effects, including programming errors, multiplicative corrections related to effective temperature of the chip, and additive factors such as flux noise and biases, among others.

\begin{figure*}
    \centering
    \includegraphics[width = \linewidth]{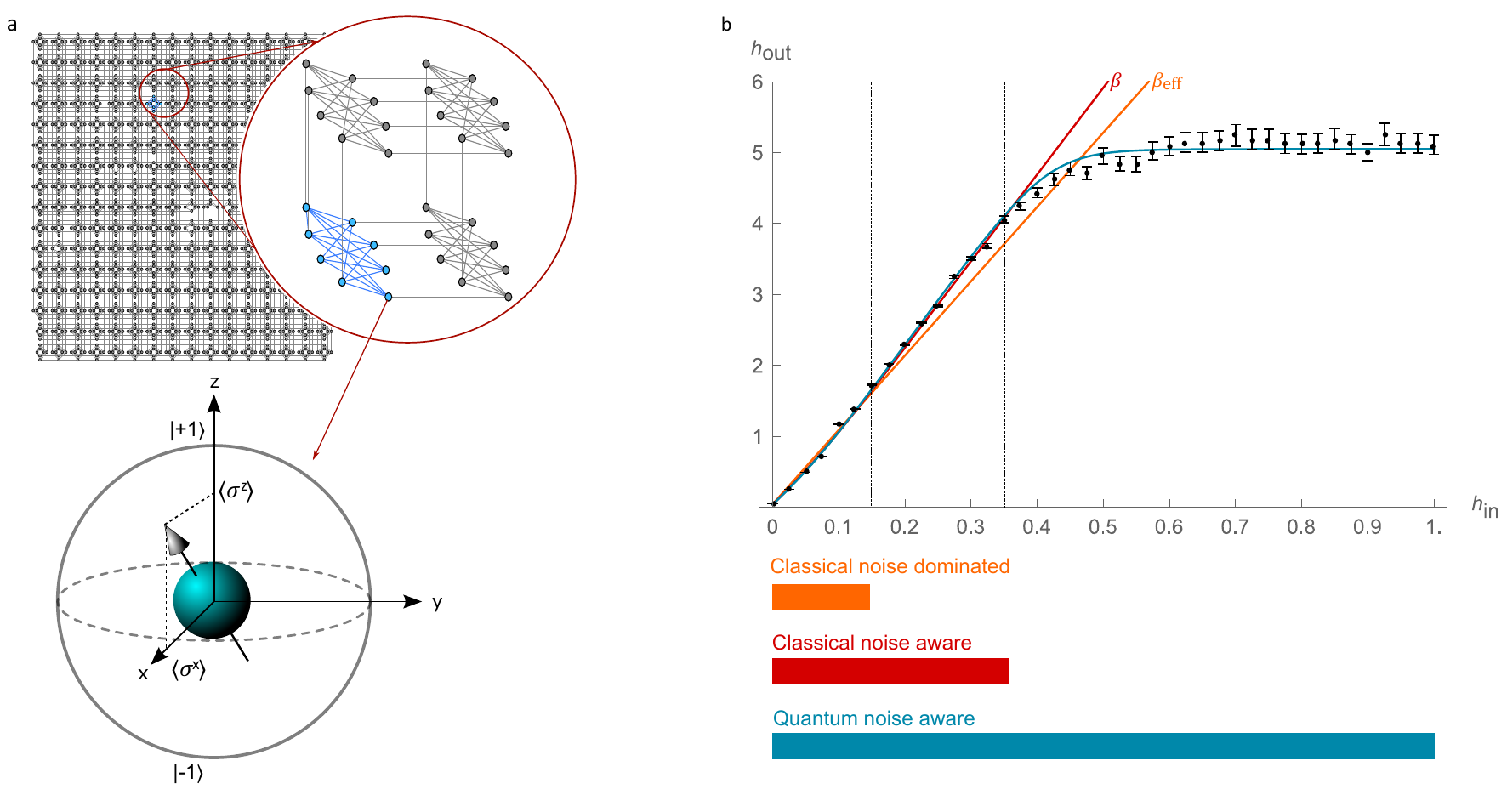}
    \caption{{\bf Noise and quantum statistics in the single-qubit output distribution.} ({\bf a}) The QPU takes input parameters  $\mathbf{J}^{\rm{in}}$ and $\mathbf{h}^{\rm{in}}$ that are specified on the chip with the chimera topology, depicted here for the 2000Q machine at Los Alamos National Laboratory. The magnifier shows details of inter- and intra-connections between qubits in four \emph{cells} composed of eight qubits each. In the end of the anneal, a classical binary projection $\sigma^{z}$ is read out for each qubit. In blue, we highlight the qubits that were used for the reported \emph{single-cell} experiments. ({\bf b}) Dependence of the field $h^{\text{out}}$ describing the output statistics of $\sigma^{z}$ in a single-qubit experiment is plotted here as a function of the positive input parameter $h^{\text{in}}$. Error bars represent statistical fluctuations up to $3$ standard deviations. The statistics observed for $h_i^{\rm{in}} > 0.35$ significantly deviates from the linear behavior expected for a classical Gibbs distribution; In the \emph{Supplementary Information}, we show that this behavior can be described by a quantum model with a residual transverse field. In the regime of input parameters $h_i^{\rm{in}} \leq 0.35$, the output distribution can be described by a classical distribution with a linear dependence $h^{\text{out}} \propto h_i^{\rm{in}}$. We show that the change of the slope around $h_i^{\rm{in}} \approx 0.15$ can be explained by fast-fluctuating noise on the local field: The region of $h_i^{\rm{in}} \leq 0.15$ is dominated by noise which creates a reduced effective response $\beta_{\text{eff}}$, while the noise plays a less pronounced role in the intermediate region where the response coefficient $\beta$ emerges as the inverse-temperature of the model (see \emph{Supplementary Information} for details). Notably, this single-qubit experiment provides a reliable estimation of fast field fluctuations that are too rapid to be measured directly.
    }
    \label{fig:1}
\end{figure*}

The premise of adiabatic quantum computation \cite{albash2018adiabatic} in an isolated setting consists in a sufficiently slow interpolation of the system Hamitonian towards the target one, $H_{\text{Ising}}$. The adiabatic theorem prescribes that when initially prepared in the lowest-energy configuration (ground state) of the starting Hamiltonian, the system is always found in the ground state of the interpolation Hamiltonian; this principle allows one to retrieve ground states of a non-trivial target Ising model, which is useful for optimization applications. However, due to finite temperature \cite{venuti2016adiabaticity}, decoherence \cite{zurek2003decoherence} and other sources of infidelity such as flux qubit noise \cite{D-Wave_documentation}, available quantum annealers, such as those produced by D-Wave Systems, do not consistently find ground states of the target models, but instead behave as non-isolated quantum systems, ending up in excited states. In other words, these quantum devices act as samplers from an unknown distribution, which is commonly expressed as a \emph{Gibbs distribution}: A probability measure that expresses the probability of measuring a certain state $\bm{\sigma} \equiv \{ \sigma_i \}_{i \in V}$ as a function of that state's energy, $\mu(\bm{\sigma}) \propto \exp (H(\bm{\sigma}))$, where $H(\bm{\sigma})$ is some energy function evaluated at $\bm{\sigma}$. This handicap for optimization applications can be turned into an advantage when the annealer is viewed as a sampling device, provided that it is possible to predict the distribution of configurations output by the quantum annealer based on the specified input model. This prediction constitutes the primary objective of this work.

Our investigation of the form and the nature of the distributions produced by quantum annealers begins with a study of the statistics of a single qubit. Consider an experiment that estimates the parameters of the output distribution of a single qubit $i$ represented by a binary variable $\sigma_i$ in the form of a Gibbs distribution, $\mu_{\rm{effective}}(\sigma_i) \propto \exp(\sigma_i h_i^{\rm{out}})$. One can estimate the effective local field $h_i^{\text{out}}$ for different values of input fields $h_i^{\text{in}}$ by looking at the empirical count of positive observations $\sigma_i = +1$. The resulting dependence of $h_i^{\text{out}}$ as a function of $h_i^{\text{in}}$ is depicted in Fig.~1b. We use the maximum likelihood approach (see \emph{Supplementary Information}) to infer parameters that best describe the output statistic in terms of classical and quantum Gibbs distributions. Our results show that while the output distribution of the quantum machine is well described by an effective classical Gibbs distribution for the range of parameters $\vert h_i^{\rm{in}} \vert \leq 0.35$, quantum Gibbs statistics is required to adequately describe the statistics of the effective local field $h_i^{\rm{out}}$ for larger input parameters $\vert h_i^{\rm{in}} \vert > 0.35$, see Fig.~1b. In both regimes, the effective inverse-temperature $\beta$ of the output distribution is very high, in the range $\beta \in [12.4,13.3]$ in the programming units for the considered qubits. This corresponds to a low-temperature regime for most classical distributions, which are notoriously challenging to sample from efficiently. For example, the low-temperature phase of Ising spin glasses occurs at $(\beta J)_{c} = 0.44$ \cite{lokhov2018optimal}, which translates to values of field and coupling magnitudes around $0.035$ in the quantum annealing programming units. Finally, and most importantly, a detailed study of the resulting distributions in Fig.~1b provides strong evidence that rapidly fluctuating noise in the residual local fields plays an essential role in describing the observed statistics of $h_i^{\rm{out}}$. These observations at the single-qubit level naturally lead to the central proposal of this study: Within a suitable input parameter range, D-Wave's quantum annealers act as low-temperature \emph{noisy Gibbs samplers}. In other words, the annealers sample from a Gibbs distribution with energy function parameters that fluctuate due to noise. This proposal is a notable departure from an ideal non-zero temperature quantum annealer, which is expected to sample from the Gibbs distribution of the input Hamiltonian at the annealer's temperature \cite{venuti2016adiabaticity}.

\begin{figure*}
    \centering
    \includegraphics[width = \linewidth]{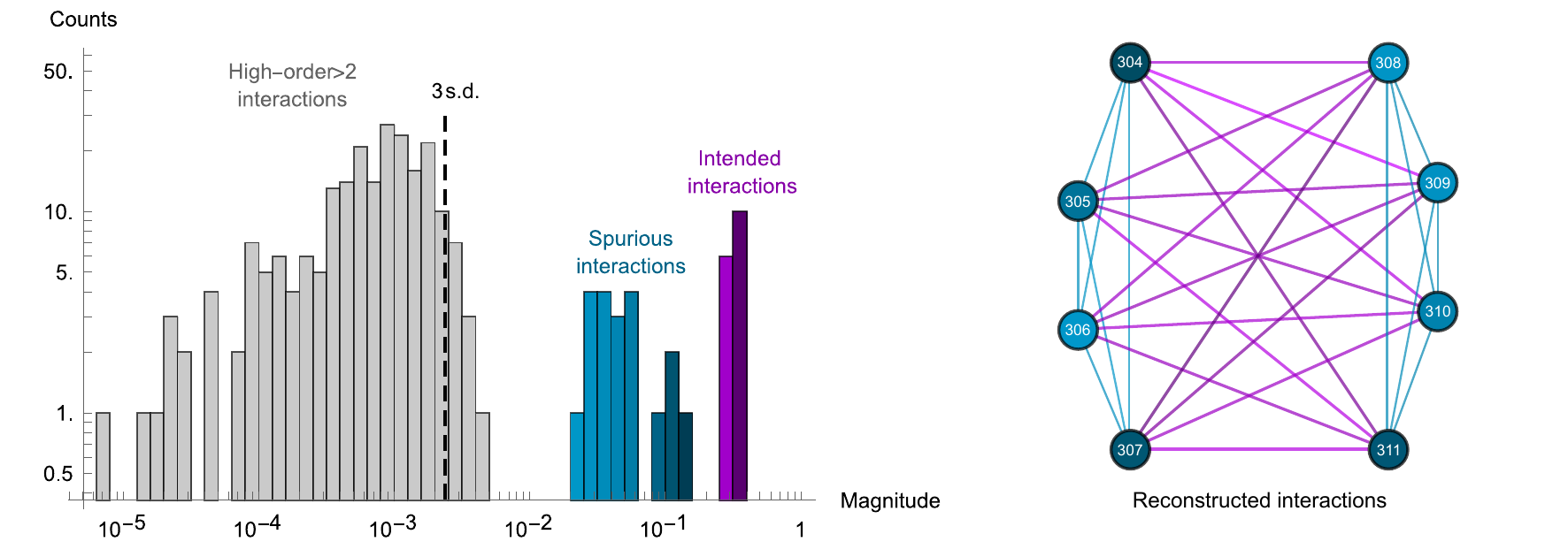}
    \caption{{\bf Characterization of the output distribution on eight qubits.} For a given input model on eight qubits forming a single cell of the chip and characterized by parameters $J^{\rm{in}}_{ij}=0.025, h^{\rm{in}}_{i}=0$, we reconstruct the most general Gibbs distribution on eight binary spins, with an energy function containing all interactions up to the maximum possible order, eight. We repeat the experiment for 50 sets of independent samples to quantify the statistical significance of the reconstructed values leading to the shown 3 standard deviations (s.d.) tolerance. Our results indicate that the second order Ising model provides an adequate description of the emerging output distribution, while higher-order couplings are not statistically significant and can be explained by statistical fluctuations. We find that among statistically significant interactions in the Ising model that best describes the output distribution, the strongest ones (purple) are in correspondence with the input couplings, while the weaker ones (blue) are the \emph{spurious couplings} that are are absent in the chip topology, as well as the \emph{spurious fields} that are also not present in the input problem.
    }
    \label{fig:2}
\end{figure*}

The single-qubit experiment provides strong evidence of the \emph{noisy Gibbs sampler} hypothesis for individual qubits, however it is not clear if similar properties will generalize to larger multi-qubit systems. To further investigate this hypothesis in the context of multi-qubit systems, we conduct a comprehensive characterization of the output distribution on eight qubits that form a single cell of the quantum annealing chip, described in Fig.~1a. In order to remain in a regime described by classical Gibbs distributions, we chose for input parameter values $J^{\rm{in}}_{ij} = 0.025$ for all edges $(i,j)$ inside the cell and $h^{\rm{in}}_{i} = 0$ for all qubits. We note that all discrete distributions on 8 spins with interaction orders up to eight can be fully specified by an exponential family distribution with 255 parameters. We reconstruct these parameters by generalizing our Interaction Screening estimator for learning of Ising models \cite{vuffray2016interaction, lokhov2018optimal} to the case of models with multi-body interactions (see \emph{Supplementary Information}). We would like to stress that the resulting estimator is exact up to statistical fluctuations.

The statistical significance of the reconstructed parameters is empirically estimated by conducting $50$ independent replicates of the algorithm and measuring the variance in the solutions, as described in the \emph{Supplementary Information}. 
Our results show that the output statistics of binary configurations is well described by a classical Gibbs distribution with an energy function structure presented in Fig.~2. In particular, the results indicate that multi-body terms beyond pairwise interactions are statistically insignificant, and hence the distribution output by the quantum annealer is well described by a Gibbs distribution on a model of the Ising type. Surprisingly, the resulting model reveals the presence of additional \emph{spurious} couplings that do not appear in the input model, and correspond to couplings that are not present in the hardware's implementation, see Fig.~2. We will later see that these spurious links are an unexpected consequence of local field noise, similar to the of change in effective inverse-temperature that noise causes in Fig.~1b.

\begin{figure*}[!htb]
    \centering
    \includegraphics[width = \linewidth]{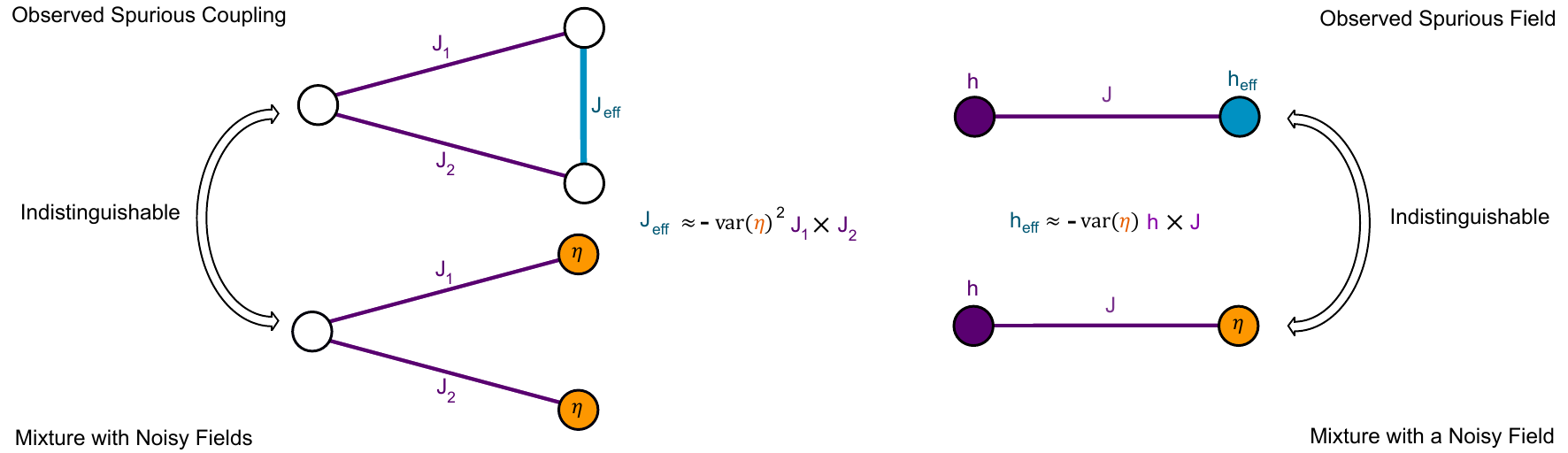}
    \caption{{\bf Noise-aware model of spurious couplings and fields.}
    Inspection of the general input-output quadratic response function indicates that spurious couplings and spurious fields emerge as a quadratic response to the adjacent input parameters. Importantly, the purely quadratic part of the response is described by negative-sign susceptibilities; For instance, interactions favoring same-spin alignment in the input model would lead to spurious interactions that favor opposite-spin alignment in the output model. On the pictured minimal Ising models, we observe that instantaneous qubit noise, introduced here as a zero-mean random variable $\eta$ with a variance denoted $\text{var}(\eta)$, provides a mechanism by which effective spurious couplings and local fields can appear in the output distribution. This shows that the observed output distribution with spurious effects is indistinguishable from a mixture of distributions that differ by a realization of the random noise in local fields. We derived analytical expressions for these models that show that the resulting spurious couplings and spurious fields respond quadratically to the input parameters of the model, with a negative-sign susceptibility, and with the same geometrical pattern as observed in the experimental response function (see \emph{Supplementary Information} for a detailed derivation). The spurious link effect is notably sensitive to the noise level as its strength increases with the square of the field noise variance, $\text{var}(\eta)$.
    }
    \label{fig:3}
\end{figure*}

To better characterize the nature of these spurious links, we employ a data-driven approach to learn how these effects depend on the input parameters. Specifically, we learn a response function that links the input parameters to the effective output parameters that describe the distribution, $\{\mathbf{J}^{\rm{out}},\mathbf{h}^{\rm{out}}\} = f(\mathbf{J}^{\rm{in}},\mathbf{h}^{\rm{in}})$, by regressing on $250$ pairs of input-output models. Input models have been independently sampled for parameters in the range $J^{\rm{in}}_{ij} \in [-0.05,0.05]$ and $h^{\rm{in}}_{i} \in [-0.05,0.05]$. As a result, we show that the input-output function is well described by a general quadratic response. This input-output function reveals that a linear scaling of the input model is the primary driver of the output distribution, but this linear model is distorted by spurious links and fields that depend quadratically on specific input parameters (see Fig. 3). This non-linearity in the response function may explain why previous studies based on a linear response assumption \cite{bian2010ising, benedetti2016estimation, perdomo2016determination, raymond2016global, marshall2017thermalization, li2020limitations} found that effective temperatures inferred under this hypothesis were instance-dependent, while the response function we construct is universal across all input models.

The response function analysis provides a valuable insight that functions of specific combinations of input parameters, i.e. negative feedback from specific edge-edge and field-edge pairs, are the drivers of the spurious effects in the output distribution. Inspired by the observation that local-field noise has a significant role in the single-qubit model, we further investigate how noise may impact the output distributions of multi-qubit models. Our first observation is that due to near-instantaneous qubit noise each sample produced by the quantum annealer represents a \emph{unique} realization of the input model. Consequently, the output distribution represents a mixture of models rather than independent runs from a consistent model. Focusing on the edge-edge and field-edge pairs highlighted by the response function analysis, our second observation (presented in Fig.~3) is that these spurious effects can arise from reconstructing a single effective model from samples of a mixture of models with random local field values. Finally, we observe that spin-reversal transforms, a common persistent bias mitigation technique, cannot eliminate the emergence of spurious effects due to instantaneous noise (see \emph{Supplementary Information}). Altogether, these observations provide a qualitative evidence that instantaneous noise on local fields represents the underlying feature yielding the spurious effects observed in D-Wave's output distribution.

To further validate our noisy Gibbs distribution hypothesis,
we conduct a comprehensive comparison to the statistics of a simulated noisy Gibbs sampler and the output distribution of D-Wave's quantum annealer. Specifically, we replicate the quadratic response function analysis with a simulator that generates samples from a mixture of noisy Ising models calibrated with the noise and scaling parameters extracted from the single-qubit analysis. As shown in Fig.~4, the input-output response function from the simulated distribution provides a strong agreement with the measured susceptibilities, providing compelling evidence that output statistics of D-Wave quantum annealer can be modeled as a noisy Gibbs distribution. Our noisy Gibbs sampler model has been validated across three generations of quantum annealers, and is also indirectly confirmed through replicating this response function analysis on the \emph{lower-noise} version of 2000Q machine \cite{D-Wave_low_noise}, where we observe that the intensity of spurious effects seen in the output distribution are significantly reduced (see \emph{Supplementary Information}).

\begin{figure*}
    \centering
    \includegraphics[width = \linewidth]{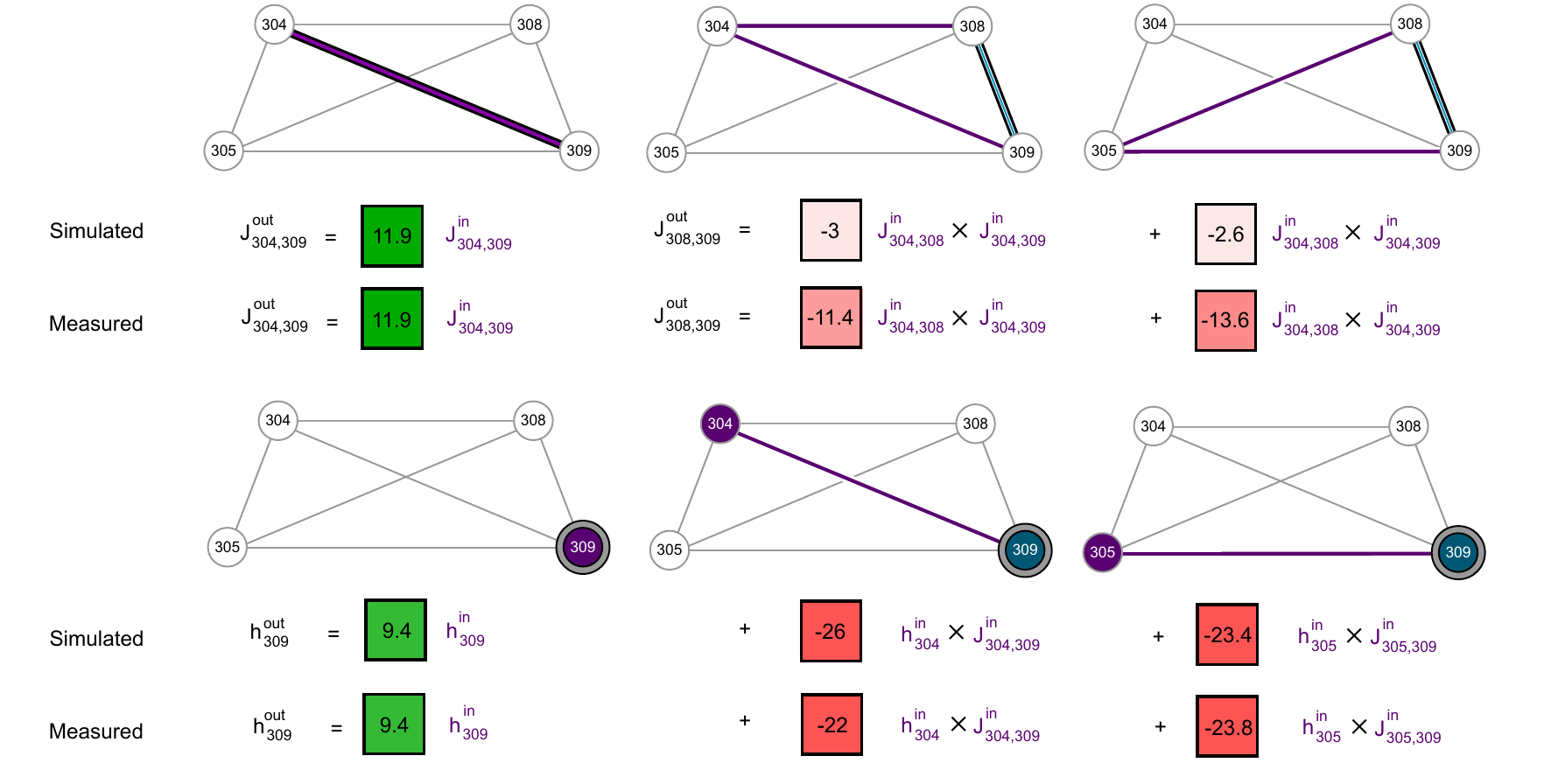}
    \caption{{\bf Leading terms in the quadratic response of effective output parameters.} We show the dominant linear (in green) and quadratic (in red) response terms for the effective parameters $J^{\rm{out}}_{304,305}$, $J^{\rm{out}}_{308,309}$, and $h^{\rm{out}}_{309}$ from Fig.~2. We present a comparison of the response measured in the experiment to the simulated response based on the mechanism explained in Fig.~3 using the noise values extracted from the single-qubit experiment presented in Fig.~1. The response for the native chimera coupling $J^{\rm{out}}_{304,305}$ is driven by a linear self-response term or effective temperature. For the spurious coupling $J^{\rm{out}}_{308,309}$, a linear response is nonexistent but a quadratic response comes from terms involving adjacent couplings forming a triangle with the spurious coupling, in agreement with observations in Fig.~2 and Fig.~3. The response for field $h^{\rm{out}}_{309}$ primarily consists of a linear part driven by the effective temperature but also has a quadratic response involving a neighboring coupling and the adjacent connected field. The complete quadratic response with all terms can be found in the Fig.~S13 of the \emph{Supplementary Information}. The spurious coupling response deviates sensibly more from simulated predictions than the spurious field response. Among the main causes for this discrepancy, we find that measurements of spurious couplings are more prone to statistical fluctuations being significantly weaker than spurious fields, and the sensitivity to noise variation is higher according to the dependence presented in Fig.~3. Remarkably, the simulated  quadratic response shows a strong agreement with the measured susceptibilities for spurious fields.}
    \label{fig:4}
\end{figure*}

We anticipate that methods presented in this study will be broadly used in characterization of analog devices that produce binary samples. Our work opens many avenues for future research, such as studying critical and low-temperature behavior of Ising spin glasses, especially with anticipated increased connectivity \cite{D-Wave_pegasus} or extension to non-stoquastic Hamitonians \cite{ozfidan2020demonstration} in future realizations of quantum annealers that would enable the study of a richer class of problems. The learned response function that maps effective output parameters to the input parameters can be used for calibration of analog machines, which would be useful for practical sampling applications. The concept of noisy Gibbs sampler is also promising in its own right for accelerated solution of robust optimization and sampling problems within hardware-in-the-loop approaches.

{\it Acknowledgements:} M.V., C.C., and A.Y.L. acknowledge support from the Laboratory Directed Research and Development program of Los Alamos National Laboratory under Project 20210114ER.

\bibliographystyle{naturemag}
\bibliography{scibib}

\begin{thebibliography}{10}
\expandafter\ifx\csname url\endcsname\relax
  \def\url#1{\texttt{#1}}\fi
\expandafter\ifx\csname urlprefix\endcsname\relax\def\urlprefix{URL }\fi
\providecommand{\bibinfo}[2]{#2}
\providecommand{\eprint}[2][]{\url{#2}}

\bibitem{lecun2015deep}
\bibinfo{author}{LeCun, Y.}, \bibinfo{author}{Bengio, Y.} \&
  \bibinfo{author}{Hinton, G.}
\newblock \bibinfo{title}{Deep learning}.
\newblock \emph{\bibinfo{journal}{Nature}} \textbf{\bibinfo{volume}{521}},
  \bibinfo{pages}{436--444} (\bibinfo{year}{2015}).

\bibitem{temme2011quantum}
\bibinfo{author}{Temme, K.}, \bibinfo{author}{Osborne, T.~J.},
  \bibinfo{author}{Vollbrecht, K.~G.}, \bibinfo{author}{Poulin, D.} \&
  \bibinfo{author}{Verstraete, F.}
\newblock \bibinfo{title}{Quantum metropolis sampling}.
\newblock \emph{\bibinfo{journal}{Nature}} \textbf{\bibinfo{volume}{471}},
  \bibinfo{pages}{87--90} (\bibinfo{year}{2011}).

\bibitem{biamonte2017quantum}
\bibinfo{author}{Biamonte, J.} \emph{et~al.}
\newblock \bibinfo{title}{Quantum machine learning}.
\newblock \emph{\bibinfo{journal}{Nature}} \textbf{\bibinfo{volume}{549}},
  \bibinfo{pages}{195} (\bibinfo{year}{2017}).

\bibitem{ladd2010quantum}
\bibinfo{author}{Ladd, T.~D.} \emph{et~al.}
\newblock \bibinfo{title}{Quantum computers}.
\newblock \emph{\bibinfo{journal}{Nature}} \textbf{\bibinfo{volume}{464}},
  \bibinfo{pages}{45--53} (\bibinfo{year}{2010}).

\bibitem{das2008colloquium}
\bibinfo{author}{Das, A.} \& \bibinfo{author}{Chakrabarti, B.~K.}
\newblock \bibinfo{title}{Colloquium: Quantum annealing and analog quantum
  computation}.
\newblock \emph{\bibinfo{journal}{Reviews of Modern Physics}}
  \textbf{\bibinfo{volume}{80}}, \bibinfo{pages}{1061} (\bibinfo{year}{2008}).

\bibitem{johnson2011quantum}
\bibinfo{author}{Johnson, M.~W.} \emph{et~al.}
\newblock \bibinfo{title}{Quantum annealing with manufactured spins}.
\newblock \emph{\bibinfo{journal}{Nature}} \textbf{\bibinfo{volume}{473}},
  \bibinfo{pages}{194--198} (\bibinfo{year}{2011}).

\bibitem{sinclair1989approximate}
\bibinfo{author}{Sinclair, A.} \& \bibinfo{author}{Jerrum, M.}
\newblock \bibinfo{title}{Approximate counting, uniform generation and rapidly
  mixing markov chains}.
\newblock \emph{\bibinfo{journal}{Information and Computation}}
  \textbf{\bibinfo{volume}{82}}, \bibinfo{pages}{93--133}
  (\bibinfo{year}{1989}).

\bibitem{jerrum1986random}
\bibinfo{author}{Jerrum, M.~R.}, \bibinfo{author}{Valiant, L.~G.} \&
  \bibinfo{author}{Vazirani, V.~V.}
\newblock \bibinfo{title}{Random generation of combinatorial structures from a
  uniform distribution}.
\newblock \emph{\bibinfo{journal}{Theoretical computer science}}
  \textbf{\bibinfo{volume}{43}}, \bibinfo{pages}{169--188}
  (\bibinfo{year}{1986}).

\bibitem{jerrum1993polynomial}
\bibinfo{author}{Jerrum, M.} \& \bibinfo{author}{Sinclair, A.}
\newblock \bibinfo{title}{Polynomial-time approximation algorithms for the
  ising model}.
\newblock \emph{\bibinfo{journal}{SIAM Journal on computing}}
  \textbf{\bibinfo{volume}{22}}, \bibinfo{pages}{1087--1116}
  (\bibinfo{year}{1993}).

\bibitem{inagaki2016coherent}
\bibinfo{author}{Inagaki, T.} \emph{et~al.}
\newblock \bibinfo{title}{A coherent ising machine for 2000-node optimization
  problems}.
\newblock \emph{\bibinfo{journal}{Science}} \textbf{\bibinfo{volume}{354}},
  \bibinfo{pages}{603--606} (\bibinfo{year}{2016}).

\bibitem{arute2019quantum}
\bibinfo{author}{Arute, F.} \emph{et~al.}
\newblock \bibinfo{title}{Quantum supremacy using a programmable
  superconducting processor}.
\newblock \emph{\bibinfo{journal}{Nature}} \textbf{\bibinfo{volume}{574}},
  \bibinfo{pages}{505--510} (\bibinfo{year}{2019}).

\bibitem{farhi2001quantum}
\bibinfo{author}{Farhi, E.} \emph{et~al.}
\newblock \bibinfo{title}{A quantum adiabatic evolution algorithm applied to
  random instances of an np-complete problem}.
\newblock \emph{\bibinfo{journal}{Science}} \textbf{\bibinfo{volume}{292}},
  \bibinfo{pages}{472--475} (\bibinfo{year}{2001}).

\bibitem{king2018observation}
\bibinfo{author}{King, A.~D.} \emph{et~al.}
\newblock \bibinfo{title}{Observation of topological phenomena in a
  programmable lattice of 1,800 qubits}.
\newblock \emph{\bibinfo{journal}{Nature}} \textbf{\bibinfo{volume}{560}},
  \bibinfo{pages}{456} (\bibinfo{year}{2018}).

\bibitem{harris2018phase}
\bibinfo{author}{Harris, R.} \emph{et~al.}
\newblock \bibinfo{title}{Phase transitions in a programmable quantum spin
  glass simulator}.
\newblock \emph{\bibinfo{journal}{Science}} \textbf{\bibinfo{volume}{361}},
  \bibinfo{pages}{162--165} (\bibinfo{year}{2018}).

\bibitem{mott2017solving}
\bibinfo{author}{Mott, A.}, \bibinfo{author}{Job, J.},
  \bibinfo{author}{Vlimant, J.-R.}, \bibinfo{author}{Lidar, D.} \&
  \bibinfo{author}{Spiropulu, M.}
\newblock \bibinfo{title}{Solving a higgs optimization problem with quantum
  annealing for machine learning}.
\newblock \emph{\bibinfo{journal}{Nature}} \textbf{\bibinfo{volume}{550}},
  \bibinfo{pages}{375--379} (\bibinfo{year}{2017}).

\bibitem{amin2018quantum}
\bibinfo{author}{Amin, M.~H.}, \bibinfo{author}{Andriyash, E.},
  \bibinfo{author}{Rolfe, J.}, \bibinfo{author}{Kulchytskyy, B.} \&
  \bibinfo{author}{Melko, R.}
\newblock \bibinfo{title}{Quantum boltzmann machine}.
\newblock \emph{\bibinfo{journal}{Physical Review X}}
  \textbf{\bibinfo{volume}{8}}, \bibinfo{pages}{021050} (\bibinfo{year}{2018}).

\bibitem{amin2015searching}
\bibinfo{author}{Amin, M.~H.}
\newblock \bibinfo{title}{Searching for quantum speedup in quasistatic quantum
  annealers}.
\newblock \emph{\bibinfo{journal}{Physical Review A}}
  \textbf{\bibinfo{volume}{92}}, \bibinfo{pages}{052323}
  (\bibinfo{year}{2015}).

\bibitem{benedetti2016estimation}
\bibinfo{author}{Benedetti, M.}, \bibinfo{author}{Realpe-G{\'o}mez, J.},
  \bibinfo{author}{Biswas, R.} \& \bibinfo{author}{Perdomo-Ortiz, A.}
\newblock \bibinfo{title}{Estimation of effective temperatures in quantum
  annealers for sampling applications: A case study with possible applications
  in deep learning}.
\newblock \emph{\bibinfo{journal}{Physical Review A}}
  \textbf{\bibinfo{volume}{94}}, \bibinfo{pages}{022308}
  (\bibinfo{year}{2016}).

\bibitem{benedetti2017quantum}
\bibinfo{author}{Benedetti, M.}, \bibinfo{author}{Realpe-G{\'o}mez, J.},
  \bibinfo{author}{Biswas, R.} \& \bibinfo{author}{Perdomo-Ortiz, A.}
\newblock \bibinfo{title}{Quantum-assisted learning of hardware-embedded
  probabilistic graphical models}.
\newblock \emph{\bibinfo{journal}{Physical Review X}}
  \textbf{\bibinfo{volume}{7}}, \bibinfo{pages}{041052} (\bibinfo{year}{2017}).

\bibitem{li2020limitations}
\bibinfo{author}{Li, R.}, \bibinfo{author}{Albash, T.} \&
  \bibinfo{author}{Lidar, D.~A.}
\newblock \bibinfo{title}{Limitations of error corrected quantum annealing in
  improving the performance of boltzmannmachines}.
\newblock \emph{\bibinfo{journal}{Quantum Science and Technology}}
  (\bibinfo{year}{2020}).

\bibitem{albash2019analog}
\bibinfo{author}{Albash, T.}, \bibinfo{author}{Martin-Mayor, V.} \&
  \bibinfo{author}{Hen, I.}
\newblock \bibinfo{title}{Analog errors in ising machines}.
\newblock \emph{\bibinfo{journal}{Quantum Science and Technology}}
  \textbf{\bibinfo{volume}{4}}, \bibinfo{pages}{02LT03} (\bibinfo{year}{2019}).

\bibitem{pearson2019analog}
\bibinfo{author}{Pearson, A.}, \bibinfo{author}{Mishra, A.},
  \bibinfo{author}{Hen, I.} \& \bibinfo{author}{Lidar, D.~A.}
\newblock \bibinfo{title}{Analog errors in quantum annealing: doom and hope}.
\newblock \emph{\bibinfo{journal}{NPJ Quantum Information}}
  \textbf{\bibinfo{volume}{5}}, \bibinfo{pages}{1--9} (\bibinfo{year}{2019}).

\bibitem{D-Wave_documentation}
\bibinfo{title}{Technical description of the d-wave quantum processing unit}.
\newblock
  \bibinfo{howpublished}{\url{https://docs.dwavesys.com/docs/latest/_downloads/09-1109A-V_Technical_Description_of_DW_QPU.pdf}}.
\newblock \bibinfo{note}{Accessed: 2020-11-12}.

\bibitem{albash2018adiabatic}
\bibinfo{author}{Albash, T.} \& \bibinfo{author}{Lidar, D.~A.}
\newblock \bibinfo{title}{Adiabatic quantum computation}.
\newblock \emph{\bibinfo{journal}{Reviews of Modern Physics}}
  \textbf{\bibinfo{volume}{90}}, \bibinfo{pages}{015002}
  (\bibinfo{year}{2018}).

\bibitem{venuti2016adiabaticity}
\bibinfo{author}{Venuti, L.~C.}, \bibinfo{author}{Albash, T.},
  \bibinfo{author}{Lidar, D.~A.} \& \bibinfo{author}{Zanardi, P.}
\newblock \bibinfo{title}{Adiabaticity in open quantum systems}.
\newblock \emph{\bibinfo{journal}{Physical Review A}}
  \textbf{\bibinfo{volume}{93}}, \bibinfo{pages}{032118}
  (\bibinfo{year}{2016}).

\bibitem{zurek2003decoherence}
\bibinfo{author}{Zurek, W.~H.}
\newblock \bibinfo{title}{Decoherence, einselection, and the quantum origins of
  the classical}.
\newblock \emph{\bibinfo{journal}{Reviews of modern physics}}
  \textbf{\bibinfo{volume}{75}}, \bibinfo{pages}{715} (\bibinfo{year}{2003}).

\bibitem{lokhov2018optimal}
\bibinfo{author}{Lokhov, A.~Y.}, \bibinfo{author}{Vuffray, M.},
  \bibinfo{author}{Misra, S.} \& \bibinfo{author}{Chertkov, M.}
\newblock \bibinfo{title}{Optimal structure and parameter learning of ising
  models}.
\newblock \emph{\bibinfo{journal}{Science Advances}}
  \textbf{\bibinfo{volume}{4}}, \bibinfo{pages}{e1700791}
  (\bibinfo{year}{2018}).

\bibitem{vuffray2016interaction}
\bibinfo{author}{Vuffray, M.}, \bibinfo{author}{Misra, S.},
  \bibinfo{author}{Lokhov, A.} \& \bibinfo{author}{Chertkov, M.}
\newblock \bibinfo{title}{Interaction screening: Efficient and sample-optimal
  learning of ising models}.
\newblock In \emph{\bibinfo{booktitle}{Advances in Neural Information
  Processing Systems}}, \bibinfo{pages}{2595--2603} (\bibinfo{year}{2016}).

\bibitem{bian2010ising}
\bibinfo{author}{Bian, Z.}, \bibinfo{author}{Chudak, F.},
  \bibinfo{author}{Macready, W.~G.} \& \bibinfo{author}{Rose, G.}
\newblock \bibinfo{title}{The ising model: teaching an old problem new tricks}.
\newblock \emph{\bibinfo{journal}{D-wave systems}} \textbf{\bibinfo{volume}{2}}
  (\bibinfo{year}{2010}).

\bibitem{perdomo2016determination}
\bibinfo{author}{Perdomo-Ortiz, A.}, \bibinfo{author}{O’Gorman, B.},
  \bibinfo{author}{Fluegemann, J.}, \bibinfo{author}{Biswas, R.} \&
  \bibinfo{author}{Smelyanskiy, V.~N.}
\newblock \bibinfo{title}{Determination and correction of persistent biases in
  quantum annealers}.
\newblock \emph{\bibinfo{journal}{Scientific reports}}
  \textbf{\bibinfo{volume}{6}}, \bibinfo{pages}{18628} (\bibinfo{year}{2016}).

\bibitem{raymond2016global}
\bibinfo{author}{Raymond, J.}, \bibinfo{author}{Yarkoni, S.} \&
  \bibinfo{author}{Andriyash, E.}
\newblock \bibinfo{title}{Global warming: Temperature estimation in annealers}.
\newblock \emph{\bibinfo{journal}{Frontiers in ICT}}
  \textbf{\bibinfo{volume}{3}}, \bibinfo{pages}{23} (\bibinfo{year}{2016}).

\bibitem{marshall2017thermalization}
\bibinfo{author}{Marshall, J.}, \bibinfo{author}{Rieffel, E.~G.} \&
  \bibinfo{author}{Hen, I.}
\newblock \bibinfo{title}{Thermalization, freeze-out, and noise: Deciphering
  experimental quantum annealers}.
\newblock \emph{\bibinfo{journal}{Physical Review Applied}}
  \textbf{\bibinfo{volume}{8}}, \bibinfo{pages}{064025} (\bibinfo{year}{2017}).

\bibitem{D-Wave_low_noise}
\bibinfo{title}{Probing mid-band and broad-band noise in lower-noise d-wave
  2000q fabrication stacks}.
\newblock
  \bibinfo{howpublished}{\url{https://www.dwavesys.com/sites/default/files/14-1034A-A_Probling_noise_in_LN_2000Q_fabrication_stacks.pdf}}.
\newblock \bibinfo{note}{Accessed: 2020-11-12}.

\bibitem{D-Wave_pegasus}
\bibinfo{title}{Next-generation topology of d-wave quantum processors}.
\newblock
  \bibinfo{howpublished}{\url{https://www.dwavesys.com/sites/default/files/14-1026A-C_Next-Generation-Topology-of-DW-Quantum-Processors.pdf}}.
\newblock \bibinfo{note}{Accessed: 2020-11-12}.

\bibitem{ozfidan2020demonstration}
\bibinfo{author}{Ozfidan, I.} \emph{et~al.}
\newblock \bibinfo{title}{Demonstration of a nonstoquastic hamiltonian in
  coupled superconducting flux qubits}.
\newblock \emph{\bibinfo{journal}{Physical Review Applied}}
  \textbf{\bibinfo{volume}{13}}, \bibinfo{pages}{034037}
  (\bibinfo{year}{2020}).

\bibitem{Crow1956Confidence}
\bibinfo{author}{Crow, E.~L.}
\newblock \bibinfo{title}{Confidence intervals for a proportion}.
\newblock \emph{\bibinfo{journal}{Biometrika}} \textbf{\bibinfo{volume}{43}},
  \bibinfo{pages}{423--435} (\bibinfo{year}{1956}).

\bibitem{nguyen2017inverse}
\bibinfo{author}{Nguyen, H.~C.}, \bibinfo{author}{Zecchina, R.} \&
  \bibinfo{author}{Berg, J.}
\newblock \bibinfo{title}{Inverse statistical problems: from the inverse ising
  problem to data science}.
\newblock \emph{\bibinfo{journal}{Advances in Physics}}
  \textbf{\bibinfo{volume}{66}}, \bibinfo{pages}{197--261}
  (\bibinfo{year}{2017}).

\bibitem{bresler2015efficiently}
\bibinfo{author}{Bresler, G.}
\newblock \bibinfo{title}{Efficiently learning ising models on arbitrary
  graphs}.
\newblock In \emph{\bibinfo{booktitle}{Proceedings of the forty-seventh annual
  ACM symposium on Theory of computing}}, \bibinfo{pages}{771--782}
  (\bibinfo{year}{2015}).

\end{thebibliography}
\balancecolsandclearpage


\onecolumngrid

{\centering {\large \bf Supplementary Information\\}}
\vspace{2cm}


\appendix

\renewcommand{\thefigure}{S\arabic{figure}}

\setcounter{figure}{0}

\renewcommand{\thetable}{S\arabic{table}}

\setcounter{table}{0}

\renewcommand{\thesection}{S\arabic{section}}

\setcounter{section}{0}


\section{Primary Hardware Platform and Experimental Settings}

The primary hardware platform on which the majority of experiments have been conducted is a D-Wave 2000Q quantum annealer at Los Alamos National Laboratory, referred to as \texttt{DW\_2000Q\_LANL}. The \texttt{DW\_2000Q\_LANL} QPU chip has a so-called chimera graph structure with $C_{16}$ topology, i.e. it is composed of two dimensional lattice of 16-by-16 unit cells. Each unit cell is composed of 8 qubits connected through a complete bipartite graph structure. A very small number of faulty qubits are disabled and not available for programming. In total, this QPU has 2032 operational qubits and 5924 operational couplers. A complete topology of the \texttt{DW\_2000Q\_LANL} hardware graph is depicted in Fig.~1a of the Main Text.

Most of experiments in this paper deal with a single unit cell with 8 qubits and 16 couplers; the specific identifiers of this cell is given in Table \ref{tbl:l2k-qpu-used}. We denote the set of qubits as $V$ with $\vert V \vert = N$, and the set of couplers as $E$. This specific set of qubits was selected as it is characteristic of a typical complete unit-cell in a hardware chip.

Unless specified otherwise, in this work we set the following additional solver parameters when submitting jobs to the D-Wave hardware: \texttt{auto\_scale = False}, which ensures that the input parameters are not automatically rescaled to utilize the maximal operating range (a feature sometimes used for optimization applications); \texttt{flux\_drift\_compensation = False}, which prevents automatic corrections to input fields based on calibration procedure that is run a few times each hour; \texttt{annealing\_time = 5}, which corresponds to a single-run annealing time of 5$\mu$s; and \texttt{num\_reads = 10000}, which specifies the number of samples collected for a single programming of the chip. The impact of the specific choice of the annealing time in the regime of parameters considered in this work is negligible, as discussed in Section \ref{sec:impact_of_annealing_time}. The motivations for disabling the flux drift compensation and the impact of spin reversal transformation are thoroughly discussed in Sections \ref{sec:effect_of_spin_reversals} and \ref{sec:local_field_variability}.

\begin{table}[!b]
    \centering
    \begin{tabular}{|p{0.16\columnwidth}|p{0.84\columnwidth}|}
    \hline
    \multicolumn{2}{|c|}{Unit Cell Considered} \\
    \hline
    qubits & $V$ = \{304, 305, 306, 307, 308, 309, 310, 311\} \\
    \hline
    couplers & $E$ =
        \{(304, 308), (304, 309), (304, 310), (304, 311),
        (305, 308), (305, 309), (305, 310), (305, 311),
        (306, 308), (306, 309), (306, 310), (306, 311),
        (307, 308), (307, 309), (307, 310), (307, 311)\} \\
    \hline
    \hline
    \multicolumn{2}{|c|}{Unit Cell Subset for High Throughput Experiments} \\
    \hline
    qubits & $V'$ = \{304, 305, 308, 309\} \\
    \hline
    couplers & $E'$ =
        \{(304, 308), (304, 309), (305, 308), (305, 309)\} \\
    \hline
    \end{tabular}
    \caption{The 8 qubits and 16 couplers used on the \texttt{DW\_2000Q\_LANL} QPU to conduct the primary experiments of this work.  A subset of 4 qubits and 4 couplers is specified for use in high-throughput experiments.}
    \label{tbl:l2k-qpu-used}
\end{table}

\section{Single-qubit Experiments}
\label{sec:single_spin_exp}

As demonstrated by the results in this paper, in the regime of couplings that are interesting for sampling applications, the output statistics are perfectly described by a certain classical Boltzmann distribution. Here, we investigate at which coupling strength this classical description breaks down, and one needs to introduce a different statistics, such as quantum Boltzmann distribution, for an adequate description of the output data. We show that this regime corresponds to the couplings strengths that are order of magnitude above the intensities that we consider throughout our study.
The experiments consist in looking, for isolated spins, at the relationship between the input magnetic field $h^{\rm{in}}$ and the outcome statistic described by the effective field $h^{\rm{out}}$. The outcome statistic of a single spin being always fully expressible by a probability distribution taking the following form,
\begin{align}
    \mu_{\rm{effective}}(\sigma) = \frac{\exp(h^{\rm{out}}\sigma)}{2\cosh{(h^{\rm{out}})}} = \frac{1+\sigma \tanh(h^{\rm{out}})}{2}.\label{eq:classical_single_spin_distribution}
\end{align}
Our process to estimate $h^{\rm{out}}$ for a given value of the input magnetic field consists in the following steps. We start by collecting $M$ samples from the D-wave annealer embodied as a list of single spin realizations $\sigma^{(k)}\in \{-1,1\}$ for $k=1,\ldots,M$. The statistic $S$ that we extract from these samples is the count of positive spin realization $S=\sum_{k=1}^M \delta_{\sigma^{(k)},1}$, where $\delta$ is the Kronecker delta. Assuming that each sample is effectively independent and identically distributed from Eq.~\eqref{eq:classical_single_spin_distribution}, we observe that the statistic $S$ is a Bernoulli process with $M$ trials and probability of success $p = \frac{1+ \tanh(h^{\rm{out}})}{2}$. We estimate the probability of success $p$ using the standard unbiased estimator $\widehat{p} = S/M$ for Bernoulli processes. We compute confidence intervals $I_\alpha = \left[ \underline{p}, \overline{p}\right]$ with confidence level $\alpha$ around our estimator $\widehat{p}$ using the exact method of Crow \cite{Crow1956Confidence} leading to minimal length intervals. Finally, we invert the relationship between $\widehat{p}$ and $h^{\rm{out}}$ to find an estimate of the output effective field. Confidence intervals on $h^{\rm{out}}$ are found using the same relation since it is a monotonic mapping. In the experiments, we have collected $M=5 \times 10^6$ samples using the D-wave spin reversal transform for values of $h^{\rm{in}}$ between $-1$ and $1$. We have chosen the confidence to be $\alpha = 0.997$ corresponding to a ``3$\sigma$" confidence level. 
The classical Boltzmann distribution for a single spin at thermal equilibrium and exposed to a magnetic field $h^{\rm{in}}$ is given by the celebrated formula,
\begin{align}
\mu_{\rm{classical}}(\sigma) = \frac{\exp\left(\beta (h^{\rm{in}}+h^{\rm{res}})\sigma\right)}{2\cosh{\left(\beta (h^{\rm{in}}+h^{\rm{res}})\right)}},
\label{eq:classical_single_spin_gibbs}
\end{align}
where $h^{\rm{res}}$ is a residual magnetic field independent of $h^{\rm{in}}$ and $\beta$ is the inverse temperature times the Boltzmann constant $\beta = 1/k_{B}T$. By comparing Eq.~\eqref{eq:classical_single_spin_distribution} with Eq.~\eqref{eq:classical_single_spin_gibbs}, we find that the relationship between $h^{\rm{in}}$ and $h^{\rm{out}}$ predicted by a classical Boltzmann distribution is linear
\begin{align}
h^{\rm{out}}_{\rm{classical}} := \beta (h^{\rm{in}}+h^{\rm{res}}).
\label{eq:classical_h_classic}
\end{align}
The classical parameters $\beta$ and $h^{\rm{res}}$ are estimated using the maximum log-likelihood approach for which the log-likelihood function takes the following form,
\begin{align}
\mathcal{L}  = \sum_{h^{\rm{in}}}  \tanh(h^{\rm{out}}_{\rm{measured}})
h^{\rm{out}}_{\rm{model}} + \frac{1}{2}\ln\left(1-\tanh\left(h^{\rm{out}}_{\rm{model}})\right)^2\right),
\label{eq:classical_likelihood_c}
\end{align}
where $h^{\rm{out}}_{\rm{measured}}$ are measurements of $h^{\rm{out}}$ for different values of $h^{\rm{in}}$ and $h^{\rm{out}}_{\rm{model}}$ describes the modeled relationship between the input and output fields, e.g. $h^{\rm{out}}_{\rm{classical}}$ from Eq.~\eqref{eq:classical_h_classic}. 

\begin{figure}
\centering
\includegraphics[width=0.84\linewidth]{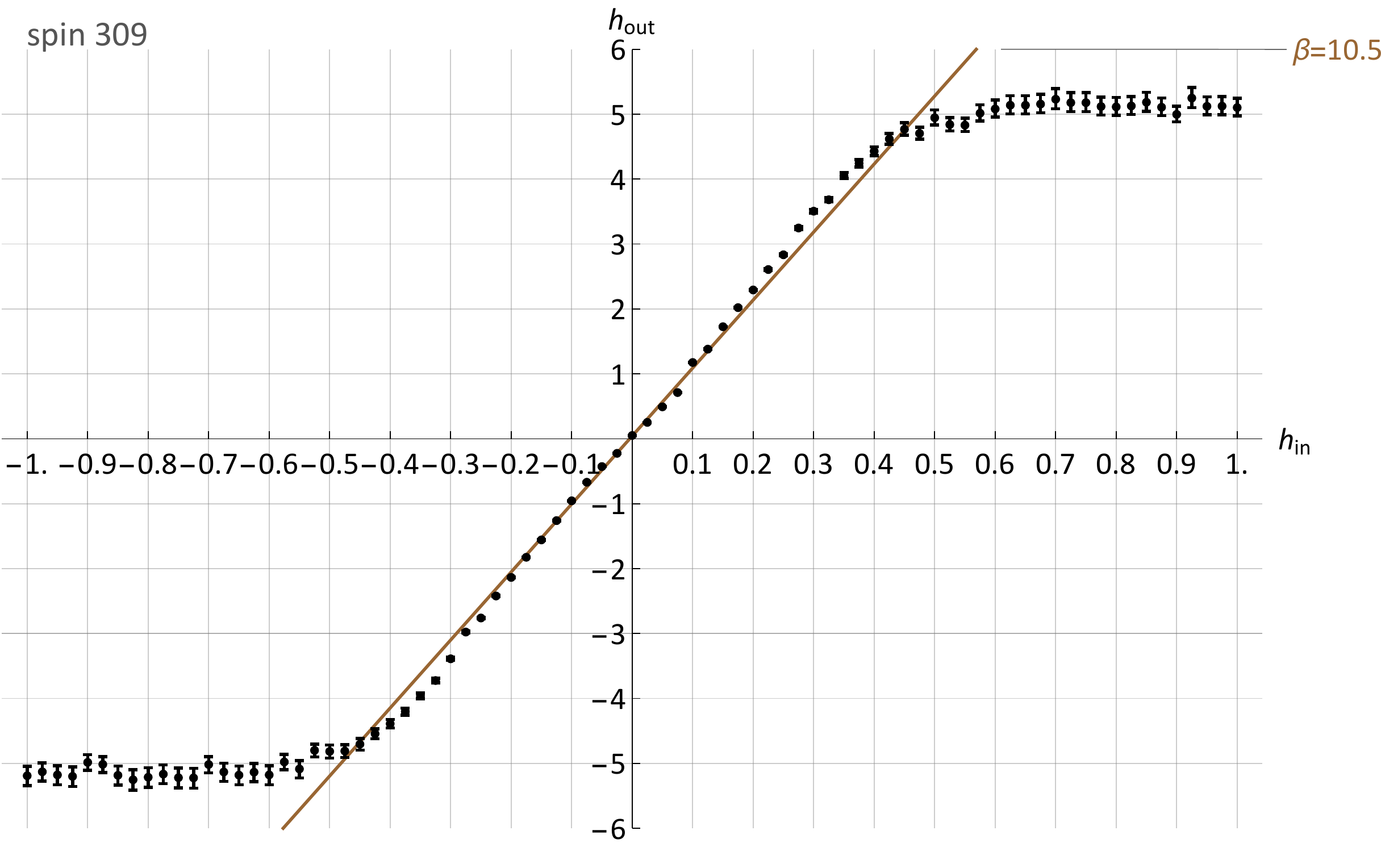}
\caption{Effective output magnetic fields $h^{\rm{out}}$ measured for different input magnetic fields $h^{\rm{in}}$ for the spin \#$309$. Each point is estimated using $M=5\times10^6$ samples with a confidence level of $\alpha=0.997$. Notice that the confidence intervals for large value of $h^{\rm{out}}$ are significantly bigger than for small value of $h^{\rm{out}}$ as the probability distribution scales exponentially with the magnitude of $h^{\rm{out}}$. The classical relationship between $h^{\rm{in}}$ and $h^{\rm{out}}$ found with the maximum log-likelihood approach is displayed in brown. The effective inverse temperature for this model is $\beta=10.5$ and the residual field is $h^{\rm{res}}=0.004$.
}
\label{fig:classical_raw_spin}
\end{figure}

Measurements of $h^{\rm{out}}$ for different values of $h^{\rm{in}}$ using the aforementioned procedure on spin \#$309$ and the classical relationship between $h^{\rm{in}}$ and $h^{\rm{out}}$ are depicted in Fig.~\ref{fig:classical_raw_spin}.
We clearly see that if the classical relationship holds for small values of $h^{\rm{in}}$ it fails to explain the flat tails of the measurements for which $|h^{\rm{in}}| \gtrapprox 0.5$. In fact, we starts to already see a deviation from the linear curve for magnitudes of the input field between $0.5 \gtrapprox |h^{\rm{in}}| \gtrapprox 0.2$. In the forthcoming analysis, we will show that the first behavior can be explained by quantum effects, and the latter by classical noise in the input field.

A quantum statistical description of a one spin system is realized through the so-called density matrix formalism. We consider a quantum spin exposed on the $z$-axis to a magnetic field $h^{\rm{in}}$ and residual field $h^{\rm{res}}$ and exposed on the $x$-axis to a transverse field $h^{\rm{trans}}$. This is a natural assumption as the Hamiltonian realized during the considered quantum annealing process is composed of these two terms \cite{johnson2011quantum}. The density matrix describing this one spin system at thermal equilibrium is the following object,

\begin{align}
\rho &= \frac{\exp\left( \beta( h^{\rm{in}} + h^{\rm{res}}) \widehat{\sigma}_z + \beta h^{\rm{trans}} \widehat{\sigma}_x\right)}{\Tr{\exp\left( \beta( h^{\rm{in}} + h^{\rm{res}}) \widehat{\sigma}_z + \beta h^{\rm{trans}} \widehat{\sigma}_x\right)}} \nonumber\\
&= \frac{1}{2}\left(I + \frac{\tanh\left( \beta \sqrt{(h^{\rm{in}}+ h^{\rm{res}})^2 +(h^{\rm{trans}})^2}\right)}{\sqrt{(h^{\rm{in}}+ h^{\rm{res}})^2 +(h^{\rm{trans}})^2}}\left((h^{\rm{in}}+ h^{\rm{res}})\widehat{\sigma}_z + h^{\rm{trans}}\widehat{\sigma}_x\right)\right),
\label{eq:classical_density_matrix}
\end{align}
where $\widehat{\sigma}_z$ and $\widehat{\sigma}_x$ are the usual Pauli matrices for the $z$ and $x$ axis respectively. The transverse field $h^{\rm{trans}}$ that appears in Eq.~\eqref{eq:classical_density_matrix} is in general a function of $h^{\rm{in}}$. Due to an observation that the experimental points seem to flatten out for large values of $h^{\rm{in}}$ in Fig.~\ref{fig:classical_raw_spin}, we choose to parametrize the transverse fields as linear transformations of the input field, namely $h^{\rm{trans}} = \xi h^{\rm{in}}$, a dependence which is consistent with a saturation in the $h^{\rm{out}}$ response.

The mean value of observing the quantum spin along the $z$-axis is given by $\Tr{\rho \widehat{\sigma}_z}$. From this relationship we deduce that the probability of observing the system taking the value $\sigma \in \{-1,1\}$ is given by the following probability distribution,
\begin{align}
\mu_{\rm{quantum}}(\sigma) = \frac{1}{2}\left(1 + \sigma (h^{\rm{in}}+ h^{\rm{res}})\frac{\tanh\left( \beta \sqrt{(h^{\rm{in}}+ h^{\rm{res}})^2 +(\xi h^{\rm{in}})^2}\right)}{\sqrt{(h^{\rm{in}}+ h^{\rm{res}})^2 +(\xi h^{\rm{in}})^2}}\right).
\label{eq:classical_single_spin_quantum}
\end{align}

With these assumptions we find that that the relationship between $h^{\rm{out}}$ and $h^{\rm{in}}$ in the quantum case can be described as follows,
\begin{align}
h^{\rm{out}}_{\rm{quantum}} := \arctanh \left((h^{\rm{in}}+ h^{\rm{res}})\frac{\tanh\left( \beta \sqrt{(h^{\rm{in}}+ h^{\rm{res}})^2 +(\xi h^{\rm{in}})^2}\right)}{\sqrt{(h^{\rm{in}}+ h^{\rm{res}})^2 +(\xi h^{\rm{in}})^2}}\right).
\label{eq:classical_h_quantum}
\end{align}
Notice that in the limit where $\xi=0$, the quantum predictions from Eq.~\eqref{eq:classical_h_quantum} converges to its classical counterpart from Eq.~\eqref{eq:classical_h_classic}.

As we will see later in Section~\ref{sec:local_field_variability}, the variability in the residual field $h^{\rm{res}}$ is non-negligible and in fact plays a key role in the explanation of the spurious link behaviors, see Section~\ref{sec:noise_and_spurious_links} for a detailed explanation.
Assuming that $h^{\rm{res}}$ is a random variable with probability density function $f(h^{\rm{res}})$, we can obtain a noisy quantum description of the measurement outcomes by applying Bayes's rule to Eq.~\ref{eq:classical_h_quantum},
\begin{align}
\mu_{\rm{qnoise}} (\sigma) = \int dh^{\rm{res}} f(h^{\rm{res}}) \mu_{\rm{quantum}}(\sigma).
\label{eq:classical_qnoise_measure}
\end{align}
We will see later that in the regime of noise relevant to our experiments, the precise knowledge about the shape of the distribution $f(h^{\rm{res}})$ plays little role. The important quantities are the mean $\mathbb{E}_f[h^{\rm{res}}] = h^{\rm{res}}_0$ and the standard deviation $\sqrt{\mathbb{E}_f[\left( h^{\rm{res}} - h^{\rm{res}}_0\right)]} = h^{\rm{res}}_{\rm{sd}}$. For simplicity, we choose $f$ to be a binomial distribution $f(h^{\rm{res}}) = \frac{1}{2}\delta(h^{\rm{res}} - h^{\rm{res}}_0 - h^{\rm{res}}_{\rm{sd}}) + \frac{1}{2}\delta(h^{\rm{res}} - h^{\rm{res}}_0 + h^{\rm{res}}_{\rm{sd}})$, where $\delta$ denotes the Dirac distribution. Combining Eq.~\eqref{eq:classical_single_spin_quantum} with Eq.~\eqref{eq:classical_qnoise_measure}, we can write down the following relationship between $h^{\rm{in}}$ and $h^{\rm{out}}$ in the noisy quantum case,
\begin{align}
h^{\rm{out}}_{\rm{qnoise}} := \arctanh &\left( (h^{\rm{in}} + h^{\rm{res}}_0 + h^{\rm{res}}_{\rm{sd}}) \frac{\tanh \left( \beta \sqrt{(h^{\rm{in}}+ h^{\rm{res}}_0 + h^{\rm{res}}_{\rm{sd}})^2 +(\xi h^{\rm{in}})^2}\right)}{2\sqrt{(h^{\rm{in}}+ h^{\rm{res}}_0 + h^{\rm{res}}_{\rm{sd}})^2 +(\xi h^{\rm{in}})^2}} \right.\nonumber\\
& \left. + (h^{\rm{in}} + h^{\rm{res}}_0 - h^{\rm{res}}_{\rm{sd}}) \frac{\tanh \left( \beta \sqrt{(h^{\rm{in}}+ h^{\rm{res}}_0 - h^{\rm{res}}_{\rm{sd}})^2 +(\xi h^{\rm{in}})^2}\right)}{2\sqrt{(h^{\rm{in}}+ h^{\rm{res}}_0 - h^{\rm{res}}_{\rm{sd}})^2 +(\xi h^{\rm{in}})^2}}\right).
\label{eq:classical_h_quantum_noise}
\end{align}
Note that for $h^{\rm{res}}_0 = 0$, the first order expansion of Eq.~\eqref{eq:classical_h_quantum_noise} for small values of $h^{\rm{in}}$ yields the simple relationship $h^{\rm{out}}_{\rm{qnoise}} \approx \beta h^{\rm{in}} \cosh(\beta h^{\rm{res}}_{\rm{sd}})^{-2}$. This shows that one effect of noise on the system consists in reducing the effective inverse temperature that one could infer from a classical (i.e. linear) regression for small values of $h^{\rm{in}}$.

\begin{figure}[!htb]
\centering
\includegraphics[width=0.84\linewidth]{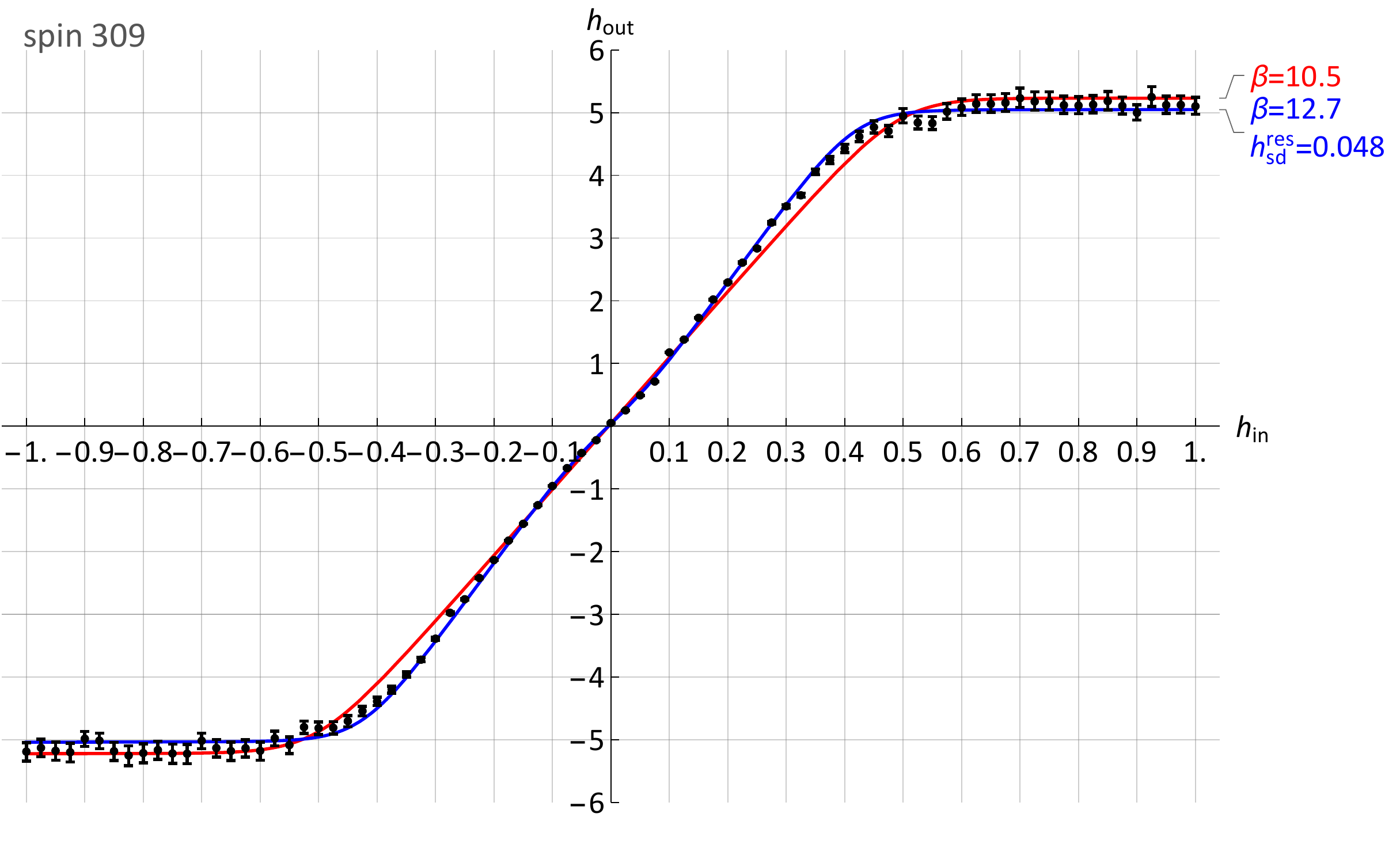}
\caption{Effective output magnetic fields $h^{\rm{out}}$ measured for different input magnetic fields $h^{\rm{in}}$ for the spin \#$309$. Each point is estimated using $M=5\times10^6$ samples with a confidence level of $\alpha=0.997$. The noiseless quantum and the noisy quantum relationships between $h^{\rm{in}}$ and $h^{\rm{out}}$ found with the maximum log-likelihood approach are displayed in red and blue respectively. The effective inverse temperature for the noisy quantum model is higher than the effective inverse temperature found with the noiseless quantum model. The noiseless quantum model and the classical models find similar temperatures. }
\label{fig:classical_noise_quantum}
\end{figure}

We infer the parameters of the proposed quantum models, i.e. $h^{\rm{res}}_0$, $\xi $ and $h^{\rm{res}}_{\rm{sd}}$, with the maximum log-likelihood approach described by Eq.~\eqref{eq:classical_likelihood_c}, replacing the quantity $h^{\rm{out}}_{\rm{model}}$ by either the functional relationship $h^{\rm{out}}_{\rm{quantum}}$ from Eq.~\eqref{eq:classical_h_quantum} or $h^{\rm{out}}_{\rm{qnoise}}$ from Eq.~\eqref{eq:classical_h_quantum_noise}.

In Fig.~\ref{fig:classical_noise_quantum}, we show the measurements acquired for spin \#$309$ along with the curves predicted by the noiseless and noisy quantum models. The flattening of the output field response for $|h^{\rm{in}}|\gtrapprox 0.5$ is correctly accounted for by the addition of the quantum transverse field. The initial deviation from the linear response for values of the input field between $0.5 \gtrapprox|h^{\rm{in}}|\gtrapprox 0.2$ are predicted accurately only by the noisy quantum model. An important feature highlighted by this study is the effect of the noise on the effective inverse temperature (or linear response) that one may infer from the data. We see from our models that the noise in the residual field effectively lowers the inverse temperature for small values of the input field of order $|h^{\rm{in}}|\lessapprox 0.2$. Indeed, the effects due to noise will be more prominent when the input field is small. In this particular illustration, the noise on the residual field accounts for more than $20\%$ of the input field for magnitudes of  $|h^{\rm{in}}|\leq 0.2$.

\begin{figure}[!htb]
\centering
\includegraphics[width=0.84\linewidth]{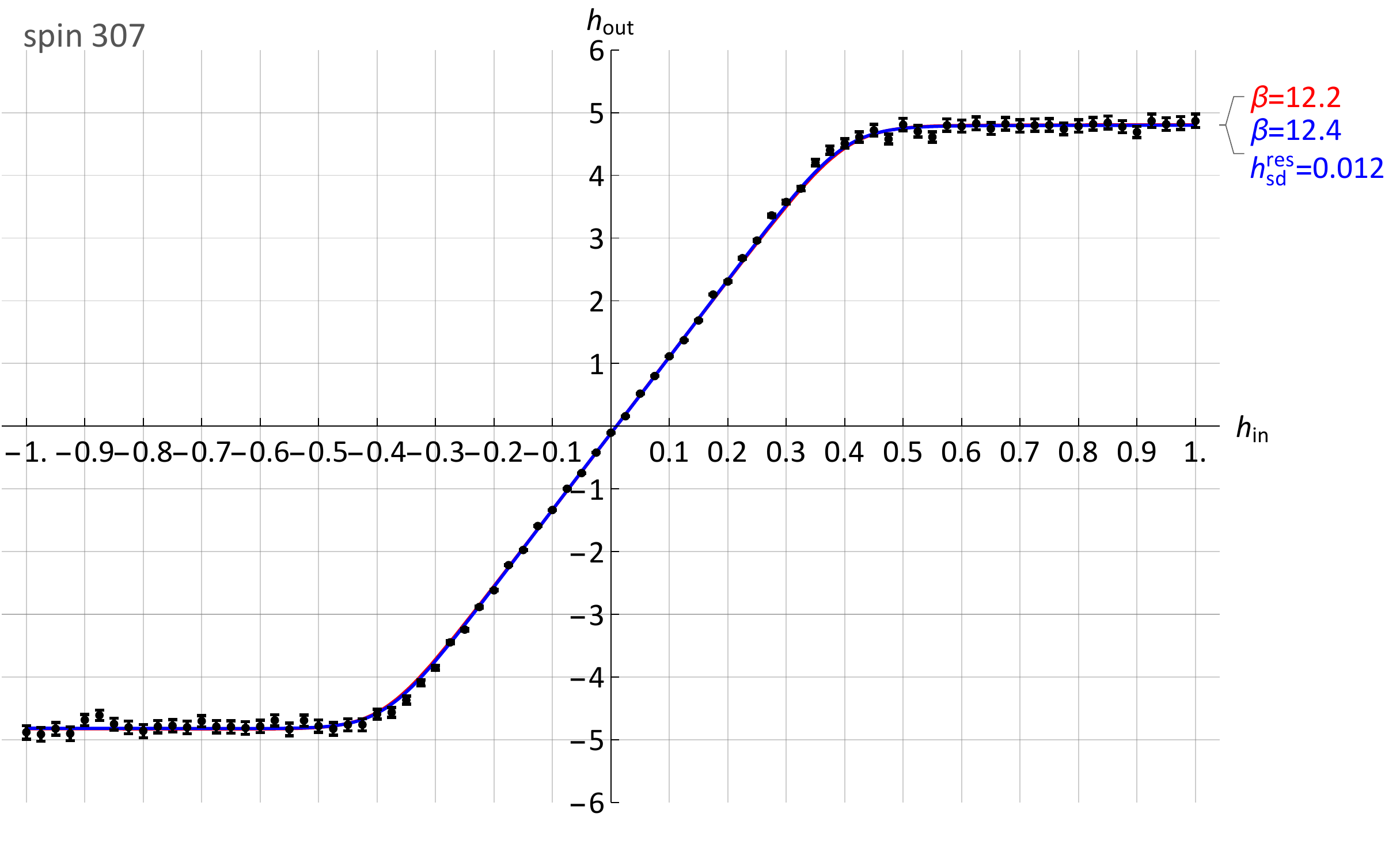}
\caption{Effective output magnetic fields $h^{\rm{out}}$ measured for different input magnetic fields $h^{\rm{in}}$ for the spin \#$307$. Each point is estimated using $M=5\times10^6$ samples with a confidence level of $\alpha=0.997$. The noiseless quantum and the noisy quantum relationships between $h^{\rm{in}}$ and $h^{\rm{out}}$ found with the maximum log-likelihood approach are displayed in red and blue respectively. This particular spin displays a low-level of residual field noise. As a consequence, the noiseless and noisy quantum models find the same inverse temperature.}
\label{fig:classical_noiseless_quantum}
\end{figure}

In Fig.~\ref{fig:classical_noiseless_quantum}, we show similar measurements collected for spin \#$307$ as long as the predictions of the proposed quantum models. Among all spins that we tested, we found that spin \#$307$ has the lowest level of residual field noise. In this particular case, we see that both noiseless and noisy quantum model predictions agree and we find an inverse temperatures consistent with the inverse temperature found for spin \#$309$ with the noisy quantum model.

For the unit cell of interest to this work, we show the regression coefficients obtained with the classical model, the noiseless quantum model and the noisy quantum model in Table~\ref{tbl:classical_reg_coeff}.

\begin{table*}
    \centering
    \begin{tabular}{|c||c|c|c|}
    \hline
    Spin  & Classical & Quantum & Noisy + Quantum \\
       \# & $(\beta, h^{\rm{res}}_0)$ & $(\beta, h^{\rm{res}}_0, \xi)$ & $(\beta, h^{\rm{res}}_0, \xi, h^{\rm{res}}_{\rm{sd}})$ \\
    \hline
    \hline
    304 & (11.3, 0.014) & (11.4, 0.014, 0.017) & (12.3, 0.014, 0.018, 0.029)  \\
    \hline
    305 &  (11.1, 0.003) & (11.2, 0.003, 0.016) & (13.1, 0.003, 0.017, 0.041) \\
    \hline
    306 & (11.4, -0.004) & (11.5, -0.004, 0.016) & (13.3, -0.004, 0.018, 0.039)  \\
    \hline
    307 & (12.1, -0.009)  & (12.2, -0.009, 0.016) & (12.4, -0.009, 0.016, 0.012)  \\
    \hline
    308 & (11.6, -0.005)  & (11.7, -0.005, 0.012) & (12.9, -0.005, 0.013, 0.032)  \\
    \hline
    309 & (10.5, 0.004) & (10.5, 0.004, 0.011) & (12.7, 0.004, 0.013, 0.048)  \\
    \hline
    310 & (11.2, -0.006) & (11.3, -0.006, 0.012) & (12.5, -0.006, 0.013, 0.035)  \\
    \hline
    311 & (11.4, 0.010) & (11.5, 0.010, 0.012) & (12.5, 0.010, 0.013, 0.031) \\
    \hline
    \end{tabular}
    \caption{List of regression coefficients for eight different spins found with the classical model, the noiseless quantum model and the noisy quantum model. The coefficients are obtained through the maximization of the log-likelihood function respective to each model. The inverse temperature found by the classical and noiseless quantum models differs from the inverse temperature obtained with the noisy quantum model for significant noise magnitudes. The other parameters are consistent between models which shows that these models form a hierarchy of increasing descriptive complexity. The transverse field responses for spins \#304 to \#307 differs significantly from the transverse field responses of spins \#308 to \#311. This suggest a difference in the hardware implementation of these two groups of spins for they are physically located on the two different sides of a chimera cell.}
    \label{tbl:classical_reg_coeff}
\end{table*}

\begin{figure}[!htb]
\centering
\includegraphics[width=0.84\linewidth]{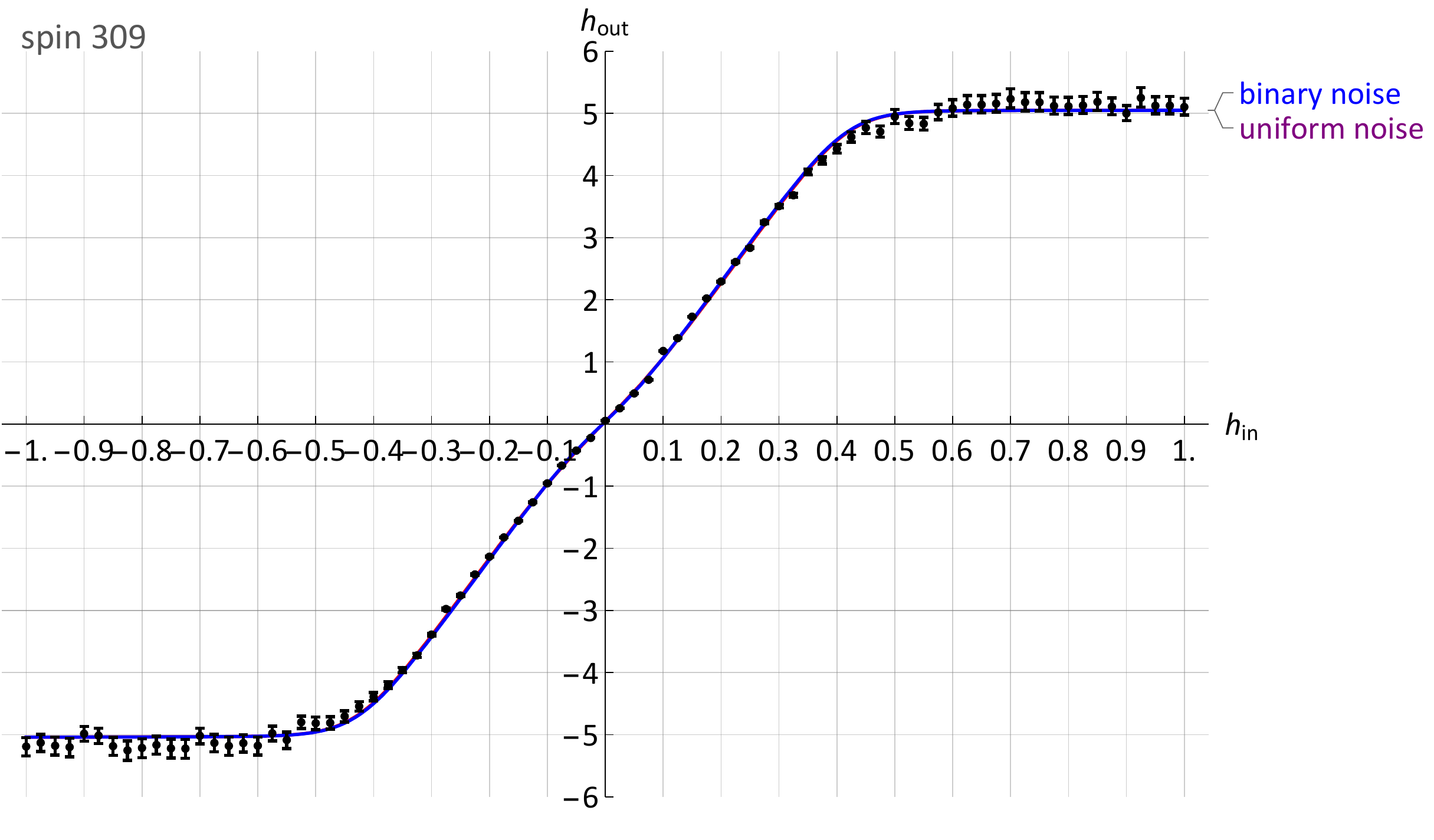}
\caption{Comparison of the effect of different noise probability distributions for noisy quantum models on spin \#$309$. Predictions obtained from noisy quantum models derived from a binary distribution and a uniform distribution are displayed in blue and purple respectively. The parameters for both models are identical and corresponds to the parameters inferred for the binary noise model. The two models are indistinguishable for this level of noise magnitude, indicating that the main contributing factors in the probabilistic description of the noise are the first and the second moments.}
\label{fig:classical_noise_comparison}
\end{figure}

Finally, we study the effect of the noise probability distribution on the model. We compare our noisy quantum model derived from a binary probability distribution with the noisy quantum model that one obtains with a uniform noise distribution, i.e.  $f(h^{\rm{res}}) = (2\sqrt{3} h^{\rm{res}}_{\rm{sd}})^{-1}$ for $h^{\rm{res}} \in \left[h^{\rm{res}}_0 - \sqrt{3} h^{\rm{res}}_{\rm{sd}}, h^{\rm{res}}_0 + \sqrt{3} h^{\rm{res}}_{\rm{sd}}\right]$ and zero otherwise. Predictions for the spin \#$309$ made by these two models of noise are displayed in Fig.~\ref{fig:classical_noise_comparison}. The parameters for both models are identical and correspond to the parameters inferred with maximum log-likelihood for the binary probability distribution. We see that for noise magnitudes of $h^{\rm{res}}_{\rm{sd}} =  0.048$ the predictions are practically indistinguishable.

\section{Learning of General Classical Distributions on Binary Variables}\label{sec:learning_general_distributions}

An arbitrary positive probability distribution on $N$ classical binary variables $\underline{\sigma} \in \{ -1, +1 \}^N$ can be represented in the the form of a Gibbs distribution with different interaction orders:
\begin{align}
    \mu(\underline{\sigma}) & = \frac{1}{Z} \exp{ \left( \underbrace{\sum_{i} h_i \sigma_i}_{\text{1st order}} + \underbrace{\sum_{ij} J_{ij} \sigma_i \sigma_j}_{\text{2d order}} + \underbrace{\sum_{ijk} J_{ijk} \sigma_i \sigma_j \sigma_k}_{\text{3d order}} + \underbrace{\sum_{ijkl} J_{ijkl} \sigma_i \sigma_j \sigma_k \sigma_l}_{\text{4th order}} + \cdots \right)},
    \label{eq:Gibbs_distribution}
\end{align}
where $Z$ denotes the normalization factor known as parition function. In general, the number of terms can be exponential. However, physical systems are typically characterized by a finite number of short-ranged multi-body interactions. For instance, the renowned Ising model corresponds to the case where only first and second orders are present. In this section, we develop an exact learning method that allows one to reconstruct an arbitrary probability distribution on binary variables, i.e. recover the parameters $\{h_i, J_{ij}, J_{ijk}, J_{ijkl}, \ldots \}$ from a number of independent spin configurations sampled from the distribution \eqref{eq:Gibbs_distribution}. For the case of Ising models with pairwise interactions only
\begin{align}
    \mu(\underline{\sigma}) = \frac{1}{Z} \exp{ \left( \sum_{i} h_i \sigma_i + \sum_{ij} J_{ij} \sigma_i \sigma_j \right)},
    \label{eq:Ising_distribution}
\end{align}
this reconstruction problem is known as \emph{inverse Ising problem}, and has been extensively studied in the past with a number of heuristic methods \cite{nguyen2017inverse}. However, the inverse Ising problem has been solved only recently, with the appearance of exact algorithms that showed that parameters of an arbitrary Ising model, including low-temperature and spin-glass models, can be recovered to an arbitrary precision with an appropriate number of samples \cite{bresler2015efficiently, vuffray2016interaction}. The state-of-the-art near-optimal performance for inverse Ising problem has been recently achieved with the estimator based on the Interaction Screening Objective (ISO) \cite{lokhov2018optimal}. For Ising model, the ISO reads
\begin{equation}
    S_i(\mathbf{J_i},h_i) = \left\langle \exp{ \left( - h_i \sigma_i - \sum_{j} J_{ij} \sigma_i \sigma_j \right)} \right\rangle_M,
    \label{eq:ISO_Ising}
\end{equation}
where $\langle f(\sigma) \rangle_M = M^{-1} \sum_{m=1}^M f(\sigma^{(m)})$ denotes the empirical average over $M$ independent samples, and $\mathbf{J_i}$ denotes the set of couplings adjacent to node $i$, i.e. $\mathbf{J_i} = \{ J_{ij} \}_{j \neq i}$ for Ising models. It is easy to see that the ISO in \eqref{eq:ISO_Ising} is a convex function of parameters $\mathbf{J_i},h_i$. Furthermore, the unique minimizer of the ISO
\begin{equation}
    (\widehat{\mathbf{J}_i},\widehat{h_i}) = \argmin S_i(\mathbf{J_i},h_i)
    \label{eq:ISO_estimation}
\end{equation}
converges to the true parameters of the distribution in the limit of a large number of samples, and yields an $O(1/\sqrt{M})$ error on the recovered model parameters for finite $M$ \cite{vuffray2016interaction}. The ISO is a local estimator, i.e. it is defined for each spin, and only involves couplings adjacent to this spin, so for reconstructing the entire model one needs to run $N$ parallel reconstruction problems \eqref{eq:ISO_estimation}. Notice that this procedure yields two estimations for the same coupling, $\widehat{J_{ij}}$ and $\widehat{J_{ji}}$, and we use their arithmetic mean $(\widehat{J_{ij}} + \widehat{J_{ji}})/2$ as the final estimate of the coupling $J_{ij}$. This estimator is used throughout the paper for reconstructing parameters of Ising models, where the required number of samples for a given precision and the associated expected confidence interval are obtained through synthetic numerical experiments, as explained below.

Here, we generalize this objective function to the case of general Gibbs distributions with multi-body interactions of the type \eqref{eq:Gibbs_distribution}. The corresponding ISO reads
\begin{equation}
    S_i(\mathbf{J_i},h_i) = \left\langle \exp{ \left( - h_i \sigma_i - \sum_{j} J_{ij} \sigma_i \sigma_j - \sum_{jk} J_{jkl} \sigma_i \sigma_j \sigma_k - \sum_{jkl} J_{jkl} \sigma_i \sigma_j \sigma_k \sigma_l - \cdots \right)} \right\rangle_M.
    \label{eq:ISO_General}
\end{equation}
Similarly to the Interaction Screening estimator for the inverse Ising problem \eqref{eq:ISO_Ising}, it is easy to see that the ISO for general models is a convex function of parameters $\mathbf{J_i},h_i$. Let us present a simple argument that illustrates the fact that in the limit of large number of samples the unique minimizer \eqref{eq:ISO_estimation} of the convex ISO objective \eqref{eq:ISO_General} is achieved at $(\widehat{\mathbf{J}_i},\widehat{h_i}) = (\mathbf{J}_i, h_i)$, meaning that the true interactions present in the model are fully ``screened''. Indeed, the ISO is an empirical average of the inverse of the factors in the Gibbs measure; if
\begin{equation}
    \mathcal{F}_i(\mathbf{J}_i, h_i) = \exp( h_i \sigma_i + \sum_{j} J_{ij} \sigma_i \sigma_j + \sum_{jk} J_{jkl} \sigma_i \sigma_j \sigma_k + \sum_{jkl} J_{jkl} \sigma_i \sigma_j \sigma_k \sigma_l + \cdots),
\end{equation}
then $S_i(\mathbf{J}_i,h_i) = \langle \mathcal{F}_i^{-1}(\mathbf{J}_i,h_i) \rangle_{M}$. In the limit of large number of samples $S_i(\mathbf{J}_i,h_i) \rightarrow S_i^*(\mathbf{J}_i,h_i) = \langle 1/\mathcal{F}_i(\mathbf{J}_i,h_i) \rangle$, where $\langle \cdots \rangle$ denotes the average over the measure \eqref{eq:Gibbs_distribution}. Let us look at the derivative of the ISO with respect to a given coupling, say $J_{ijk}$. This derivative corresponds to weighted three-body correlation, for instance $\partial S_i^{*} / \partial J_{ijk} = \langle\sigma_i \sigma_j \sigma_k / \mathcal{F}_i(\mathbf{J}_i,h_i)\rangle$, and this reflect the key property of the Interaction Screening based estimator. When $(\widehat{\mathbf{J}_i},\widehat{h_i}) = (\mathbf{J}_i, h_i)$, $\partial S_i^{*} / \partial J_{ij} \vert_{\mathbf{J}_i,h_i} = 0$, meaning that the minimum of ISO is achieved at $(\widehat{\mathbf{J}_i},\widehat{h_i}) = (\mathbf{J}_i, h_i)$ as $M \to \infty$.

Again, similarly to the case of Ising models, after running $N$ parallel local reconstructions, the resulting couplings $\mathbf{J_i} = \{ J_{ij}, J_{ijk}, J_{ijkl}, \ldots \}$ are symmetrized using all permutations of tuples $(i, j, k, l, \ldots)$. This estimator is used for probing general distributions with multi-body interactions in the next section.

It is important to highlight that the estimator given by the generalized ISO \eqref{eq:ISO_General} is exact: With a given number of samples $M$, the deviation between reconstructed $(\widehat{\mathbf{J}_i},\widehat{h_i})$ and true $(\mathbf{J}_i, h_i)$ model parameters decay as $\sim 1/\sqrt{M}$. Next, we describe an empirical procedure that we have developed to estimate the reconstruction fidelity in practice, which dictate how much data we needed in all experiments to statistically exclude the finite-sample considerations from all conclusions that we draw throughout the work.

\section{Empirical Estimation of Reconstruction Errors}
\label{sec:empirical_error_estimation}

For finite number of samples $M$, couplings estimated via \eqref{eq:ISO_estimation} are accurate up to a certain error that decays with $M$. This error can be estimated theoretically, however the resulting worst-case bounds can be loose for a given model. To the best of our knowledge, there is no standard approach to tightly quantify the finite sampling error of learning models of the form \eqref{eq:Gibbs_distribution} in practice. To address this challenge, we propose the following empirical error estimation procedure for a given model with specific parameters and a fixed number of samples. Specifically, given a black-box sample generator $B$, a finite sample set $M$, and a replicate parameter $R$, we conduct the following procedure:
\begin{enumerate}
    \item collect $M$ samples of the black-box $B$;
    \item reconstruct model $m$ from the collected samples;
    \item for $r$ from $1$ to $R$ independent replicates do the following:
    \begin{enumerate}
        \item collect $M$ synthetic samples from $m$ using an auxiliary sampling algorithm;
        \item reconstruct model $m_r$ from the collected samples;
    \end{enumerate}
    \item compute statistics over the parameters reconstructed across the reconstructed $\{m_r\}_{r \in 1,\ldots,R}$ models.
\end{enumerate}
In this work the black-box sampler $B$ is given by our \texttt{DW\_2000Q\_LANL} QPU, and the sampling algorithm is a brute-force approach that enumerates every possible state and computes the exact probably of each state. The this brute-force approach is feasible for small number of spins which will be the focus of our targeted experiments. For larger problems one could utilize more scalable sampling algorithms, including those based on Markov-Chain Monte-Carlo techniques, or Belief Propagation with decimation.

The last step will provide the information on the typical variability of reconstruction accuracy due the effect of finite samples. Given a set of $R$ replicates of the models of the type \eqref{eq:Gibbs_distribution} reconstructed from $R$ independent sets of samples. Let us define $(\mathbf{h}^r, \mathbf{J}^r) \;\; \forall r \in R$ as parameters of the models of the type \eqref{eq:Gibbs_distribution} learned from each of the $R$ sets of samples. We also define $(\delta \mathbf{h}, \delta \mathbf{J})$ representing deviations from the parameters $(\mathbf{h}, \mathbf{J})$ reference model that was used to produce synthetic samples:
\begin{align}
    \delta \mathbf{h}^r = \mathbf{h} - \mathbf{h}^r \;\; \forall r \in R \\
    \delta \mathbf{J}^r = \mathbf{J} - \mathbf{J}^r \;\; \forall r \in R
\end{align}
In the step 4 of the procedure above, we estimate the empirical mean and standard deviation on each of the parameters in the set $h_i, \mathbf{J}_i$ from $R$ values. For a sufficiently large $R$, these quantities indicate an expected scale of error coming from model recovery for a given number of samples.

\section{Probing of Multi-Body Interactions}
\label{sec:multi-body}

The objective of experiment in this section is to leverage the ISO for general distributions on binary variables \eqref{eq:ISO_General} to determine the class of models that adequately describes the output distribution produced by the \texttt{DW\_2000Q\_LANL} QPU. Although the target Hamiltonian has a form of a classical Ising model, \emph{a priori} this distribution could be more general than the Ising Gibbs distribution \eqref{eq:Ising_distribution}. In this section, we probe the existence of multi-body interactions beyond pairwise in the energy function of the output distribution. We conduct an experiment on a single chimera graph cell involving 8 qubits (see Table \ref{tbl:l2k-qpu-used}). As explained in the previous section, for an 8-spin model, the most general positive probability distribution can be written in the form \eqref{eq:Gibbs_distribution} with an energy function being a polynomial of order eight. Specifically, for $\underline{\sigma} \in \{-1, +1\}^8$, the possible distribution reads
\begin{align}
    \mu(\underline{\sigma}) = \frac{1}{Z} \exp{ \left( \sum_{i} h_i \sigma_i + \sum_{ij} J_{ij} \sigma_i \sigma_j + \sum_{ijk} J_{ijk} \sigma_i \sigma_j \sigma_k + \cdots + J_{ijklmnop} \sigma_i \sigma_j \sigma_k \sigma_l \sigma_m \sigma_n \sigma_o \sigma_p \right) }.
    \label{eq:ho_boltzmann}
\end{align}

This model has a total of 8 $h$ parameters 247 $J$ parameters.
%
Our goal will be to practically show existence or absence of multi-body interactions in the output distribution. Presence of interactions can be established if the reconstructed couplings are statistically significant, i.e. they are larger in absolute value than the reconstruction error resulting from a finite-sample reconstruction. We show that a few million D-Wave samples will be sufficient to provide an accurate reconstruction of model parameters.

\begin{table}
    \centering
    \begin{tabular}{|r||r|r|r|r|r|}
    \hline
    Model & Qubits & Couplers & $h^{\text{in}}$ & $J^{\text{in}}$ & Samples \\
    \hline
    \hline
    ferromagnet & $V$ & $E$ & 0.0 & 0.025 & 10,000,000 \\
    \hline
    anti-ferromagnet & $V$ & $E$ & 0.0 & -0.025 & 10,000,000 \\
    \hline
    \end{tabular}
    \caption{The input parameters sent to the \texttt{DW\_2000Q\_LANL} QPU to collect data for the 8-th order distribution reconstruction experiment.}
    \label{tbl:8o-models}
\end{table}
The experiment we conduct here focuses on learning 8-th order models from the samples output by the D-Wave hardware on two canonical models, a ferromagnet and an anti-ferromagnet.  The parameter details of these two models are presented in Table \ref{tbl:8o-models}. The primary objective of this experiment is to determine what model parameters are statistically significant. Specifically, what learned parameters can-and-cannot be attributed to artifacts from to finite sampling and the model reconstruction algorithm. Through a procedure that was detailed in section \ref{sec:empirical_error_estimation}, a number of supporting simulations are conducted to determine error bounds on the reconstructed model parameters. Leveraging the obtained error value, we determine the three standard deviations threshold that determine the statistical significance of the reconstructed values. Recovered couplings with absolute values above this threshold are very unlikely to be due to a reconstruction error, while values below can be explained by the finite sample noise in the reconstruction process.

\begin{figure}
\centering
\includegraphics[width=0.39\linewidth,page=6]{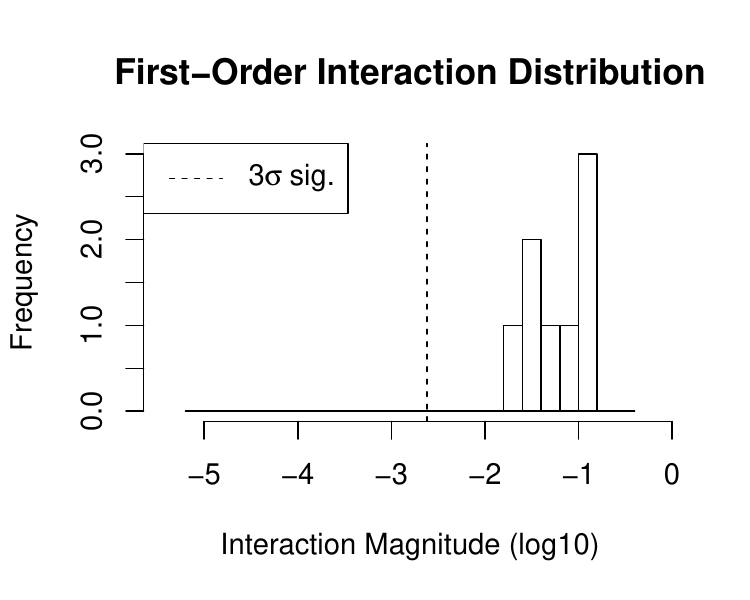}
\includegraphics[width=0.39\linewidth,page=7]{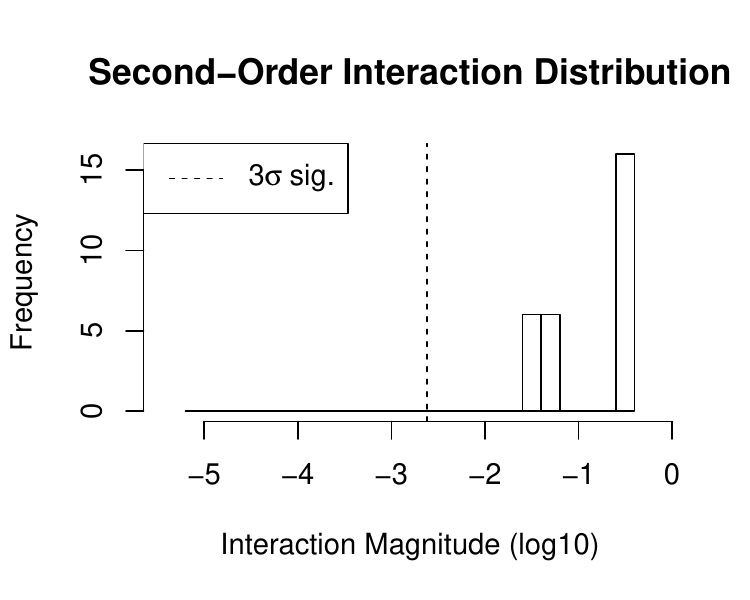} \\
\vspace{-0.3cm}
\includegraphics[width=0.39\linewidth,page=8]{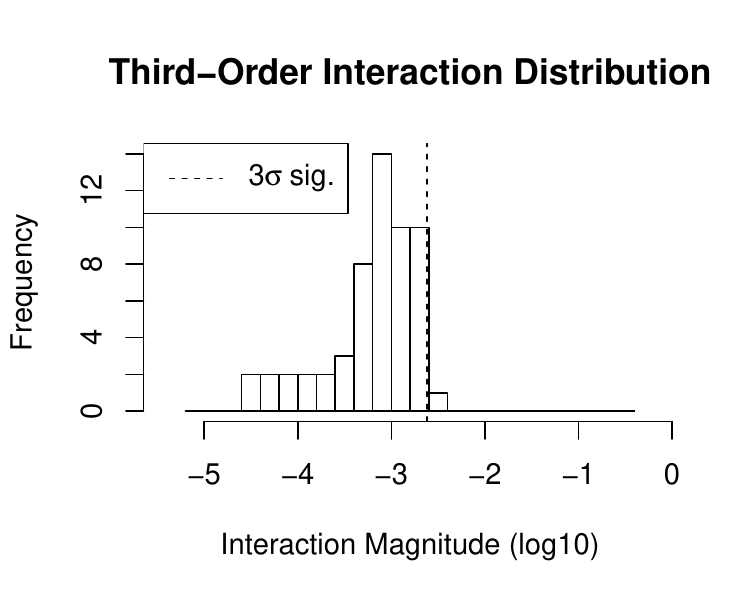}
\includegraphics[width=0.39\linewidth,page=9]{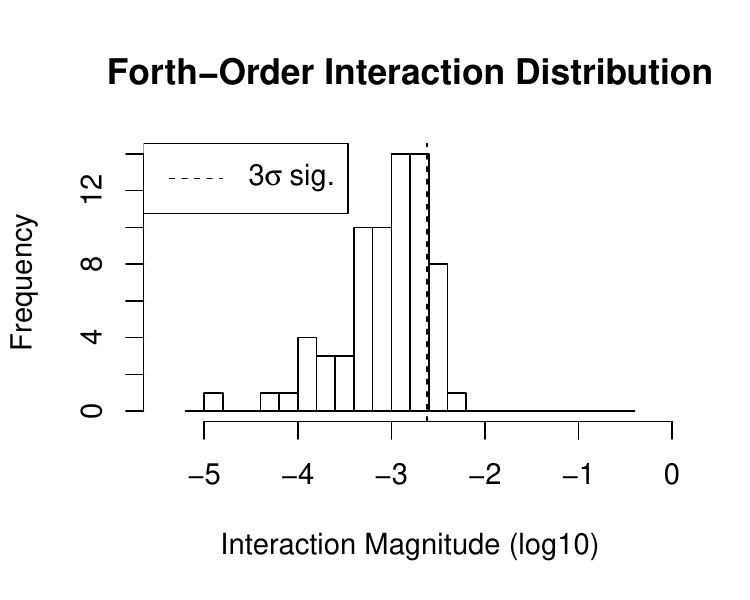} \\
\vspace{-0.3cm}
\includegraphics[width=0.39\linewidth,page=10]{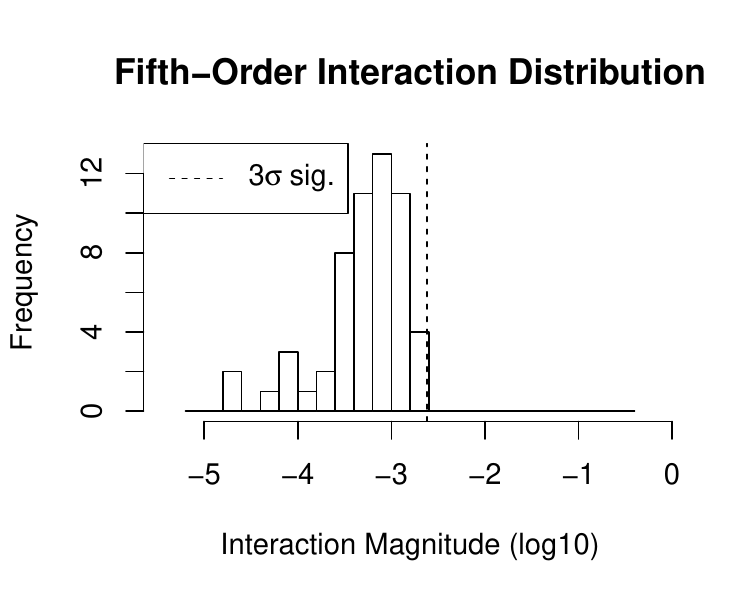}
\includegraphics[width=0.39\linewidth,page=11]{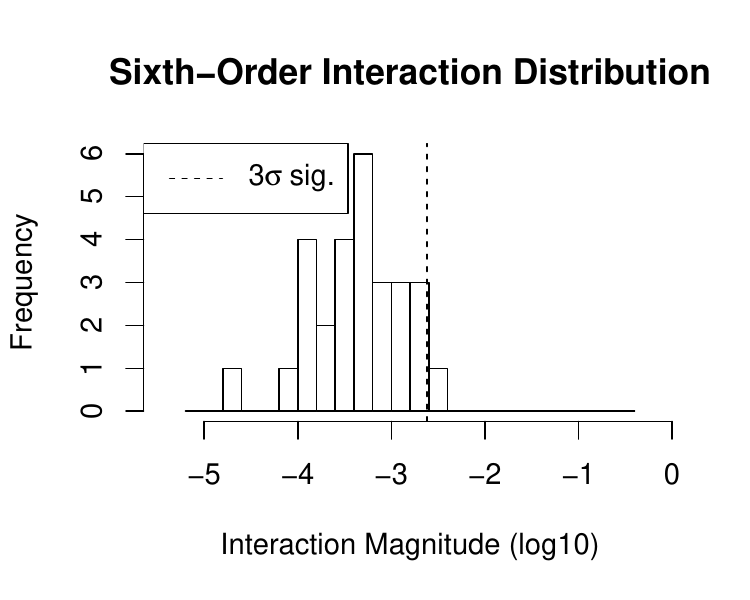} \\
\vspace{-0.3cm}
\includegraphics[width=0.39\linewidth,page=12]{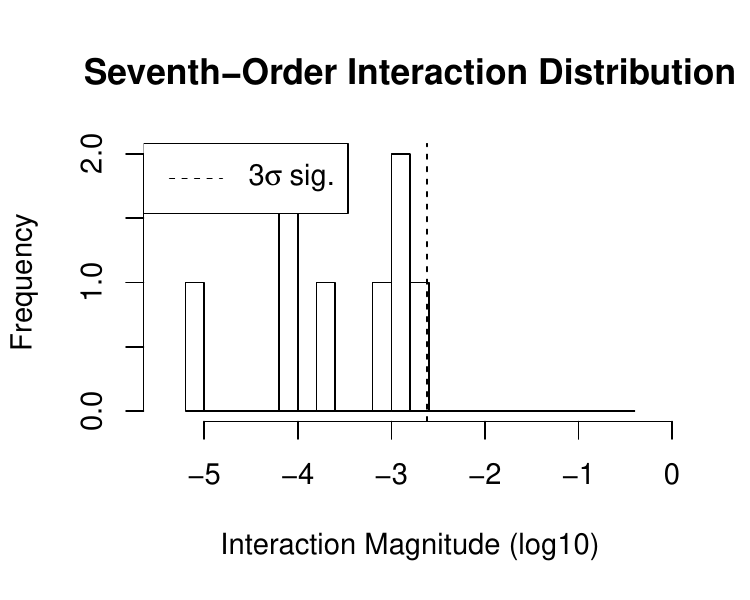}
\includegraphics[width=0.39\linewidth,page=13]{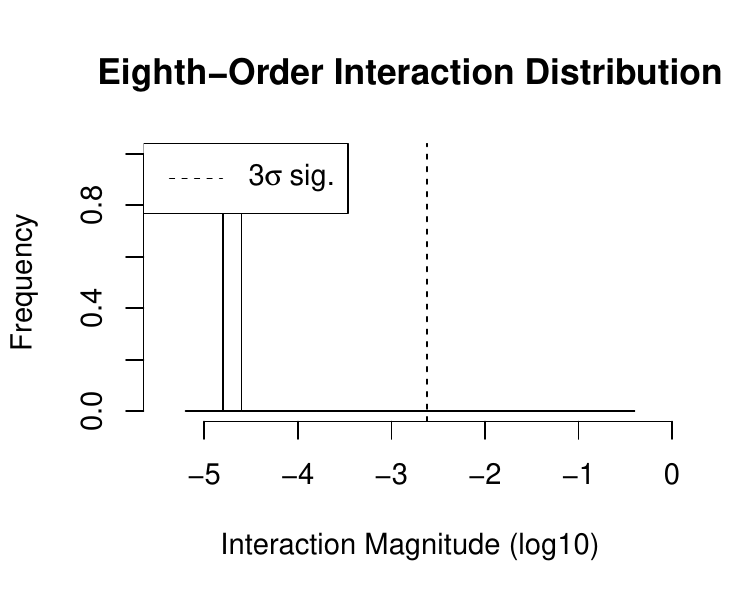}
\caption{Histograms of reconstruction parameters magnitudes from the 8-th order reconstruction experiment on the Ferromagnet model. Values above the $3\sigma$ line are considered to be statically significant, while the values below are artifacts of finite sampling. These results indicate that the distribution that the \texttt{DW\_2000Q\_LANL} hardware is sampling from is well approximated by a second-order model.}
\label{fig:fm-8o-parameter-hist}
\end{figure}

\begin{figure}
\centering
\includegraphics[width=0.39\linewidth,page=6]{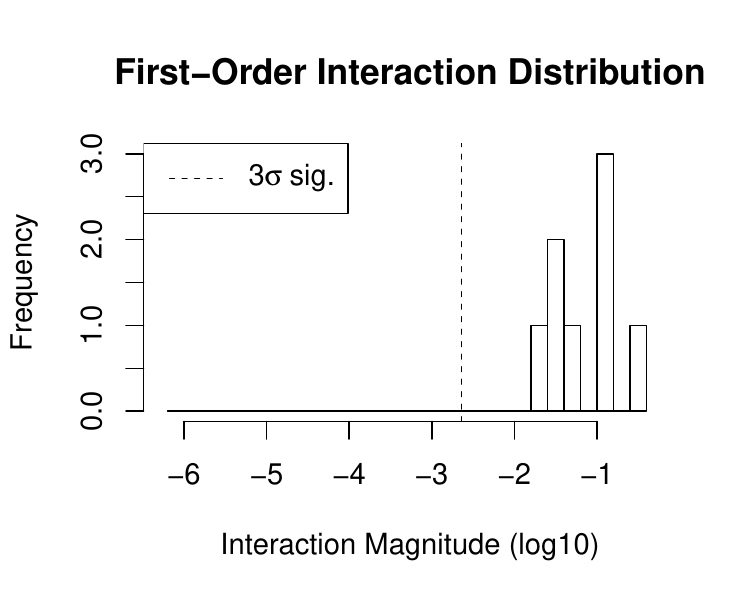}
\includegraphics[width=0.39\linewidth,page=7]{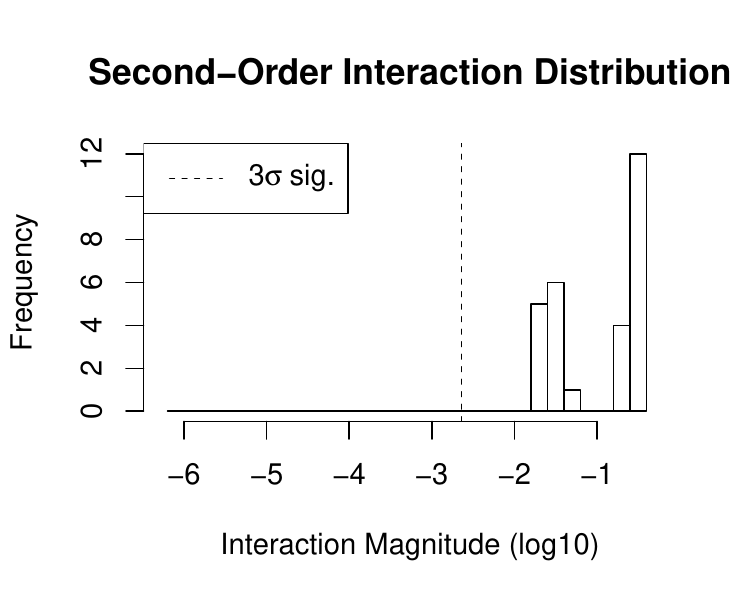} \\
\vspace{-0.3cm}
\includegraphics[width=0.39\linewidth,page=8]{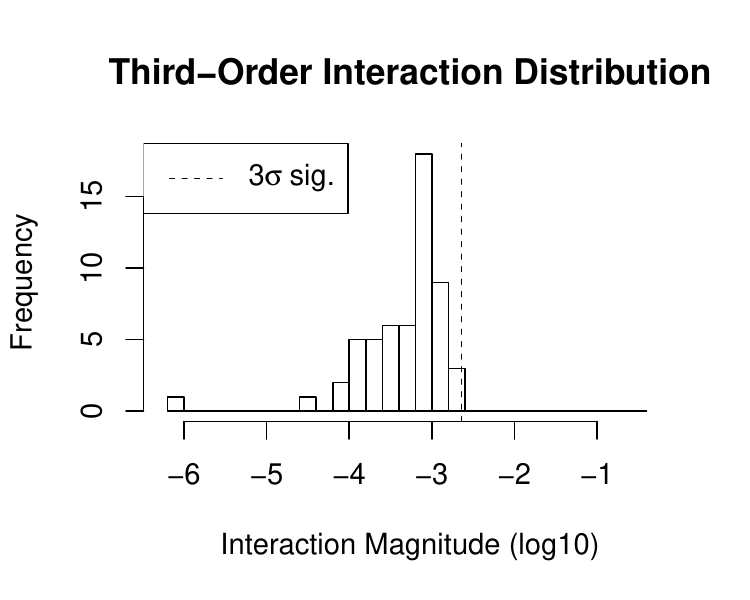}
\includegraphics[width=0.39\linewidth,page=9]{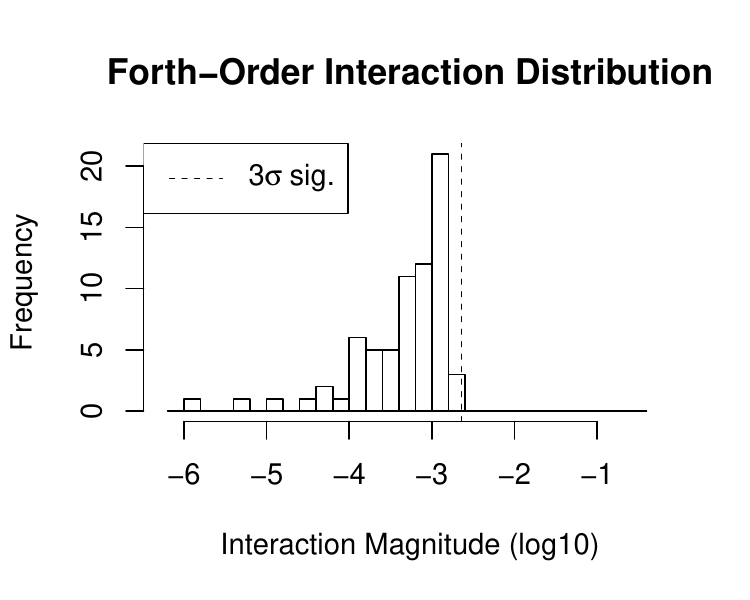} \\
\vspace{-0.3cm}
\includegraphics[width=0.39\linewidth,page=10]{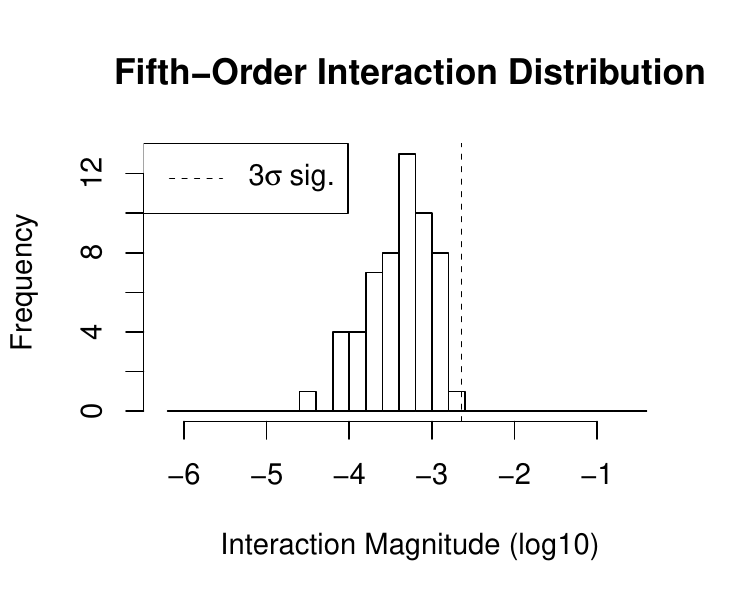}
\includegraphics[width=0.39\linewidth,page=11]{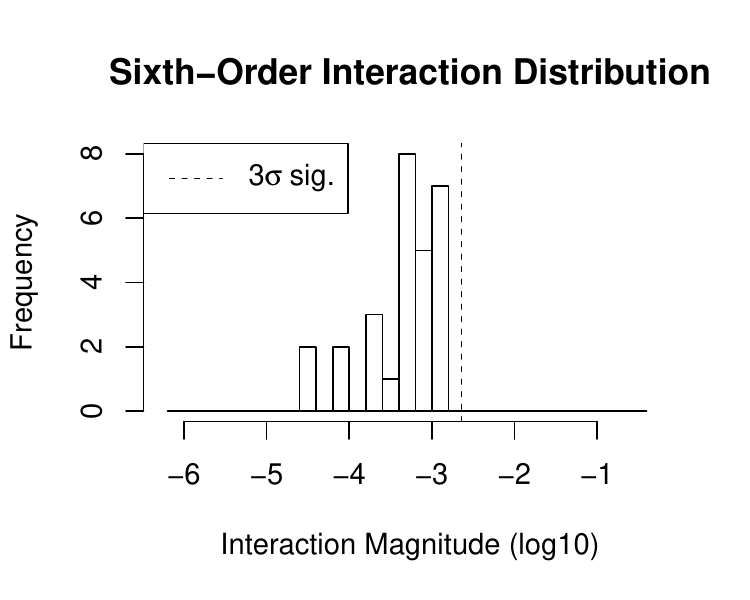} \\
\vspace{-0.3cm}
\includegraphics[width=0.39\linewidth,page=12]{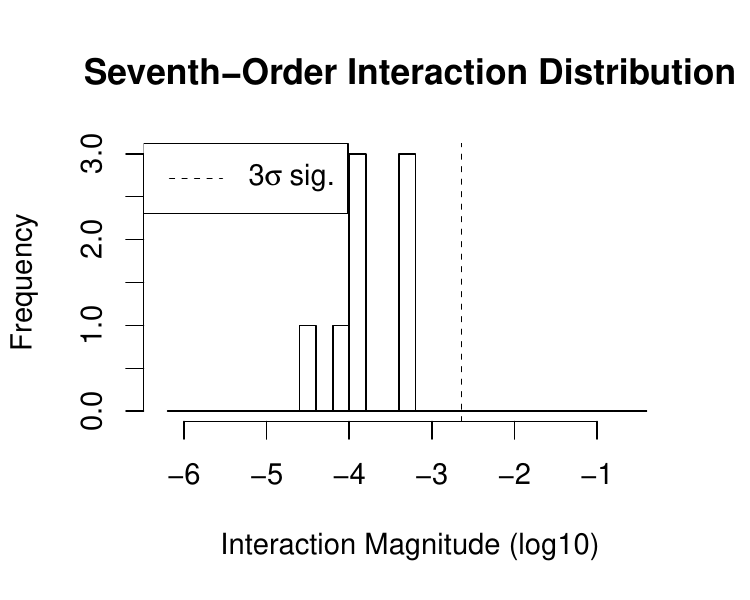}
\includegraphics[width=0.39\linewidth,page=13]{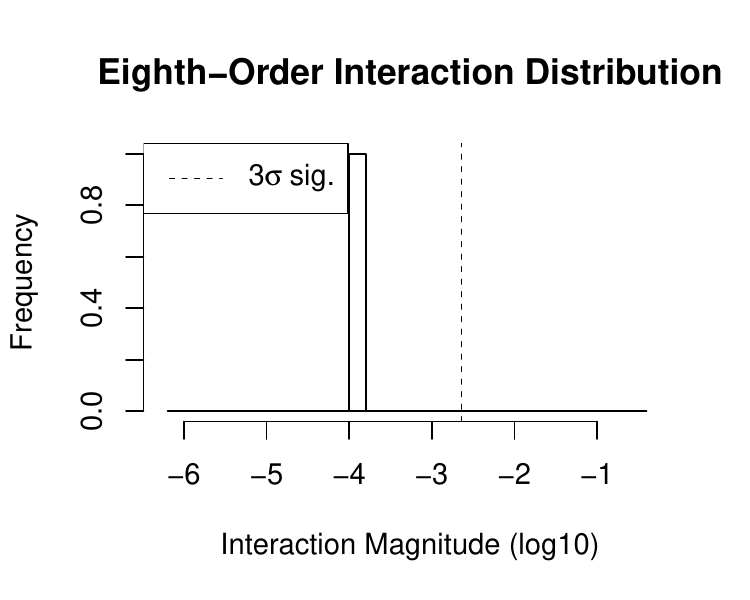}
\caption{Histograms of reconstruction parameters magnitudes from the 8-th order reconstruction experiment on the Anti-Ferromagnet model. Values above the $3\sigma$ line are considered to be statically significant, while the values below can be artifacts of finite sampling.  These results indicate that the distribution that the \texttt{DW\_2000Q\_LANL} hardware is sampling from is well approximated by a second-order model.}
\label{fig:afm-8o-parameter-hist}
\end{figure}

Figs.~\ref{fig:fm-8o-parameter-hist} and \ref{fig:afm-8o-parameter-hist} present the absolute values of the 255 model parameters broken down by the interaction-order for the ferromagnetic and the anti-ferromagnetic cases, respectively. We find that in both cases, a second-order model provides an accurate representation of the output distribution of \texttt{DW\_2000Q\_LANL}. This experiment thus provides a convincing evidence that a second-order model is sufficient for modeling the distribution that the quantum annealer samples from for a range of input parameters of interest to this work.

\begin{figure}
\centering
\includegraphics[width=0.41\linewidth,page=2]{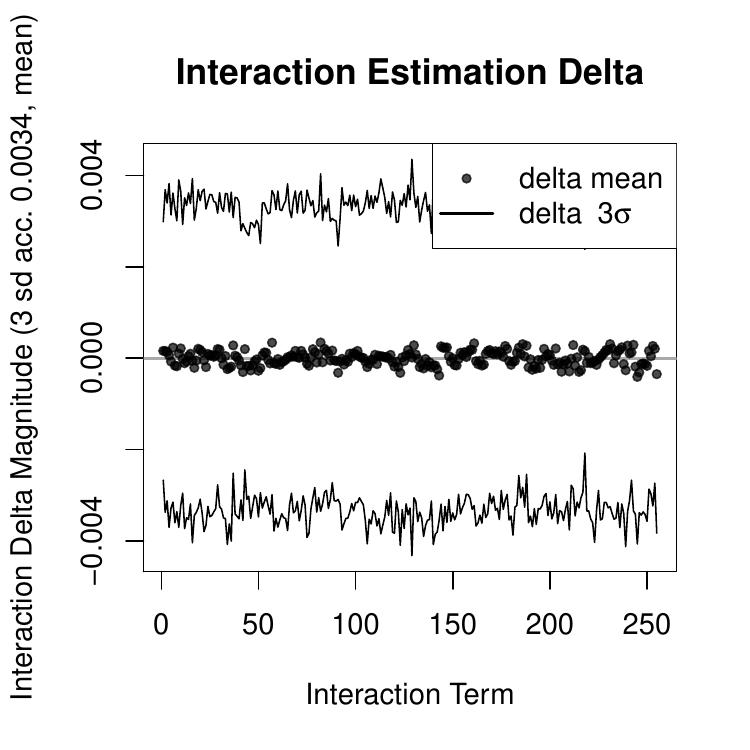}
\includegraphics[width=0.41\linewidth,page=2]{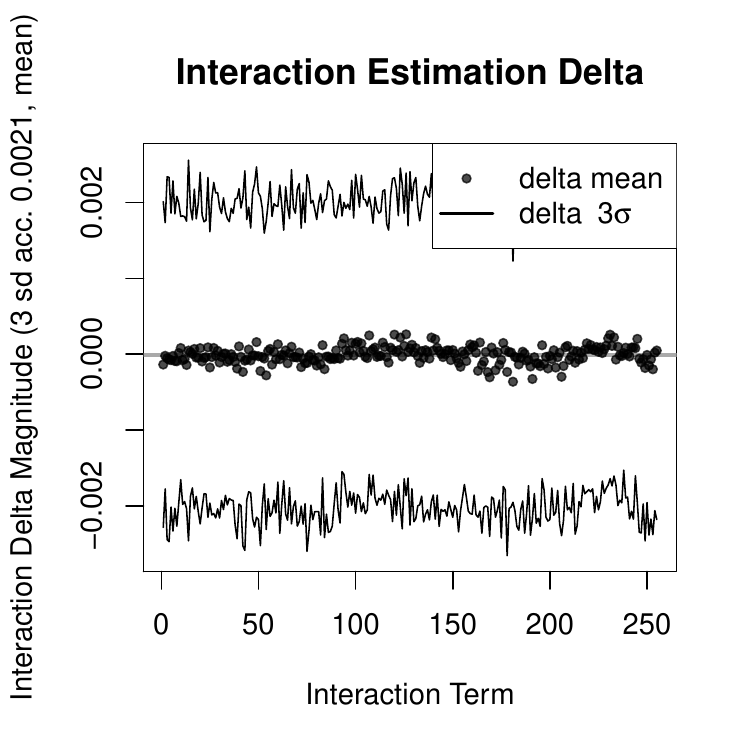}
\caption{Multi-body reconstruction variability results for the Ferromagnet model (left) and Anti-Ferromagnet (right) used in the multi-body probing experiment, using the validation protocol from section \ref{sec:empirical_error_estimation} with $R=50$ replicates. The mean values of the all 255 parameters are presented with whiskers shown at three standard deviations from the mean. The average-case deviations are 0.0034 and 0.0021, respectively.}
\label{fig:8o-parameter-variance}
\end{figure}

We further validate these results by quantifying the typucal variation of model parameters used in the 8-th order reconstruction based on finite sampling error. To that end, the validation protocol described in section \ref{sec:empirical_error_estimation} is executed with with $R=50$ replicates and $M=10^7$, to replicate the number of samples used in the data collection for most of the key experiments in this work. Fig.~ \ref{fig:8o-parameter-variance} shows data for the mean and the standard deviation for each parameter deviation $\delta h$ or $\delta J$ from the reference model, estimated from running the variance measuring procedure defined in the section \ref{sec:empirical_error_estimation} on the reconstruction models. We see that estimated variance of reconstructed values is very similar across different couplings. The average-case variance across all model parameters using three standard deviations are 0.0034 and 0.0021 for the Ferromagnet and Anti-Ferromagnet models, respectively. These values have been used for computing the threshold values that appear in Figs.~ \ref{fig:fm-8o-parameter-hist} and \ref{fig:afm-8o-parameter-hist} in the previous section.

\section{Reconstruction and Validation of Two-Body Models}
\label{sec:two-body}

The previous section conducted a 8-th order reconstruction with 10,000,000 samples and argued that a 2-nd order model provides a sufficient approximation of the distribution that the \texttt{DW\_2000Q\_LANL} hardware samples from.  To that end, the remaining experiments in this work focuses on learning 2-nd order models of the hardware's output distribution.
Leveraging the knowledge that this 2-nd order model is sufficient has a significant advantages in that the number of model parameters reduces from 255 (8-th order) to only 36 (2-nd order), which in turn reduces the amount of data required to accurately learn the associated 2-nd order model. To further lessen the data requirements, we reduce the number of qubits considered from 8 to 4, focusing on the upper-half of the cell. This effectively decreases the number of 2-nd order parameters from 36 to 16. It will later become evident that these reductions are necessary to make the experiments viable on reasonable time scales.
After conducting these model reductions, we replicate here the validation experiments from the section \ref{sec:multi-body} to establish an optimal number of samples required for reconstruction of parameters in the two-body model. Table \ref{tbl:2o-models} specifies the input parameters for two additional models, {\em strong ferromagnet} and {\em strong anti-ferromagnet}, using coupling sign convention consistent with \eqref{eq:Gibbs_distribution}.

\begin{table}
    \centering
    \begin{tabular}{|r||r|r|r|r|r|}
    \hline
    Model & Qubits & Couplers & $h^{in}$ & $J^{in}$ & Samples \\
    \hline
    \hline
    strong ferromagnet & $V'$ & $E'$ & 0.0 & 0.05 & $4\cdot10^6$ \\
    \hline
    strong anti-ferromagnet & $V'$ & $E'$ & 0.0 & -0.05 & $4\cdot10^6$ \\
    \hline
    \end{tabular}
    \caption{The input parameters sent to the \texttt{DW\_2000Q\_LANL} QPU to collect data for the 2-nd order distribution reconstruction experiment.}
    \label{tbl:2o-models}
\end{table}

\begin{figure}
\centering
\includegraphics[width=0.41\linewidth,page=2]{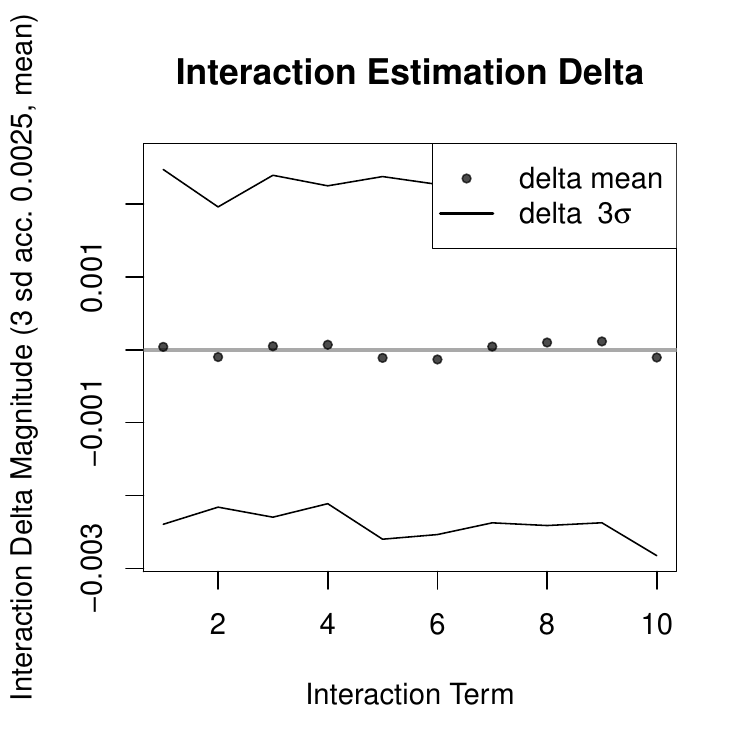}
\includegraphics[width=0.41\linewidth,page=2]{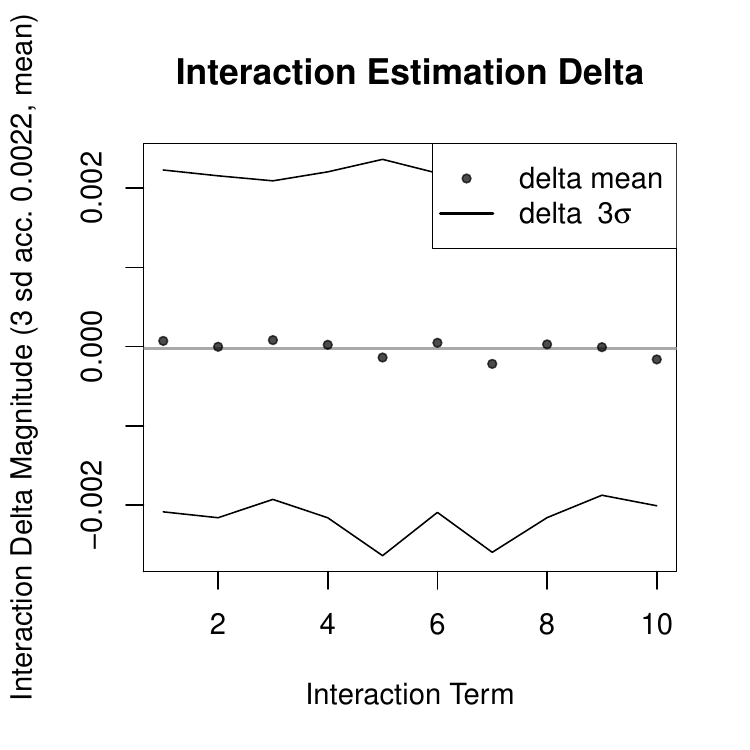}
\caption{Two-body reconstruction variability results for the Strong Ferromagnet model (left) and Strong Anti-Ferromagnet (right), using the proposed validation protocol with 50 replicates.  The mean values of the all 16 parameters are presented with whiskers shown at three standard deviations from the mean.  The average-case deviations are 0.0025 and 0.0022, respectively.}
\label{fig:2o-strong-parameter-variance}
\end{figure}

\begin{figure}
\centering
\includegraphics[width=0.41\linewidth,page=6]{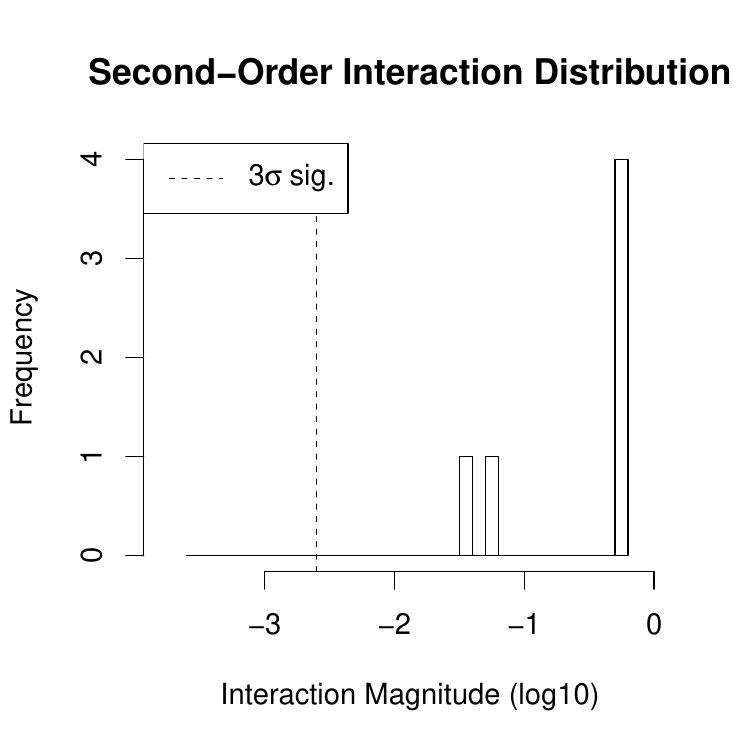}
\includegraphics[width=0.41\linewidth,page=6]{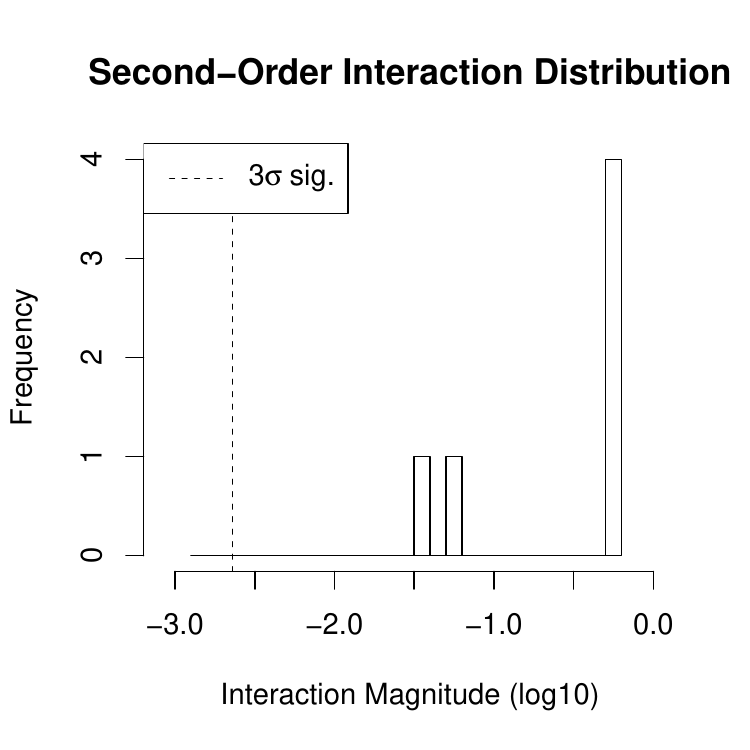}
\caption{Second-Order terms of the two-body reconstruction of the for the Strong Ferromagnet model (left) and Strong Anti-Ferromagnet (right).  The recovered values are greater than 0.03, well above the accuracy threshold of 0.0025, suggesting an accurate model.}
\label{fig:2o-strong-parameter-values}
\end{figure}

Fig.~\ref{fig:2o-strong-parameter-variance} presents the results of the validation experiment for the 2-nd order model reconstruction. 50 reconstruction replicates are used in the validation experiments conducted in this section. The results indicate that reconstruction accuracy is approximately 0.0025 and 0.0022, which is comparable to the accuracy used in the 8-th order reconstruction experiment, using a smaller number of samples. 
Fig.~\ref{fig:2o-strong-parameter-values} presents the strength of the second order terms that are recovered from the hardware data. These absolute values are well above the recovery accuracy threshold, indicating that $4 \cdot 10^6$ samples are sufficient for accurately recovering the two-body model. We use this number of samples in the remainder of experiments in this paper.

In Fig.~2 of the Main Text, we summarize the results of these experiments and show the structure of the output distribution. In particular, we find that among statistically significant couplings in the output distribution, the strongest ones are in correspondence with the input couplings, while the weakest ones are the \emph{spurious couplings} that are not present in the input problem and, moreover, are absent in the chip topology. In what follows, we construct additional experiments aimed at clarifying the nature of these spurious couplings.

\section{Impact of the Annealing Time}
\label{sec:impact_of_annealing_time}

In all presented experiments thus far, we used the annealing time of 5$\mu s$ per each sample. It is important to understand how the statistics of these reconstruction experiments might differ as the annealing time varies. Here, we replicate the model reconstruction experiment for the Strong Ferromagnet and Strong Anti-Ferromagnet models, using the following varying annealing time parameters, \texttt{annealing\_time = 1, 5, 25, 125, 625}, which corresponds to single-run annealing time of 1$\mu s$,  5$\mu s$,  25$\mu s$, 125$\mu s$, 625$\mu s$ respectively. Additionally, in this experiment, the \texttt{num\_reads} parameter was reduced from 10,000 (this work's default) to 4,000 in all cases, to adhere to the maximum job run-time limit of the \texttt{DW\_2000Q\_LANL} QPU.

\begin{figure*}
\centering
\includegraphics[width=0.88\linewidth,page=1]{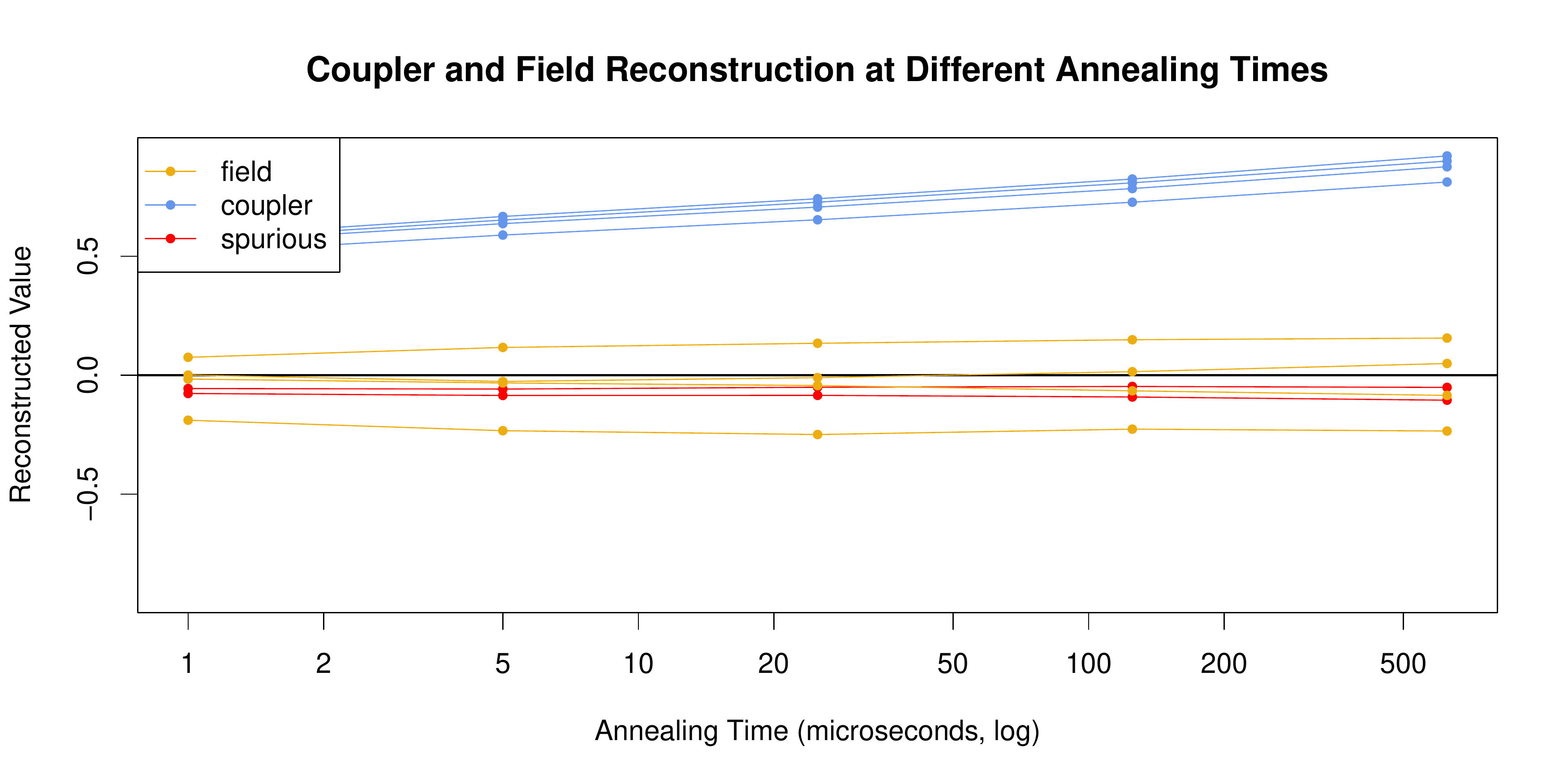}
\includegraphics[width=0.88\linewidth,page=1]{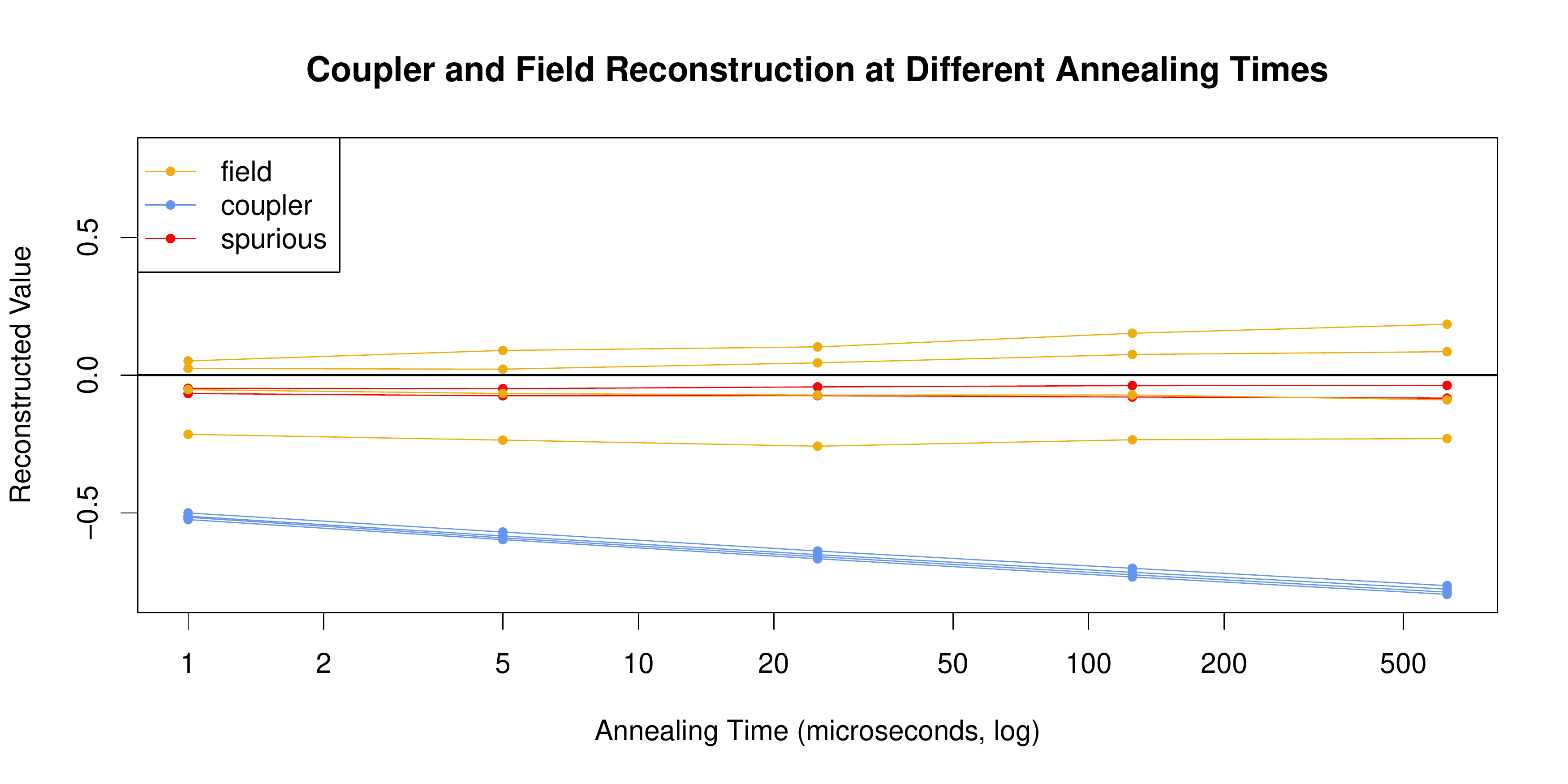}
\caption{A comparison of the second-order model reconstructions for different annealing times on the Strong Ferromagnet model (top) and Strong Anti-Ferromagnet model (bottom).  The results indicate that large increases in the annealing time results in slight reductions in the effective temperature of the reconstructed models.}
\label{fig:2o-strong-annealing-time}
\end{figure*}

Fig.~\ref{fig:2o-strong-annealing-time} presents the results of this experiment. We find that an increase of the annealing time by two orders of magnitudes results only in a slight increase of the absolute coupling values in the reconstructed model. At the same time, this minor change comes at a significant increase in data collection time. Due to this observation, in this work we chose to standardize around the annealing time of 5$\mu s$, which is essential for the high-throughput data collection, and represents the fastest available annealing time across several generations of quantum annealers.

\section{Quadratic Response}
\label{sec:quadratic_response}

In this section, we provide additional details on the study of the quadratic response experiment described in the Main Text. We assume the most general parametrization of the quadratic input-output relationship between the input parameters and the parameters of the reconstructed output distribution. Specifically, for each output model parameter we learn the quadratic $\chi^{hh}, \chi^{hJ}$, linear $\beta^h, \beta^J$ and offset $c$ coefficients of the following quadratic function,
\begin{align}
    h_i &= [h^{in} \;\; J^{in}] \begin{bmatrix} \chi^{hh}_i & \chi^{hJ}_i \\ \chi^{Jh}_i & \chi^{JJ}_i \end{bmatrix} \begin{bmatrix} h^{in} \\ J^{in} \end{bmatrix}  + [\beta^h_i \;\; \beta^J_i]  \begin{bmatrix} h^{in} \\ J^{in} \end{bmatrix} + c_i \;\; \forall i \in V' \label{eq:h_quad_model}\\
    J_{ij} &= [h^{in} \;\; J^{in}] \begin{bmatrix} \chi^{hh}_i & \chi^{hJ}_{ij} \\ \chi^{Jh}_{ij} & \chi^{JJ}_{ij} \end{bmatrix} \begin{bmatrix} h^{in} \\ J^{in} \end{bmatrix}  + [\beta^h_{ij} \;\; \beta^J_{ij}]  \begin{bmatrix} h^{in} \\ J^{in} \end{bmatrix} + c_{ij} \;\; \forall i \in V', j \in V' \label{eq:j_quad_model}
\end{align}
Given sufficient data, these quadratic functions can be recovered using a least squares regression. For example, the function form of the output parameter $h_i$ can be learned from a collection of $S$ different input-output realizations by solving the following convex optimization problem,
\begin{align}
    \argmin_{{\chi^{hh}_i}, {\chi^{hJ}_i}, {\chi^{Jh}_i}, {\chi^{JJ}_i} {\beta^h_i}, {\beta^J_i}, c} \; & \sum_{s \in S} \left( [h^{in}_s \;\; J^{in}_s] \begin{bmatrix} \chi^{hh}_i & \chi^{hJ}_i \\ \chi^{Jh}_i & \chi^{JJ}_i \end{bmatrix} \begin{bmatrix} h^{in}_s \\ J^{in}_s \end{bmatrix}  + [\beta^h_i \;\; \beta^J_i]  \begin{bmatrix} h^{in}_s \\ J^{in}_s \end{bmatrix} + c_i - h_{is} \right)^2 \label{eq:quad-opt}\\
    & {\chi^{hh}_i}, {\chi^{hJ}_i}, {\chi^{Jh}_i}, {\chi^{JJ}_i} {\beta^h_i}, {\beta^J_i}, c \in \mathbb{R} \nonumber
\end{align}
Note that similar optimization problems can be solved to learn the quadratic relation of all output parameters $h,J$. The number of data points (i.e. $|S|$) should be at least $n  \ln n$ to accurately recover the quadratic function, where $n$ is the number of unknown values in \eqref{eq:quad-opt}. The core experiment of this section consists in performing a series of 2-nd order reconstruction experiments over random input models, and then in using these pairs of input-output models to recover the coefficients in the quadratic response function. 

The primary challenge of this experiment is the time required to collect a sufficient amount of data to fit the quadratic response function. To minimize the data requirements, we focus on the 4-spin model defined as $N',E'$ in Table \ref{tbl:l2k-qpu-used}. For this specific model, the quadratic functions \eqref{eq:h_quad_model}, \eqref{eq:j_quad_model} have 57 parameters; we consider 250 input-output model pairs to recover these parameters. Each of the 250 input models is selected i.i.d. from the following input parameter distribution,
\begin{align}
    h^{in}_i & \in \{-0.05:0.01:0.05\} \;\; \forall i \in N' \label{eq:quadresp_field_pertubation_set},\\
    J^{in}_{ij} & \in \{-0.05:0.01:0.05\} \;\;  \forall i,j \in E'.\label{eq:quadresp_coupling_pertubation_set}
\end{align}
Following the validation study for the {\em strong ferromagnet} and {\em strong anti-ferromagnet} models, $4 \cdot 10^6$ samples are sufficient to accurately reconstruct 2-nd order models with parameters as large as 0.05. Altogether, this experiment reconstructs the outputs for 250 input models using 4,000,000 samples for each model, which results in a total of a \emph{billion} samples collected.

Characteristic dominant terms in the recovered quadratic response function is presented in the Fig.~4 of the Main Text (data on all measured quadratic response terms is given below, in Sections~\ref{sec:noise_and_spurious_links} and \ref{sec:three_generations}.). The zero order terms in the response functions are interpreted as residual fields and couplings; the first-order terms are related to the native couplings present in the chip; and finally, the second-order terms are responsible for the spurious couplings. In previous work, the primary hypothesis behind the response function was formulated in terms of a particular case of a linear assumption \cite{bian2010ising, benedetti2016estimation, perdomo2016determination, raymond2016global, marshall2017thermalization, li2020limitations}, where each parameter would be multiplied by a single effective temperature. The non-linearity of the general response function that we construct here may explain why this effective temperature was found to be instance-dependent: This corresponds to a linear approximation of a non-linear function.

Spurious couplings identified under a careful statistical analysis indicate that a simple linear model is not sufficient for an accurate characterization of the D-Wave's input-output relationship. It is important to note that the second-order response that we find here is different from the previously observed next-nearest-neighbour couplings in the strong input regime, where a quadratic cross-talk relation with an \emph{opposite sign} susceptibility has been suggested (see the section ``Compensation of qubit nonidealities'' in the \emph{Methods} of \cite{king2018observation}). We conjecture that the emergence of the next-nearest-neighbour couplings observed in the strong regime has a quantum nature and is due to the induced effects of the transverse field; a detailed exploration of this phenomenon is beyond the scope of our study that focuses on the classical regime of the output distribution in a multi-qubit setting.

The discovery of strong and structured quadratic response functions for the output distribution that our D-Wave \texttt{DW\_2000Q\_LANL} QPU samples is invaluable to applications such as hardware calibration, problem embedding and accurate sampling. However, identifying the root cause of these unexpected output parameters can provide valuable information about how to design better quantum annealers and can provide novel analytics for evaluating the performance of quantum annealers. To that end, the next section provides the theoretical grounds to explain the quadratic response as side effects of instantaneous qubit noise.

\section{Characterisation of the Local Field Variably}
\label{sec:local_field_variability}

In preparation for constructing a model that would explain the form of the quadratic response function, in this section we investigate possible drifts of the reconstructed model parameters. To this end, we perform reconstructions over several days, and monitor the stability of the recovered model.

\begin{figure}
\centering
\includegraphics[width=0.41\linewidth,page=2]{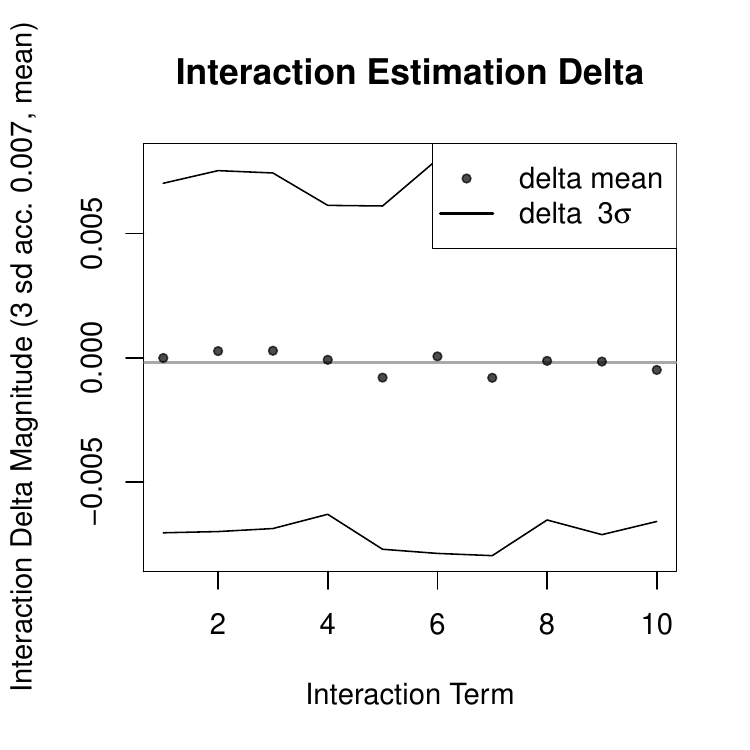}
\includegraphics[width=0.41\linewidth,page=4]{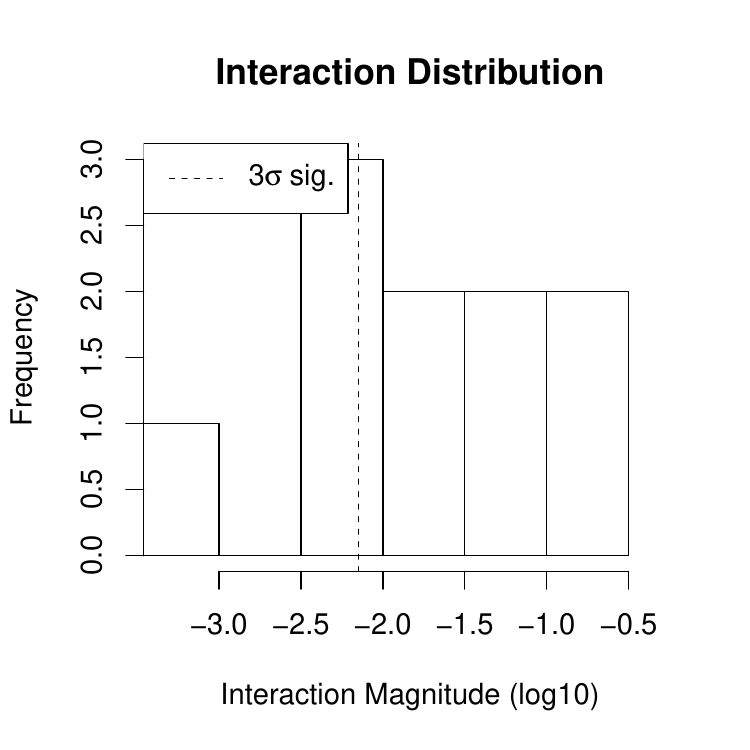}
\caption{The reconstruction variability (left) and recovered interaction magnitudes (right) of the zero problem.  The results suggest indicate that 200,000 samples is sufficient to accurately measure the qubit bias occurring in the \texttt{DW\_2000Q\_LANL} QPU.}
\label{fig:2o-zero-parameters}
\end{figure}

The foundation of this variability study is the reconstruction of the output distribution of the zero value problem, that is $h^{in}, J^{in} = 0$ as shown in Table \ref{tbl:2o-zero-model}. Furthermore, we would like to perform this reconstruction accurately but with a minimum number of samples, so that the possible flux drift dynamics can be observed in the time between multiple reconstructions. We begin by calibrating a 2-body reconstruction specifically for the zero-value problem by proposing that only 200,000 samples are required for an accurate reconstruction, which is 20 times less data than what is used for a typical 2-body reconstruction. Repeating the previous reconstruction variability analysis, Fig.~\ref{fig:2o-zero-parameters} presents both the variance and recovery accuracy of this zero-valued model. The reconstruction accuracy is approximately 0.007, which is about two times less accurate than the previously considered reconstruction experiments.  However, we find this accuracy is still acceptable as our primary interest in this experiment are model values that are above 0.100. Indeed, Fig.~\ref{fig:2o-zero-parameters} indicates that the \texttt{DW\_2000Q\_LANL} QPU has a number of biases that are on the order of magnitude larger than reconstruction accuracy threshold of 0.008.

\begin{table}
    \centering
    \begin{tabular}{|r||r|r|r|r|r|}
    \hline
    Model & Qubits & Couplers & $h^{in}$ & $J^{in}$ & Samples \\
    \hline
    \hline
    Zero-value & $V'$ & $E'$ & 0.0 & 0.000 & 200,000 \\
    \hline
    \end{tabular}
    \caption{The input parameters sent to the \texttt{DW\_2000Q\_LANL} QPU to collect data for the 2-nd order reconstruction experiment.}
    \label{tbl:2o-zero-model}
\end{table}

\begin{figure}
\centering
\includegraphics[width=0.66\linewidth]{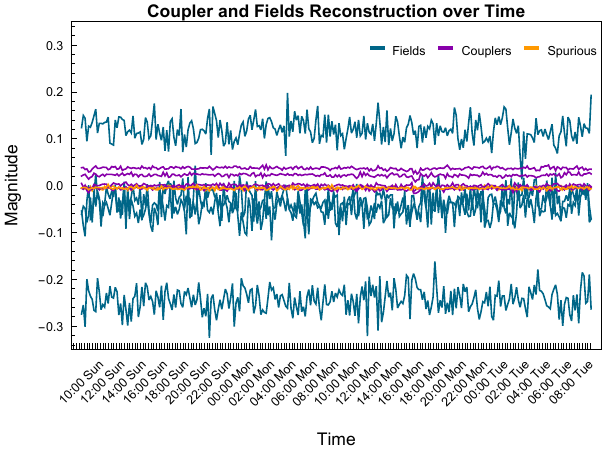}
\caption{Reconstruction of the zero problem repeated over several days. A clear persistent bias in the local fields is evident from offset mean values. The fluctuations in the reconstructed field are significantly larger than the reconstruction error, suggesting fluctuation in the local fields over time.}
\label{fig:zero-overtime}
\end{figure}

Next, we investigate how the reconstructed values of the zero problem change over time. The objective is to understand how the low frequency noise in the hardware changes over time and how that can impact data collection over the span of minutes to hours. In this experiment, the data for the proposed zero problem is collected at 10 minute intervals over a period of 48 hours and then the 2-body reconstruction is used to recovery a model from the observed samples. Fig.~\ref{fig:zero-overtime} shows the reconstructed values over time and whiskers around the points indicate the error bounds on the model reconstruction values. Considering the mean values of these time series one can observe that there is a persistent bias on both the reconstructed fields and couplers, which is most likely an artifact of the initial hardware's calibration. Looking at the variance of the time series highlights a high variance in the fields terms and a much lower variance in the coupler terms. Overall, the results of these experiments suggest that all parameters of the output distribution at the exception of local fields remain stable over time.

\section{Explaining Quadratic Response via Instantaneous Qubit Noise}
\label{sec:noise_and_spurious_links}

In the previous section~\ref{sec:local_field_variability}, we saw that magnetic field is affected by comparably large fluctuations. However, this analysis was conducted by performing the reconstruction over a certain time window, which may average out the fluctuations. On the other hand, the analysis in section~\ref{sec:single_spin_exp} allowed us to estimate \emph{instanteneous} fluctuations of the residual random magnetic field for each qubit individually. Here, we show that these rapid fluctuations in the individual qubit fields can be responsible for spurious effective interactions with non-trivial quadratic-type responses in the input quantities. We start by showing theoretically the type of spurious interactions one may expect to reconstruct on toy models for which we can derive close form formulas. Then, using numerical simulations, we will quantify the quadratic response caused by noise on a four qubit system and compare these predictions with the quadratic response measured on the D-Wave quantum annealer.  

\subsection{Spurious Magnetic Field Response}\label{subsec:theory_spurious_fields}
We first consider a simple system consisting of two classical spins $\sigma_1,\sigma_2 \in \{-1,1\}$ linked by a coupling $J\in \mathbb{R}$. The first spin is subject to a noisy magnetic field $h^{\rm{sd}}_{1} s_1$ whose direction varies according to the uniform random variable $s_1\in \{-1,1\}$, whereas the second spin is subject to a constant magnetic field $h_2$. The Bolzmann distribution of this two spins system at inverse temperature $\beta$ for a particular noise realization $s_1$ is given by the following expression,
\begin{equation}
    \mu(\sigma_1,\sigma_2 \mid s_1) = \frac{\exp{\left( \beta (J \sigma_1 \sigma_2 + h^{\rm{sd}}_{1} s_1\sigma_1 +h_2 \sigma_2 )\right)}}{Z(s_1)}, \label{eq:quadth_one_coupling_dist}
\end{equation}
where the partition function $Z(s_1)$ depends on the noise realization $s_1$. Suppose now that we want to perform an Ising model reconstruction using a collection of iid. samples from the distribution in Eq.~\eqref{eq:quadth_one_coupling_dist} where the noise realization changes randomly on a sample to sample basis. Our collection of samples end up arising from a mixture of models as around half of the configurations comes from  $\mu(\sigma_1,\sigma_2 \mid s_1 = +1)$ and the other half comes from $\mu(\sigma_1,\sigma_2 \mid s_1 = -1)$. Therefore, the effective model that we can reconstruct with this heterogeneous collection of samples is the following mixture of Ising models, 
\begin{equation}
    \mu_{\rm{effective}}(\sigma_1,\sigma_2) = \sum_{s_1\in \{-1,1\}} \frac{1}{2} \mu(\sigma_1,\sigma_2 \mid s_1).
\end{equation}
This effective model is also an Ising model and, after some algebra, it can be explicitly formulated with respect to the initial coupling and fields,
\begin{equation}
     \mu_{\rm{effective}}(\sigma_1,\sigma_2) = \frac{\exp{\left( \beta (J \sigma_1 \sigma_2 + h^{\rm{effective}}_{1} \sigma_1 +h_2 \sigma_2 )\right)}}{Z_{\rm{effective}}}, \label{eq:quadth_one_coupling_mixture}
\end{equation}
where
\begin{align}
      h^{\rm{effective}}_{1}
      = -\frac{1}{\beta}\arctanh{\left(\tanh{(\beta J)} \tanh{(\beta h_2)} \tanh{(\beta h^{\rm{sd}}_{1})}^2\right)}. \label{eq:quadth_one_coupling_response}
\end{align}
For small coupling and field magnitudes $J$ and $h_2$, the expression in Eq.~\eqref{eq:quadth_one_coupling_response} reduces to $h^{\rm{effective}}_{1} \approx - \beta J h_2 \tanh{(\beta h^{\rm{sd}}_{1})}^2$. We see with this toy model that fast fluctuating magnetic field noise induces an effective magnetic response. The samples coming from a mixture of models with field noise are indistinguishable from samples coming from the single model with an effective response. The key qualitative features of this response is 1) its intensity is roughly proportional to the product of the coupling and opposite (constant) field intensity, and 2) the sign of the response is negative. This effect being roughly proportional to the square of the standard deviation of the noise, it is negligible for low noise values but becomes much more pronounced when the noise becomes large.

\subsection{Spurious Coupling Response}\label{subsec:theory_spurious_couplings}
To illustrate how spurious couplings can occur from a mixture of noisy Ising models, we look at a chain of three spins $\sigma_1$, $\sigma_2$ and $\sigma_3$ connected via couplings $J_{12}$ and $J_{23}$. We assume that the spin at the extremity of the chain are subject to a noisy magnetic field $h^{\rm{sd}}_{1} s_1$ and $h^{\rm{sd}}_{3} s_3$ where the random variable $s_1, s_3 \in \{-1,1\}$ are independent and uniformly distributed. The magnetic field on the middle spin is assumed to be zero. For a particular noise realization, the Boltzmann distribution of this chain of spins is given by the following conditional distribution,
\begin{align}
    \mu(\sigma_1,\sigma_2, \sigma_3 \mid s_1, s_3)
    = \frac{\exp{\left( \beta (J_{12} \sigma_1 \sigma_2 + J_{23} \sigma_2 \sigma_3 + h^{\rm{sd}}_{1} s_1\sigma_1 + h^{\rm{sd}}_{3} s_3\sigma_3\right)}}{Z(s_1,s_3)}, \label{eq:quadth_two_coupling_dist}
\end{align}
where the partition function $Z(s_1,s_3)$ is a function of the noise realization. Similarly to the previous subsection, we consider fast fluctuating noise that changes randomly on a sample to sample basis. In this case, a collection of samples coming from the mixture of noisy Ising models described by Eq.~\eqref{eq:quadth_two_coupling_dist} becomes indistinguishable from iid. samples coming from the effective model,
\begin{equation}
      \mu_{\rm{effective}}(\sigma_1,\sigma_2,\sigma_3) = \sum_{s_1,s_3\in \{-1,1\}} \frac{1}{4} \mu(\sigma_1,\sigma_2,\sigma_3 \mid s_1,s_3).
\end{equation}

This effective model appears to be an Ising model as well with no magnetic fields and with an additional coupling between $\sigma_1$ and $\sigma_3$,

\begin{align}
    \mu_{\rm{effective}}(\sigma_1,\sigma_2, \sigma_3)
     = \frac{\exp{\left( \beta (J_{12} \sigma_1 \sigma_2 + J_{23} \sigma_2 \sigma_3 + J^{\rm{effective}}_{13} \sigma_1 \sigma_3) \right)}}{Z_{\rm{effective}}}, \label{eq:quadth_two_coupling_mixture}
\end{align}
where the effective coupling $J^{\rm{effective}}_{13}$ can be explicitly written with respect to the mixture parameters,

\begin{equation}
      J^{\rm{effective}}_{13} = -\frac{1}{\beta}\arctanh{\left(\tanh{(\beta J_{12})} \tanh{(\beta J_{23})} \tanh{(\beta h^{\rm{sd}}_{1})}^2 \tanh{(\beta h^{\rm{sd}}_{3})}^2\right)}. \label{eq:quadth_two_coupling_response}
\end{equation}
For small coupling values, Eq.~\eqref{eq:quadth_two_coupling_response} reduces to $J^{\rm{effective}}_{13} \approx -\beta J_{12} J_{23} \tanh{(\beta h^{\rm{sd}}_{1})}^2 \tanh{(\beta h^{\rm{sd}}_{3})}^2$. We immediately see that this coupling response induced by field noise retains the main qualitative features observed in the previous subsection. The intensity of the response is quadratic in the couplings $J_{13}$ and $J_{23}$ for it is proportional to their product and the sign of the response is negative. Note that this coupling response, which involves three spins, is predicted to be weaker than the magnetic field response discussed in the previous subsection as it is roughly proportional to the square of both noise standard deviations $h^{\rm{sd}}_{1}$ and $h^{\rm{sd}}_{3}$.

\subsection{Simulations and Predictions Using Single Spin Measurements}
In the previous subsections, we have seen on simple toy models that field noise leads to effective field and spurious coupling responses. For small coupling and field magnitudes, these ``spurious" responses were mainly quadratic in the input parameters. We now want to quantify the quadratic responses cause by field noise on a realistic four spin system and compare them to the type of quadratic responses found experimentally in Section~\ref{sec:quadratic_response}. This four spin system being already too complex to obtain a closed form formula, we resort to using numerical simulations to extract the quadratic response coefficients. We model the system by a classical Boltzmann distribution conditioned on a noise realization $\underline{s}\in \{-1,1\}^4$ of the field noise parameters. The probability to obtain a configuration $\underline{\sigma}\in\{-1,1\}^4$ given $\underline{s}$ reads as follows,
\begin{equation}
    \mu(\underline{\sigma} \mid \underline{s}) = \frac{\exp{\left(H_{\rm{fields}}\left(\underline{\sigma} \mid \underline{s}\right) +H_{\rm{couplings}}\left(\underline{\sigma}\right)\right)}}{Z(\underline{s})},\label{eq:quadth_four_spins}
\end{equation}
where the partition functions $Z(\underline{s})$ is noise dependant. The Hamiltonian describing the magnetic field interaction contains terms $h_i$ for input fields, $h^{\rm{bias}}$ for permanent biases and $h^{\rm{sd}}_{i}$ for the standard deviation of the noise as described in Section~\ref{sec:single_spin_exp}. An individual temperature $\beta_i$ is also assigned for each spin,
\begin{align}
 H_{\rm{fields}}\left(\underline{\sigma} \mid \underline{s}\right) &= \sum_{i=1,\ldots,4}\beta_{i}(h^{\rm{sd}}_{i} s_i + h^{\rm{bias}}_{i} + h_{i})\sigma_{i}.
\end{align}
The coupling Hamiltonian contains terms $J_{ij}$ for the input coupling strengths that are only along physical couplers and possessing their individual temperatures $\beta_{ij}$. Motivated by considerations from Section~\ref{sec:local_field_variability}, we assume that the interactions are noiseless and without biases,
\begin{align}
    H_{\rm{couplings}}\left(\underline{\sigma}\right)
    = \beta_{12} J_{12} \sigma_1 \sigma_2 + \beta_{14}J_{14} \sigma_1 \sigma_4 + \beta_{23}J_{23} \sigma_2 \sigma_3 +\beta_{34}J_{34} \sigma_3 \sigma_4.
\end{align}
The effective model describing the probability distribution of the four spin system is obtained after averaging Eq.~\eqref{eq:quadth_four_spins} over the uniform and independant noise realizations,
\begin{align}
    \mu_{\rm{effective}}\left(\underline{\sigma}\right) = \frac{1}{4}\sum_{\underline{s}\in\{-1,1\}^4} \mu(\underline{\sigma} \mid \underline{s}).
    \label{eq:quadth_four_effective_model}
\end{align}
The numerical procedure to reconstruct from Eq.~\eqref{eq:quadth_four_effective_model} a quadratic response as a function of the input couplings $J_{ij}$ and fields $h_{i}$ is reminiscent of the experimental protocol described in Sections~\ref{sec:quadratic_response}. 
We start by randomly selecting 20000 input coupling and field configurations whose values lies in the set $\{-0.05, -0.04, \cdots, 0.05\}$, see Eq.~\eqref{eq:quadresp_field_pertubation_set} and Eq.~\eqref{eq:quadresp_coupling_pertubation_set}. Then for each of these configurations, we compute numerically the effective frequencies of the $2^4=16$ spin configurations using Eq.~\eqref{eq:quadth_four_effective_model} and summing over the $2^4=16$ possible noise realizations. These frequencies are used in our reconstruction procedure, described in Section~\ref{sec:learning_general_distributions}, to infer an effective Ising model with Hamiltonian,
\begin{align}
    H_{\rm{effective}}(\underline{\sigma}) = \hspace{-0.39cm} \sum_{i,j\in \{1,2,3,4\}}J^{\rm{effective}}_{ij} \sigma_i \sigma_j + \hspace{-0.39cm} \sum_{i\in\{1,2,3,4\}} h^{\rm{effective}}_i \sigma_i.\label{eq:quadth_effective_hamiltonian}
\end{align}
Note that the Hamiltonian in Eq.~\eqref{eq:quadth_effective_hamiltonian} contains all pairwise interactions between four spins and the spurious interactions are represented by the effective couplings $J^{\rm{effective}}_{13}$ and ${J^{\rm{effective}}_{24}}$. Finally, we fit a quadratic response model between the inputs configurations and their corresponding inferred effective couplings, as described by Eq.~\eqref{eq:h_quad_model} and Eq.~\eqref{eq:j_quad_model}, following the optimization procedure in Eq.~\eqref{eq:quad-opt}.
The spins $\sigma_1, \sigma_2, \sigma_3, \sigma_4$ in our model are identified with the hardware spins $\#304, \#308, \#305, \#309$ respectively. The values of field temperatures $\beta_i$, field biases $h^{\rm{bias}}_i$ and field noise standard deviations $h^{\rm{sd}}_i$ are chosen to be those measured using the single spin quantum experiments described in Section~\ref{sec:single_spin_exp}. These values can be found in the last column of Table~\ref{tbl:classical_reg_coeff}. The values of the coupling temperatures have been adjusted such that the simulated and measured effective temperatures coincide, i.e. $\beta_{12}=12.1, \beta_{14}=12.2, \beta_{23}=12.5$ and $\beta_{24}=12.6$.

\begin{figure*}
    \centering
    \includegraphics{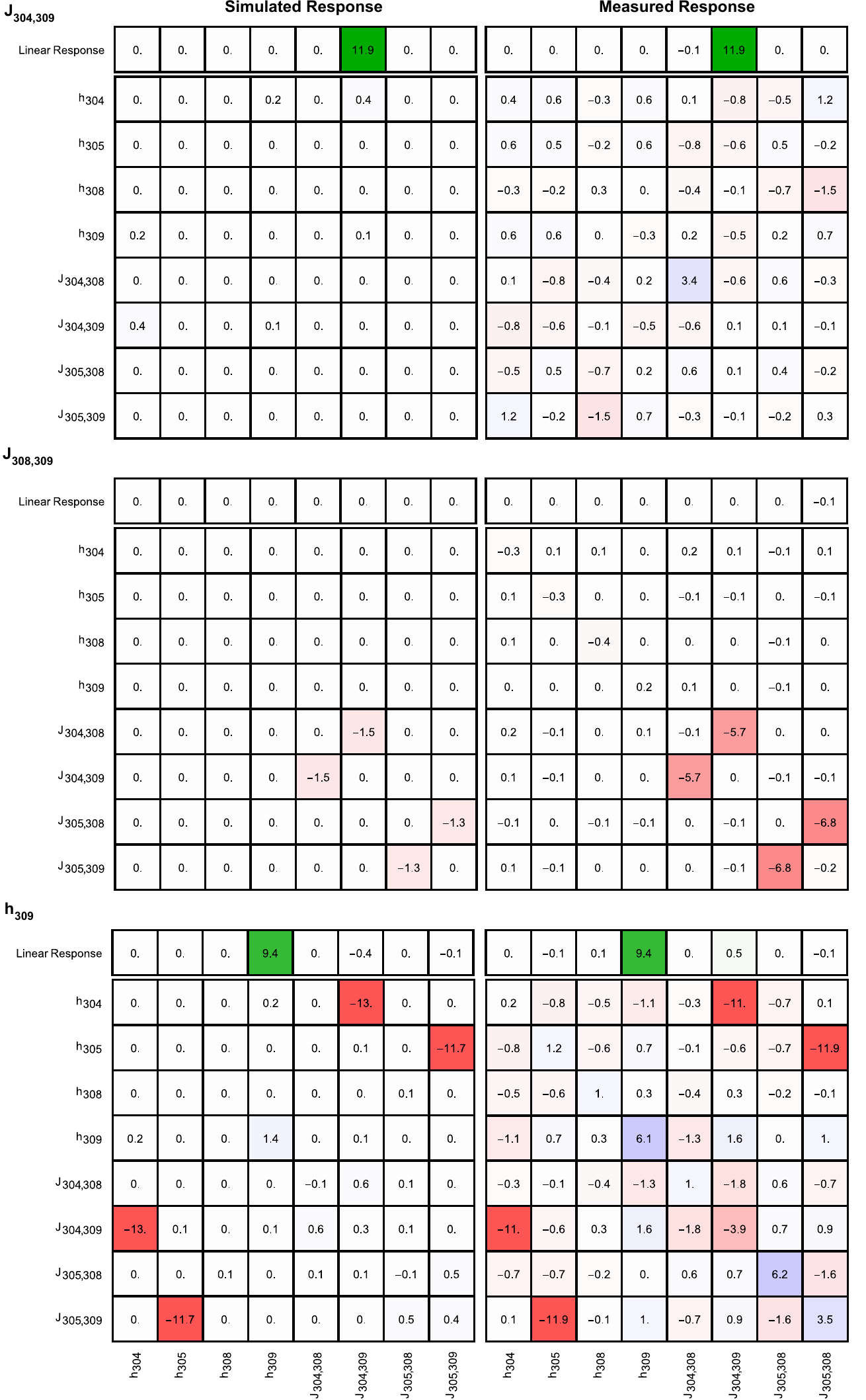}
    \caption{General quadratic response of effective output parameters. A comparison of the simulated and true responses at the quadratic level are given for the effective parameters $J^{\rm{out}}_{304,305}$, $J^{\rm{out}}_{308,309}$, and $h^{\rm{out}}_{309}$. The dominant terms in the general quadratic response presented here are discussed in the Fig.~4 of the Main Text. Notice that here and below, we use a symmetric matrix representation for the quadratic response, which results in a factor 2 difference for matrix elements compared to the results presented in the Fig.~4 of the Main Text.}
    \label{fig:full_response_simulated_vs_true}
\end{figure*}

The typical simulated and measured response for existing couplings, fields, and spurious couplings are depicted in
the Fig.~\ref{fig:full_response_simulated_vs_true}. We see that the patterns are qualitatively following the predictions from the simple theoretical models: The response for existing couplings is a linear self-transform or effective temperature, the response for fields consists in an effective temperature and a negative quadratic response from adjacent couplings and connecting neighboring fields, and finally the spurious couplings are formed by a negative quadratic response from adjacent couplings that span a triangle with the spurious coupling. The main noticeable difference with the theoretical predictions is the lowering of the coupling effective temperatures from its model temperature due to the presence of field noise. The comparison between the measured and simulated effective temperature can be found in Table~\ref{tbl:quadth_lin_response}. The leading off-diagonal coefficients of the quadratic response, both simulated and measured, are displayed in Table~\ref{tbl:quadth_quad_response}. The effective temperatures found through the simulation remarkably matches the effective temperatures measured in the hardware with a maximum difference of at most $7\%$. The negative sign of the quadratic response and the type of interactions involved is in perfect agreement with the theoretical model. There exists a discrepancy in the magnitudes predicted by the simulation and those found experimentally. The predictions are up to two times weaker for the susceptibility of the fields and up to four times weaker for the spurious couplings. This can be explained by the strength of the noise induced response as our theoretical models predicts that the field response is a second order effect in the noise parameters, see Subsection~\ref{subsec:theory_spurious_fields}, and the spurious links response is a fourth order effect in the field noise intensity, see Subsection~\ref{subsec:theory_spurious_couplings}. Therefore, only a $40\%$ difference in the single spin noise standard deviation can explain such differences. Note that the the quadratic response of the spurious coupling $J_{304,305}$ being weaker that the spurious coupling $J_{308,309}$ is correctly predicted by the simulation and with a similar ratio.

\begin{table}
\centering
\begin{tabular}{|c||c|c|}
    \hline
    \multicolumn{3}{|c|}{Linear Self-Response}\\
    \hline
    Output/Input & Simulated & Measured\\
    \hline
    \hline
    $h_{304}$ & $11.1$ & $10.5$ \\
    \hline
    $h_{305}$ & $10.4$& $10.1$  \\
    \hline
    $h_{308}$ & $11.2$& $10.5$  \\
    \hline
    $h_{309}$ & $9.4$& $9.4$  \\
    \hline
    $J_{304,308}$ & $12.0$ & $12.0$   \\
    \hline
    $J_{304,309}$ & $11.9$ & $11.9$ \\
    \hline
    $J_{305,308}$ & $12.2$ & $12.2$   \\
    \hline
    $J_{305,309}$ & $12.0$ & $12.0$  \\
    \hline
\end{tabular}
 \caption{Effective simulated and measured temperatures for existing couplings and magnetic fields. The individual coupling temperatures have been adjusted in the model such that the effective simulated temperatures coincide with the effective measure temperatures.}
\label{tbl:quadth_lin_response}
\end{table}

\begin{table}
\centering
\begin{tabular}{ |c||c|c|c| } 
\hline
\multicolumn{4}{|c|}{Main Quadratic Response}\\
\hline
Output& Input & Simulated & Measured \\
\hline
\hline
$h_{304}$    & $h_{308}, J_{304,308}$ & -4.9 & -10.5\\
                                & $h_{309}, J_{304,309}$ & -3.8 & -7.4 \\ 
\hline
$h_{305}$    & $h_{308}, J_{305,308}$ & -10.7 & -9.2\\ 
                                & $h_{309}, J_{305,309}$ & -8.2 & -11.6\\ 
\hline
$h_{308}$    & $h_{304}, J_{304,308}$ & -6.5 & -12.6\\
                                & $h_{305}, J_{305,308}$ & -5.9 & -11.1\\ 
\hline
$h_{309}$    & $h_{304}, J_{304,309}$ & -13.0 & -11.0\\
                                & $h_{305}, J_{305,309}$ & -11.7 & -11.9\\ 
\hline
$J_{304,305}$ & $J_{304,308}, J_{305,308}$ & -0.9 & -4.1\\
                                & $J_{304,309}, J_{305,309}$ & -0.7 & -4.0\\ 
\hline
$J_{308,309}$ & $J_{304,308}, J_{304,309}$ & -1.5 & -5.7\\
                                & $J_{305,308}, J_{305,309}$ & -1.3 & -6.8\\ 
\hline

\end{tabular}
\caption{Main simulated and measured components of the quadratic response. The leading terms are off-diagonal and correspond to an adjacent couplings and connected field for magnetic fields and correspond to two adjacent couplings forming a triangle for spurious links.}
\label{tbl:quadth_quad_response}
\end{table}

\section{Impacts of Spin Reversal Transformations}
\label{sec:effect_of_spin_reversals}

\subsection{Theoretical Considerations}

We are looking at distributions of Ising models on $N$ spins $\underline{\sigma} = \{\sigma_1,\ldots\,\sigma_{N}\}$ depending on input couplings $\underline{J}$ and magnetic fields $\underline{h}$. If we consider that the fields are potentially noisy and there exists individual residual fields $b_i$, random or deterministic, the Boltzmann distribution takes the following form,
\begin{align}
    \mu(\underline{\sigma}\mid \underline{J}, \underline{h}) = \frac{\exp{\left(\sum_{ij} \beta_{ij} J_{ij} \sigma_i \sigma_j + \sum_{i} \beta_i (h_i + b_i) \sigma_i\right)}}{Z},\label{eq:srt_ising_model}
\end{align}
where $\beta_{ij}$ and $\beta_i$ are effective individual temperatures for couplings and fields respectively. The residual fields may have potentially strong undesirable effects such as favoring particular spin configurations among others that were initially designed to be equiprobable. There exists a heuristical method that aims at mitigating this problem called the spin reversal transform (SRT). This method consists at looking at $2^N$ possible remapping of the model, each of them indexed by a ``gauge" which is a binary configuration $\tau = \{-1,1\}^N$. For a given configuration $\tau$, this gauge transform maps a spin configuration, input couplings and input fields to the values $\underline{\sigma}^\tau$, $\underline{h}^\tau$ and $\underline{J}^\tau$ in the following way,

\begin{align}
    \sigma^{\tau}_i = \sigma_i \tau_i,  \quad h^{\tau}_{i} = h_i \tau_i, \quad  J^{\tau}_{ij}= J_{ij} \tau_i \tau_j.\label{eq:srt_gauge_transform}
\end{align}
The particularity of the transformation~\eqref{eq:srt_gauge_transform} is that it creates an equivalence relationship between Hamiltonians \emph{without} residual fields as for any gauge $\tau$,
\begin{align}
   \sum_{ij} \beta_{ij} J^{\tau}_{ij} \sigma_i \sigma_j + \sum_{i} \beta_i h^{\tau}_i \sigma_i = \sum_{ij} \beta_{ij} J_{ij} \sigma_i \sigma_j + \sum_{i} \beta_i h_i \sigma_i.
\end{align}
For system without biases, it implies that samples generated from any set of gauge transformed inputs $\underline{h}^\tau$ and $\underline{J}^\tau$ are identical after a remapping of the samples using the same gauge $\underline{\sigma}^\tau$. The SRT method consists in generating samples from a mixtures of randomly selected gauge transformed models with residual fields, and which are therefore no longer equivalent, in order to empirically average over the residual field values. The effective model describing this mixture is given by the average of the Bolztmann distribution \eqref{eq:srt_ising_model} over all possible gauge transforms and reads,
\begin{align}
    \mu^{\rm{effective}}(\underline{\sigma}\mid \underline{J}, \underline{h}) = \frac{1}{2^{N}} \sum_{\tau \in \{-1,1\}^N} \mu(\underline{\sigma}^{\tau}\mid \underline{J}^{\tau}, \underline{h}^{\tau}),
\end{align}
where the gauge transformed models is explicitly expressed as follows,

\begin{align}
    \mu(\underline{\sigma}^{\tau}\mid \underline{J}^{\tau}, \underline{h}^{\tau})
    = \frac{\exp{\left(\sum_{ij} \beta_{ij} J_{ij} \sigma_i \sigma_j + \sum_{i} \beta_i (h_i + \tau_i b_i) \sigma_i\right)}}{Z^{\tau}}.\label{eq:srt_gauge_model}
\end{align}
Note that the partition function in Eq.~\eqref{eq:srt_gauge_model} depends on the gauge transforms through the residual field values. We see with Eq.~\eqref{eq:srt_gauge_model} that the sign of the residual fields are effectively randomly flipped with the SRT method. Thus, the SRT removes the undesirable effects of permanent residual fields but transforms it into magnetic field noise with other potentially unwanted effects such as lower effective temperature, spurious links and field quadratic response described in Section~\ref{sec:noise_and_spurious_links}. For the D-wave hardware where the fields noise is more important compared to permanent biases, see Section~\ref{sec:single_spin_exp}, the SRT removes permanent biases for only a limited increase in the noise. Therefore, the Ising model reconstructions appears less fluctuating over time with SRT than without.
To illustrate this last point, consider an Ising model with permanent residual fields $b_i$ and no input fields i.e. $\underline{h} \equiv 0$. In this model, the average of value of the spins are non-zero in general and are a non trivial function of the residual fields and couplings. However, in the effective model produced by the SRT $\mu^{\rm{effective}}(\underline{\sigma}\mid \underline{J}, 0)$, the average value of each spin is identically zero. This implies that the effective model has zero effective magnetic field regardless of the value of the residual biases. To see this, we first note that when $\underline{h} = 0$, the partition function of a gauge transformation is invariant under a global sign change  $Z^{\tau} = Z^{ - \tau}$. This further implies the following equivalence between two probabilities of gauge transformed spin configurations $\mu(\underline{\sigma}^{\tau}\mid \underline{J}^{\tau},0) = \mu(\underline{\sigma}^{-\tau}\mid \underline{J}^{-\tau},0)$ for all $\tau$. Since $\underline{\sigma}^{-\tau} = \underline{\sigma}^{\tau}$, it shows that the average value of any spin $\sigma_u$ vanishes as,
\begin{align}
    \mathbb{E}_{\rm{SRT}}\left[\sigma_u\right] &= \sum_{\underline{\sigma}\in\{-1,1\}^N} \sigma_u \mu^{\rm{effective}}(\underline{\sigma}\mid \underline{J}, \underline{h}) \nonumber\\
    &= \frac{1}{2^{N}} \sum_{\underline{\sigma}\in\{-1,1\}^N} \sigma_u \sum_{\tau \in \{-1,1\}^N} \mu(\underline{\sigma}^{\tau}\mid \underline{J}^{\tau}, \underline{h}^{\tau}),\nonumber\\
    &= \frac{1}{2^{N}} \sum_{\underline{\sigma}\in\{-1,1\}^N} \sigma_u \sum_{\tau \in \{-1,1\}^N} \frac{1}{2}\left(\mu(\underline{\sigma}^{\tau}\mid \underline{J}^{\tau}, \underline{h}^{\tau}) + \mu(\underline{\sigma}^{-\tau}\mid \underline{J}^{-\tau}, \underline{h}^{-\tau})\right),\nonumber\\
    &= \frac{1}{2^{N}}  \sum_{\tau \in \{-1,1\}^N} \sum_{\underline{\sigma}\in\{-1,1\}^N}\frac{1}{2}\left(\sigma_u \mu(\underline{\sigma}^{\tau}\mid \underline{J}^{\tau}, \underline{h}^{\tau}) + \sigma_u \mu(\underline{\sigma}^{-\tau}\mid \underline{J}^{-\tau}, \underline{h}^{-\tau})\right),\nonumber\\
    &= \frac{1}{2^{N}}  \sum_{\tau \in \{-1,1\}^N} \sum_{\underline{\sigma}\in\{-1,1\}^N}\frac{1}{2}\left(\sigma_u \mu(\underline{\sigma}^{\tau}\mid \underline{J}^{\tau}, \underline{h}^{\tau}) - \sigma_u \mu(\underline{\sigma}^{\tau}\mid \underline{J}^{\tau}, \underline{h}^{\tau})\right) =0.
\end{align}

\subsection{Mitigating Persistent Bias and Flux Drift}
Section \ref{sec:local_field_variability} highlighted persistent biases that the hardware exhibits. 
A useful feature of the spin reversal symmetry group is that combining data collected from symmetric models has the effect of averaging out persistent biases.

\begin{figure*}
    \centering
    \includegraphics{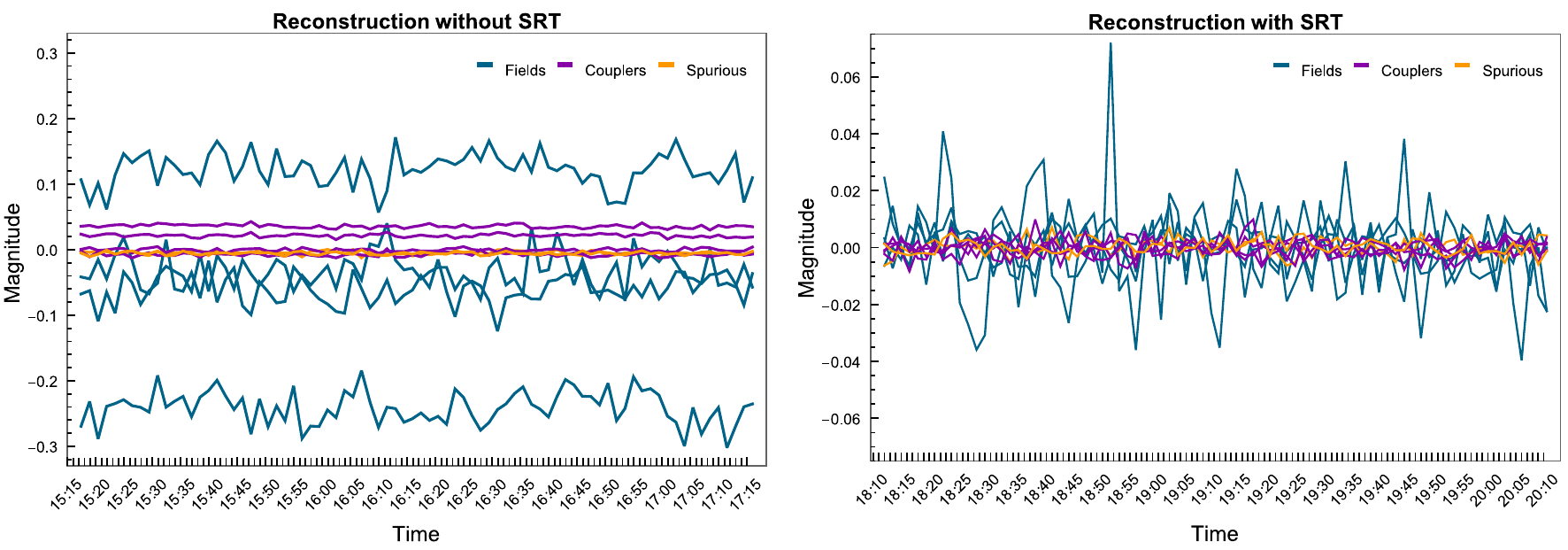}
    \caption{Illustration of the impact of spin reversal transforms on the persistent biases and flux drifts. Reconstruction of the parameters of the zero input problem was repeated over several hours. Reconstructed output parameters are not exactly centered around zero, and instead are fluctuating around values that can be interpreted as persistent biases. Local fields present much larger fluctuations compared to both native couplers and spurious couplings, that are significantly larger than the reconstruction error, a signature of the instantaneous qubit noise averaged over the reconstruction period. The left plot features normal raw data, and the right plot utilizes data obtained with random spin-reversal transformations applied to the input problems, which to a large extent eliminates the impact of persistent bias and reduces the variance of fluctuations.}
    \label{fig:biases_SRT}
\end{figure*}

This property is highlighted by replicating the field variability experiment with and without D-Wave's spin reversal transform feature.
In this revised experiment, the data for the zero problem is collected at 1.5 minute intervals over a period of 2 hours and then the 2-body reconstruction is used to recovery a model from the observed samples.
This experiment is conducted with two settings, {\em raw data} using this work's standard setting of \texttt{num\_spin\_reversal\_transforms = 0} and spin reversal transform data using \texttt{num\_spin\_reversal\_transforms = 10} (this setting results in a total of 200 transforms across the 200,000 samples collected for the zero-problem).
Fig.~\ref{fig:biases_SRT} shows the reconstructed values over time.
Table \ref{tbl:slow-field-variation} presents the mean and variance of each of the field values in these time series.
These results highlight how the spin reversal transforms can provide a drastic mitigation of the hardware's persistent bias.
An unexpected result from conducing these spin reversal transforms is a notable reduction in the variance of the field values.
Although the root cause for this reduction is not clear, we hypothesize that it is a side effect of an increased number of QPU programming cycles, which present another source of biases during the data collection process.
In any case, the notable bias mitigating impacts of spin reversal transforms can have a significant positive impact on  applications where the user would like an unbiased output distribution.

\begin{table}
    \centering
    \begin{tabular}{|r||r|r||r|r|}
    \hline
         & \multicolumn{2}{c||}{Raw Data} & \multicolumn{2}{c|}{Spin Reversals} \\
    \hline
    Spin & $\mu(h)$ & $\sigma(h)$ & $\mu(h)$ & $\sigma(h)$ \\
    \hline
    \hline
    304 & -0.241 & 0.025 & -0.001 & 0.019 \\
    \hline
    305 & -0.060 & 0.024 & 0.000 & 0.006 \\
    \hline
    308 &  0.121 & 0.025 & 0.000 & 0.010 \\
    \hline
    309 & -0.033 & 0.025 & 0.001 & 0.007 \\
    \hline
    \end{tabular}
    \begin{tabular}{|r||r|r||r|r|}
    \hline
         & \multicolumn{2}{c||}{Raw Data} & \multicolumn{2}{c|}{Spin Reversals} \\
    \hline
    Coupler & $\mu(J)$ & $\sigma(J)$ & $\mu(J)$ & $\sigma(J)$ \\
    \hline
    \hline
    304-308 & 0.0215 & 0.0024 & -0.0004 & 0.0030 \\
    \hline
    304-309 & -0.0076 & 0.0028 & -0.0007 & 0.0029 \\
    \hline
    305-308 & 0.0357 & 0.0027 & 0.0002 & 0.0035 \\
    \hline
    305-309 & -0.0005 & 0.0026 & 0.0000 & 0.0026 \\
    \hline
    \end{tabular}
    \caption{A comparison of the mean and variance of reconstructed field values over time with and without spin reversal transforms during data collection.  The use of spin reversal transforms corrects for persistent bias and also reduces the variance of reconstructed values.}
    \label{tbl:slow-field-variation}
\end{table}

\subsection{Model Reconstruction with Spin Reversal Transforms}
\begin{figure}
\centering
\includegraphics[width=0.39\linewidth,page=5]{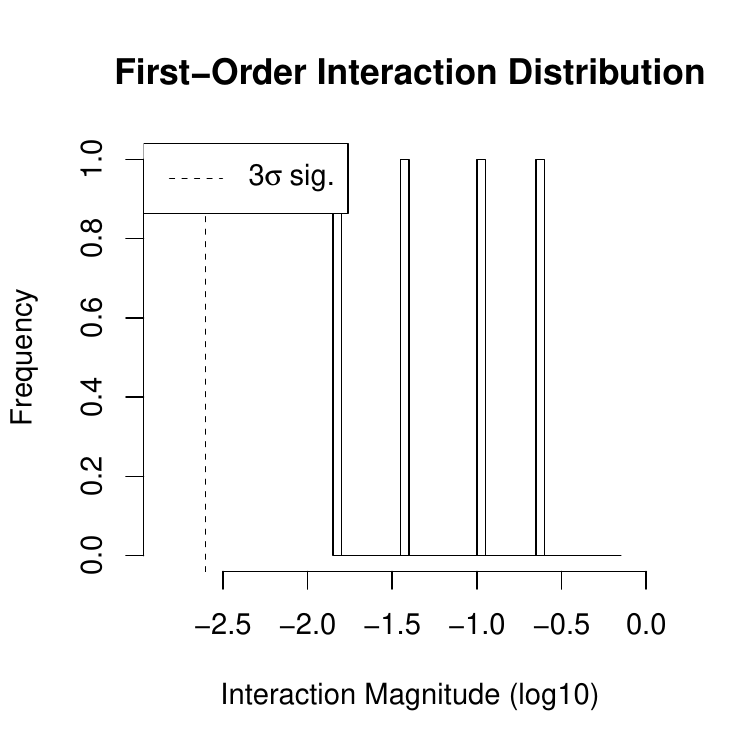}
\includegraphics[width=0.39\linewidth,page=5]{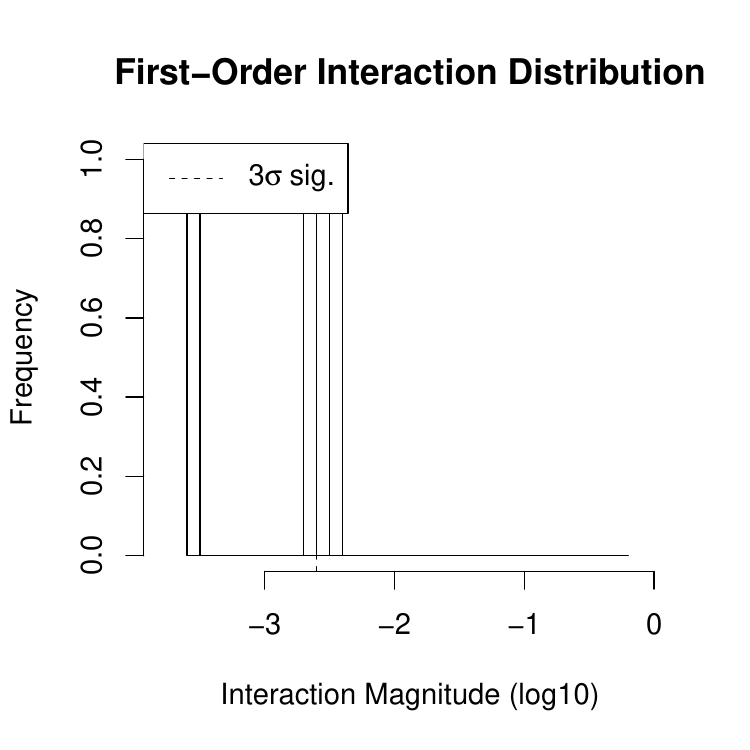}
\includegraphics[width=0.39\linewidth,page=6]{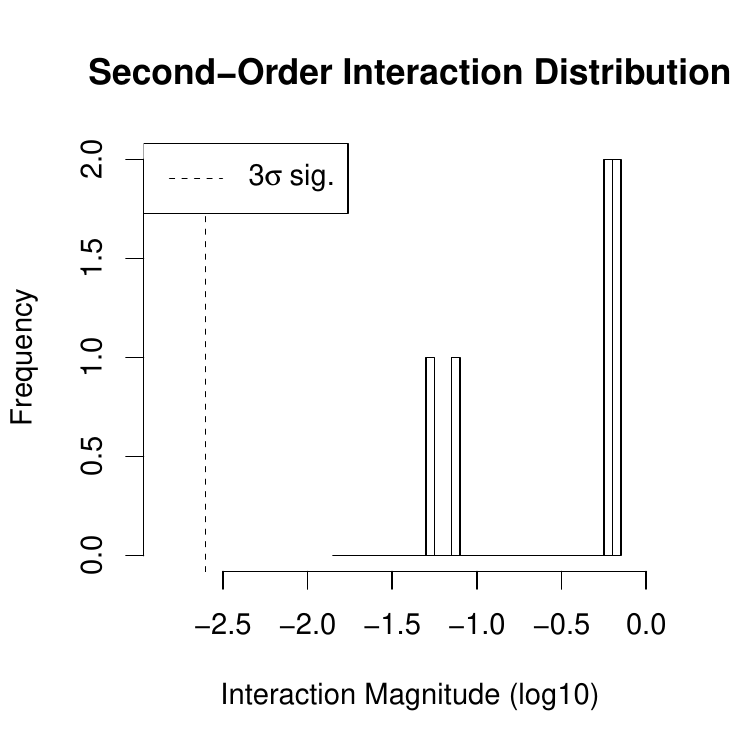}
\includegraphics[width=0.39\linewidth,page=6]{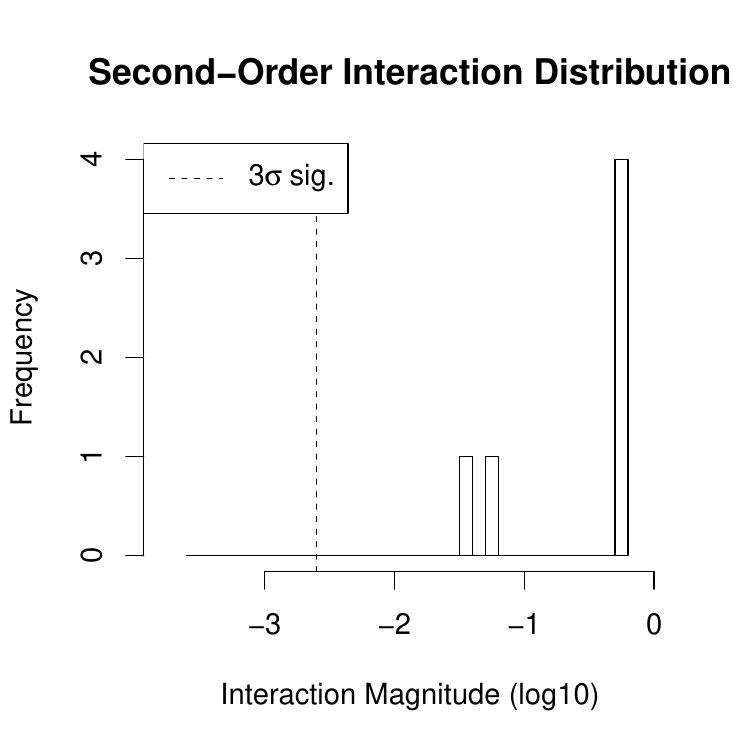}
\caption{The first-order (top) and second-order (bottom) terms of the two-body reconstruction without (left column) and with (right column) spin reversal transforms for the Strong Ferromagnet model.}
\label{fig:2o-strong-parameter-srt-values}
\end{figure}

Given the potential bias mitigating benefits of spin reversal transforms, it is natural to inquire how this feature impacts the results presented thus far.
We begin by reviewing the two-body reconstruction results presented in Fig. ~\ref{fig:2o-strong-parameter-srt-values}, which provides a side-by-side comparison of the results of Strong Ferromagnet model from Table \ref{tbl:2o-models} with and without spin reversal transforms. As the first observation, we notice that there is a notable change in the results of the zero-order terms. In the case with spin reversal transforms the zero-order terms are near zero, while statically significant non-zero values are exhibited in the raw data. The second observation is that the first-order terms do not show a notable change; in fact, these two reconstructions are remarkably consistent with and without spin reversal transforms. Both of these results indicate an absence of detrimental artifacts from utilizing this feature during data collection at the time scales that these experiments require.

\subsection{Quadratic Response with Spin Reversal Transforms}
\begin{figure*}[h]
\centering
\includegraphics[width=\linewidth]{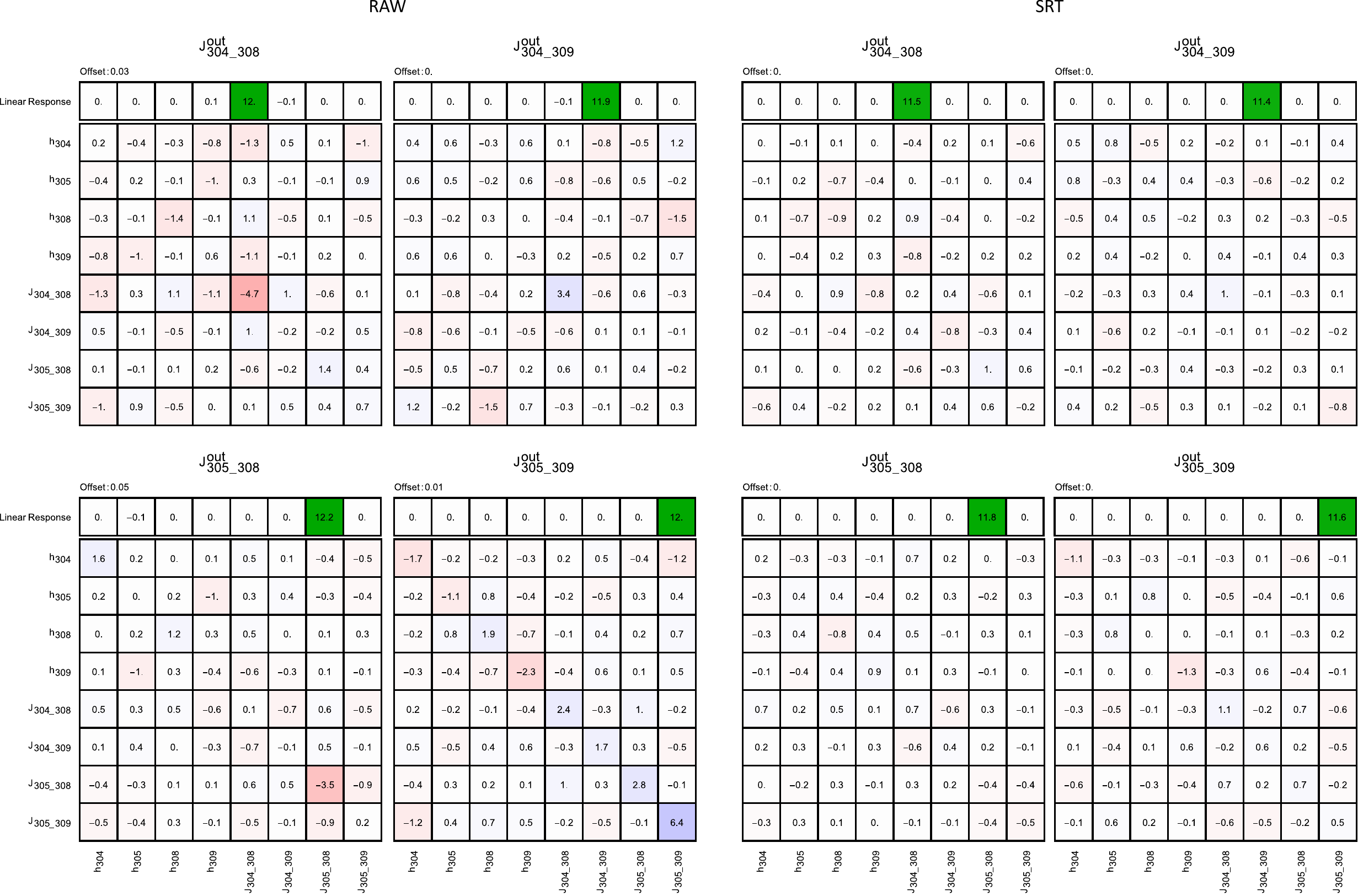}
\caption{A comparison of the key motifs of the quadratic response function (here, for the chimera couplings) without (left) and with (right) spin reversal transforms. The motifs are largely similar, however one can notice a considerable reduction in apparent noise in the quadratic response of the first-order terms. We hypothesize this is due mitigation of the flux qubit drift that occurs throughout the many hours of data collection required by this analysis.}
\label{fig:quad_comparison_part1}
\end{figure*}

\begin{figure*}[h]
\centering
\includegraphics[width=\linewidth]{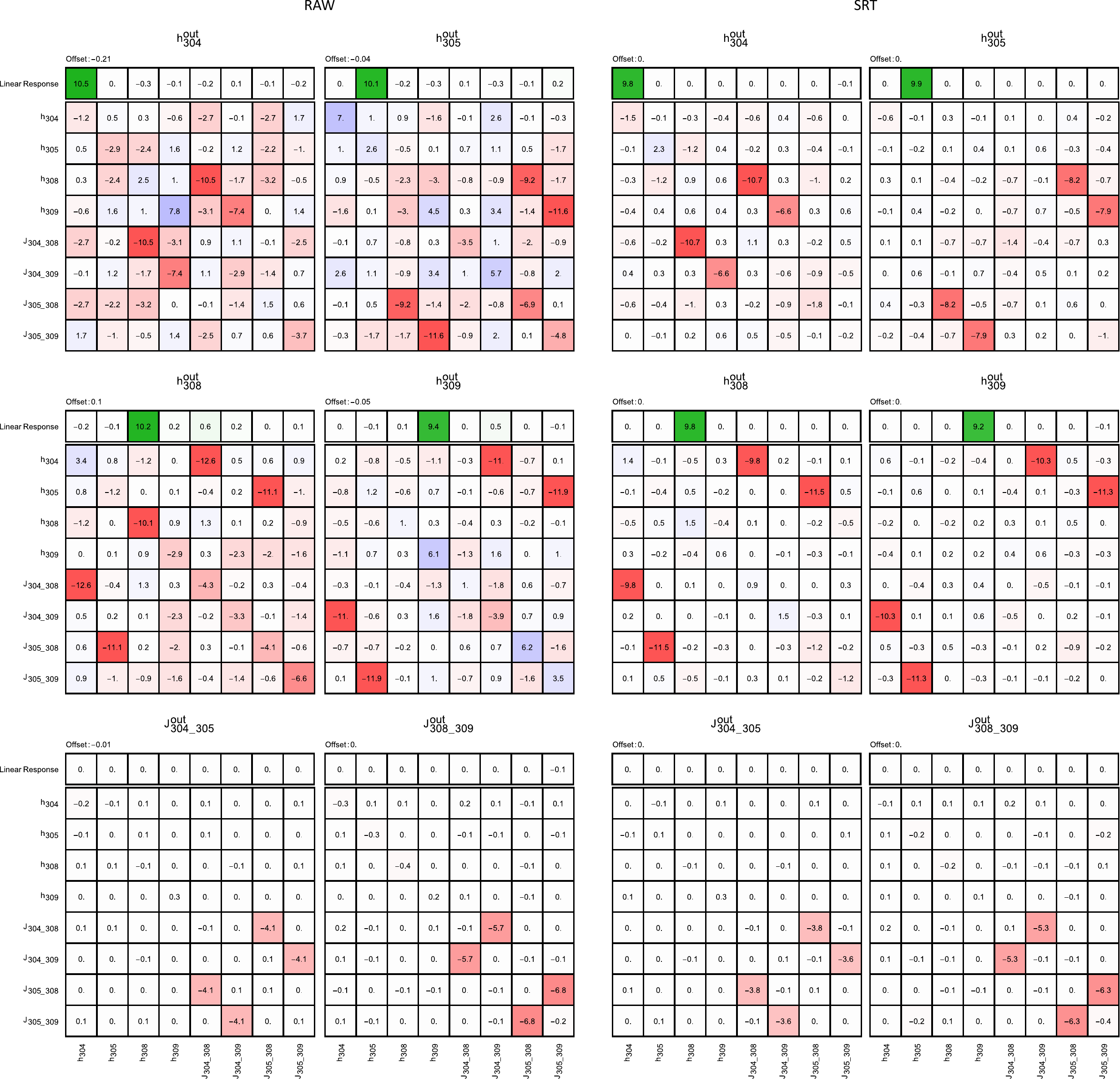}
\caption{A comparison of the key motifs of the quadratic response function (here, for the local fields and spurious couplings) without (left) and with (right) spin reversal transforms. The motifs are largely similar, however one can notice a considerable reduction in apparent noise in the quadratic response of the first-order terms. We hypothesize this is due mitigation of the flux qubit drift that occurs throughout the many hours of data collection required by this analysis.}
\label{fig:quad_comparison_part2}
\end{figure*}

Figs.~\ref{fig:quad_comparison_part1} and \ref{fig:quad_comparison_part2} replicate the quadratic response experiment from Section \ref{sec:quadratic_response} with and without spin reversal transforms. This comparison shows that the overall consistency of the quadratic response picture. In accordance with the theoretical predictions outlined in the beginning of this section, we observe that the persistent bias essentially disappears under the SRT setting, and the overall response becomes much cleaner.

\section{Tests Across Three Generations of Quantum Annealers and Validation of the Instantaneous Noise Model}
\label{sec:three_generations}

The previous sections have argued that the quadratic response model is a good valuable tool for characterizing the input-output behavior of a quantum annealer. Furthermore, the strength of spurious links can provide an indirect measurement of the instanteneous qubit noise that is occurring on a specific hardware device. We had the opportunity to collect data for the response analysis on 7 distinct QPUs spanning three generations of quantum annealing hardware. The specific device names and components used in the experiment are presented in Table \ref{tbl:other-qpus-used}.

\begin{table*}
    \centering
    \begin{tabular}{|l||l|l|}
    \hline
    QPU & $V'$ & $E'$ \\
    \hline
    \hline
    \multicolumn{3}{|c|}{2X} \\
    \hline
    LANL's \texttt{DW2X} & \{880, 881, 884, 885\}  & \{(880, 884), (880, 885), (881, 884), (881, 885)\} \\
    \hline
    \hline
    \multicolumn{3}{|c|}{2000Q} \\
    \hline
    \hline
    \texttt{DW\_2000Q\_LANL} & \{304, 305, 308, 309\}  & \{(304, 308), (304, 309), (305, 308), (305, 309)\} \\ 
    \hline
    NASA's \texttt{C16} & \{1800, 1801, 1804, 1805\}  & \{(1800, 1804), (1800, 1805), (1801, 1804), (1801, 1805)\} \\
    \hline
    \texttt{DW\_2000Q\_1} & \{1800, 1801, 1804, 1805\}  & \{(1800, 1804), (1800, 1805), (1801, 1804), (1801, 1805)\} \\
    \hline
    \texttt{DW\_2000Q\_2} & \{1800, 1801, 1804, 1805\}  & \{(1800, 1804), (1800, 1805), (1801, 1804), (1801, 1805)\} \\
    \hline
    \texttt{DW\_2000Q\_3} & \{1800, 1801, 1804, 1805\}  & \{(1800, 1804), (1800, 1805), (1801, 1804), (1801, 1805)\} \\
    \hline
    \hline
    \multicolumn{3}{|c|}{Lower-Noise 2000Q} \\
    \hline
    \texttt{DW\_2000Q\_5} & \{1688, 1689, 1692, 1693\}  & \{(1688, 1692), (1688, 1693), (1689, 1692), (1689, 1693)\} \\
    \hline
    \texttt{DW\_2000Q\_6} & \{1688, 1689, 1692, 1693\}  & \{(1688, 1692), (1688, 1693), (1689, 1692), (1689, 1693)\} \\
    \hline
    \end{tabular}
    \caption{The 4 qubits and 4 couplers used in the quadratic response experiments on a variety of QPUs.  The experiments span three generations of D-Wave hardware, the 2X (2015), 2000Q (2017), and Lower-Noise 2000Q (2019).}
    \label{tbl:other-qpus-used}
\end{table*}

\begin{figure*}
\centering
\includegraphics[width=\linewidth]{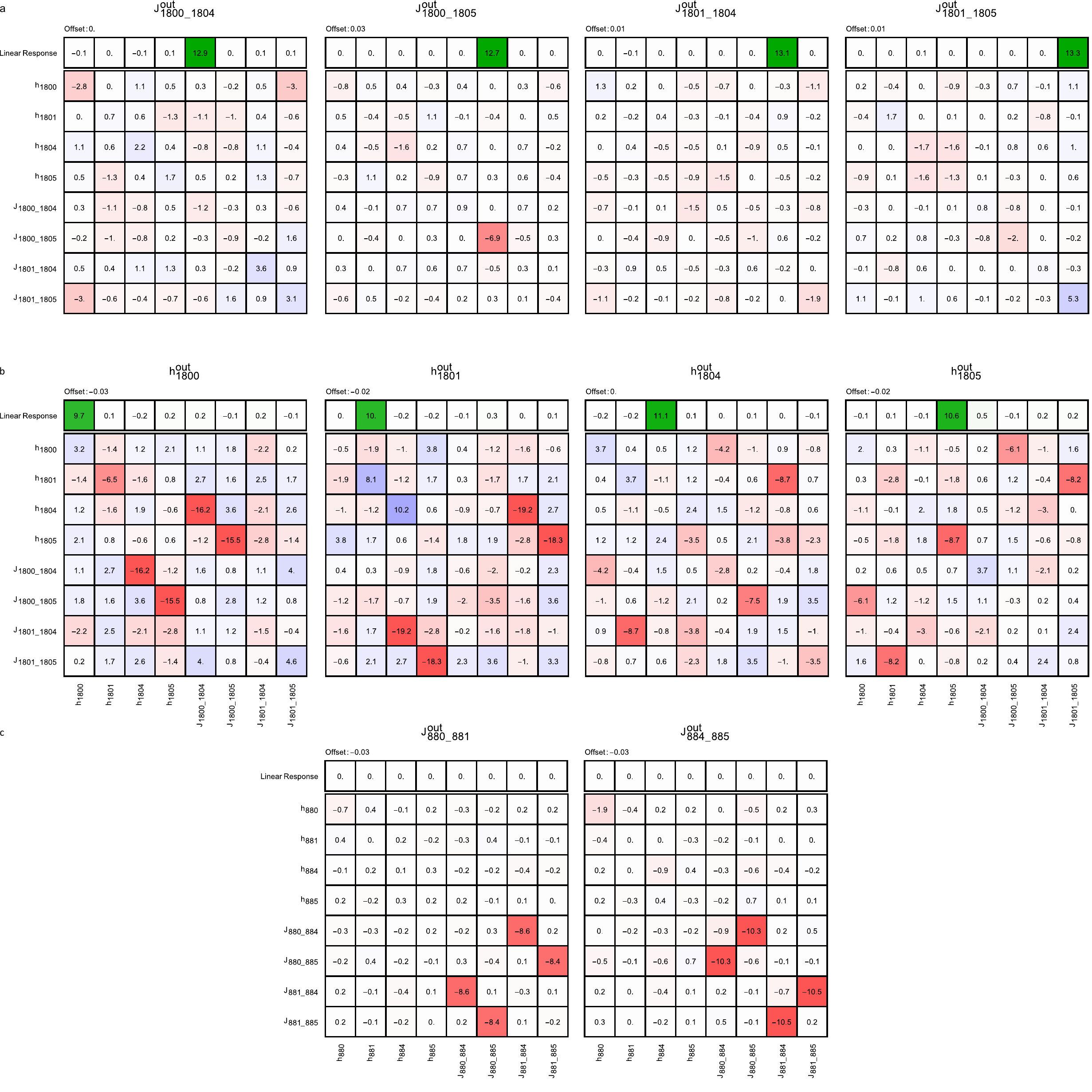}
\caption{Quadratic response experiment results for the LANL's \texttt{DW2X} chip: (a) Chimera couplings; (b) Local Fields; (c) Spurious couplings.}
\label{fig:quad_response_2x}
\end{figure*}

\begin{figure*}
\centering
\includegraphics[width=\linewidth]{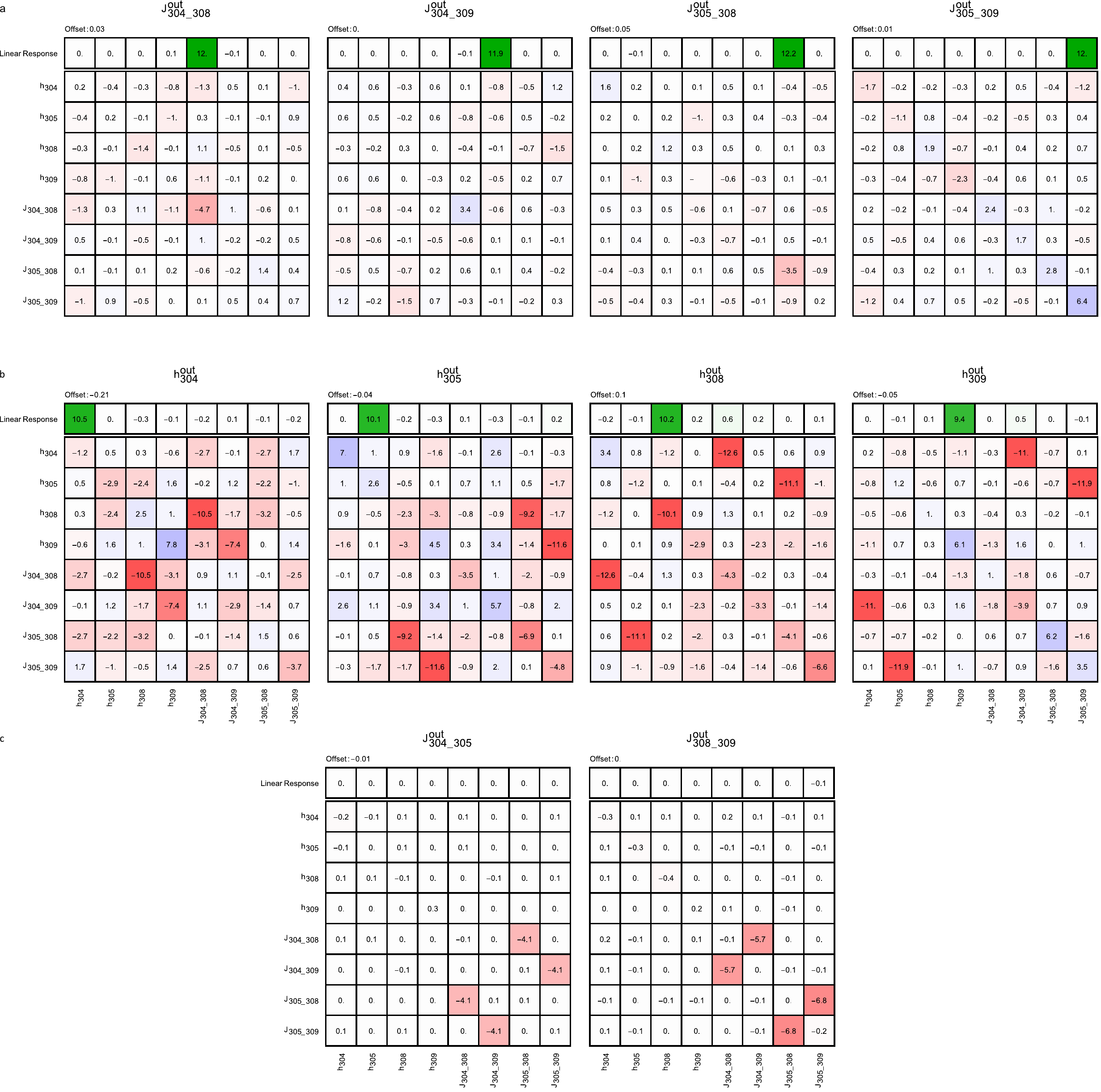}
\caption{Quadratic response experiment results for \texttt{DW\_2000Q\_LANL} chip: (a) Chimera couplings; (b) Local Fields; (c) Spurious couplings.}
\label{fig:quad_response_l2k}
\end{figure*}

\begin{figure*}
\centering
\includegraphics[width=\linewidth]{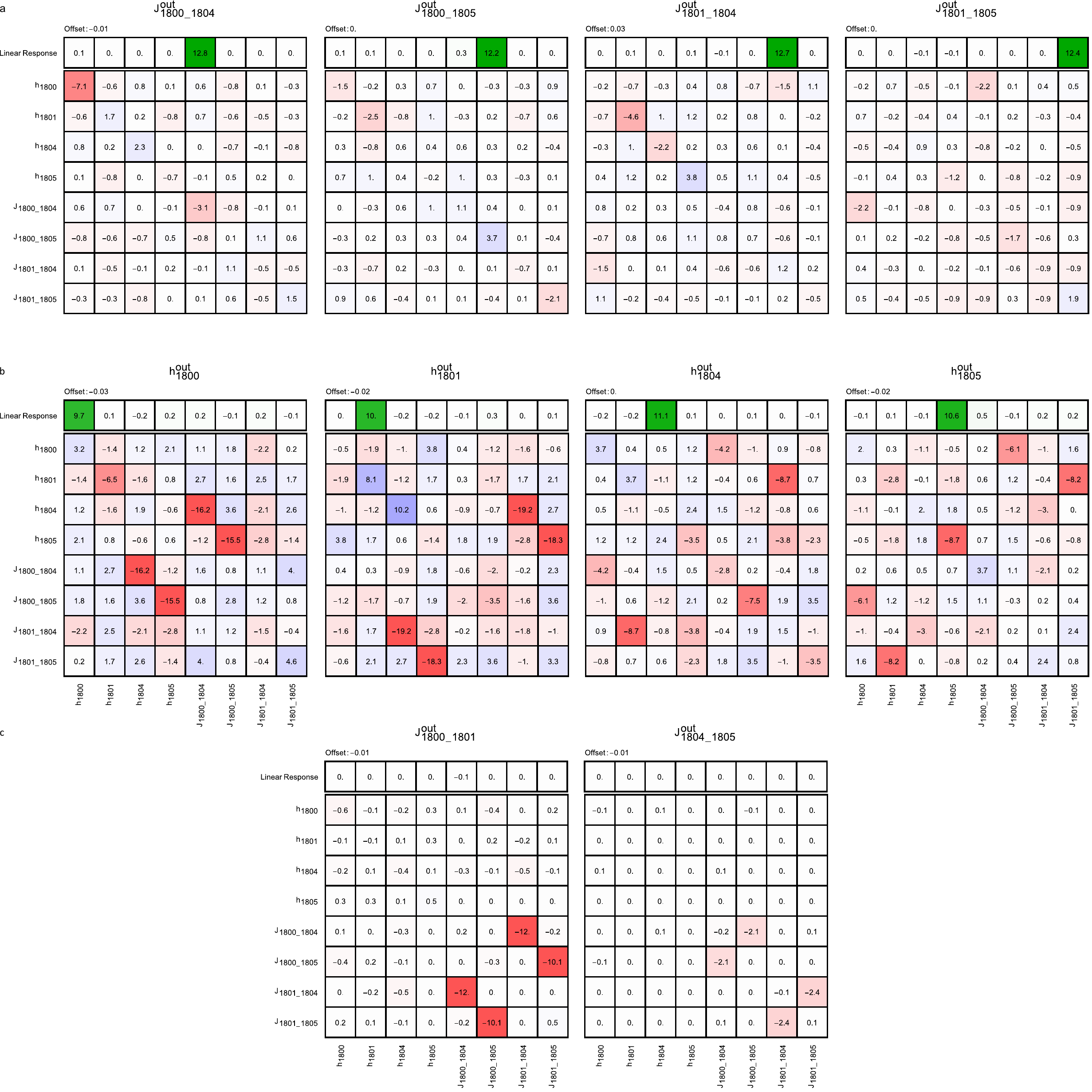}
\caption{Quadratic response experiment results for the NASA's \texttt{C16} chip: (a) Chimera couplings; (b) Local Fields; (c) Spurious couplings.}
\label{fig:quad_response_n2k}
\end{figure*}

\begin{figure*}
\centering
\includegraphics[width=\linewidth]{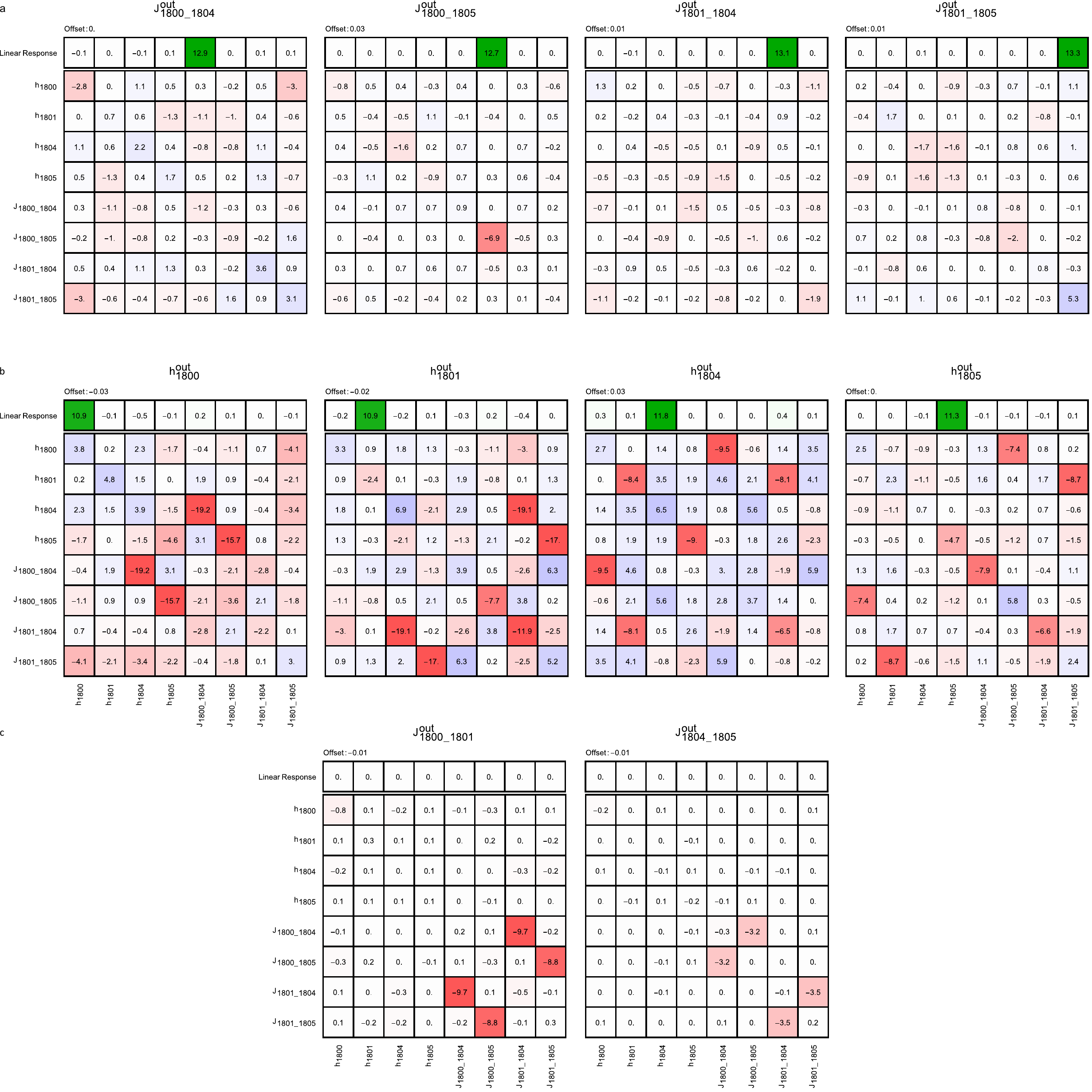}
\caption{Quadratic response experiment results for the \texttt{DW\_2000Q\_1} chip: (a) Chimera couplings; (b) Local Fields; (c) Spurious couplings.}
\label{fig:quad_response_2k}
\end{figure*}

\begin{figure*}
\centering
\includegraphics[width=\linewidth]{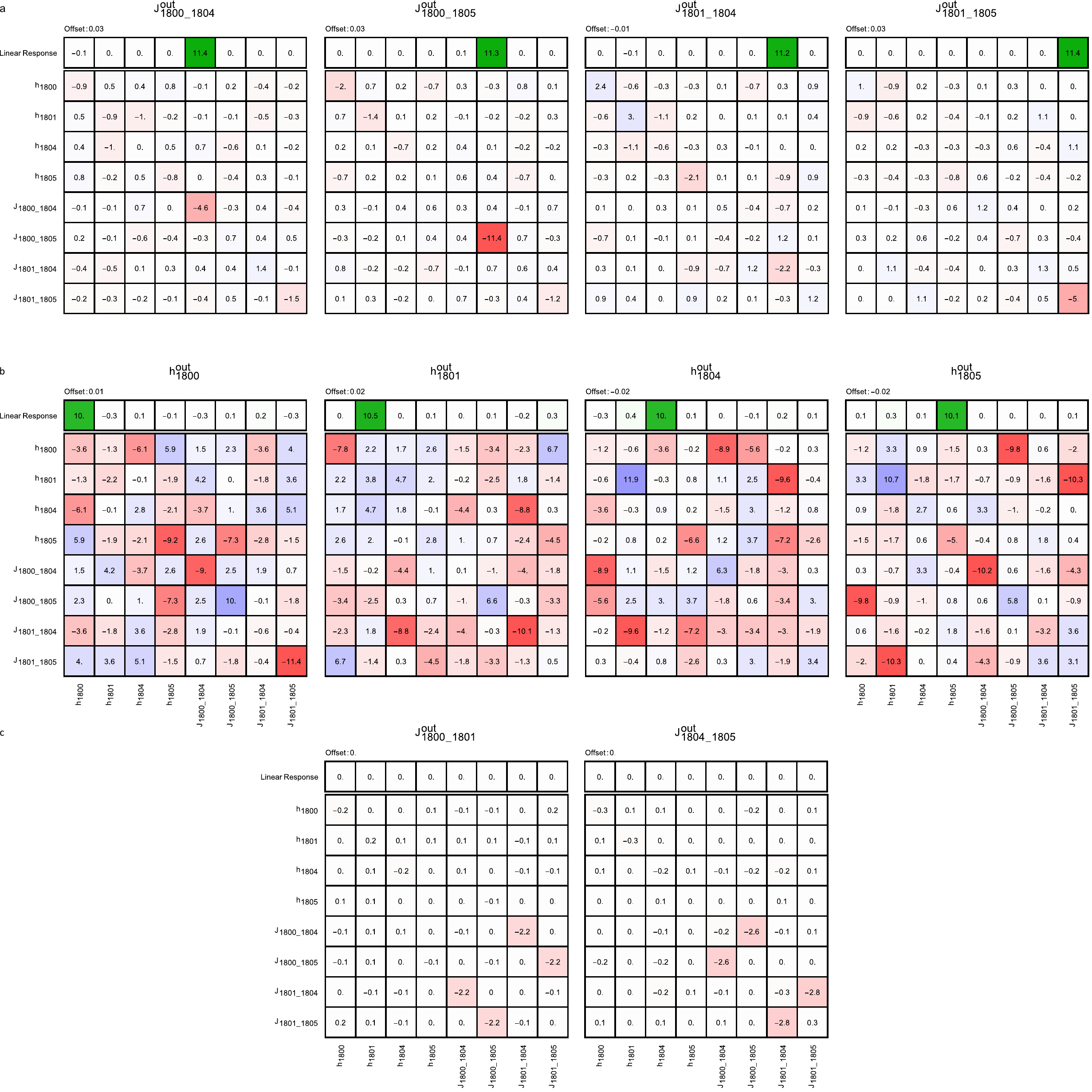}
\caption{Quadratic response experiment results for the \texttt{DW\_2000Q\_2} chip: (a) Chimera couplings; (b) Local Fields; (c) Spurious couplings.}
\label{fig:quad_response_2k2}
\end{figure*}

\begin{figure*}
\centering
\includegraphics[width=\linewidth]{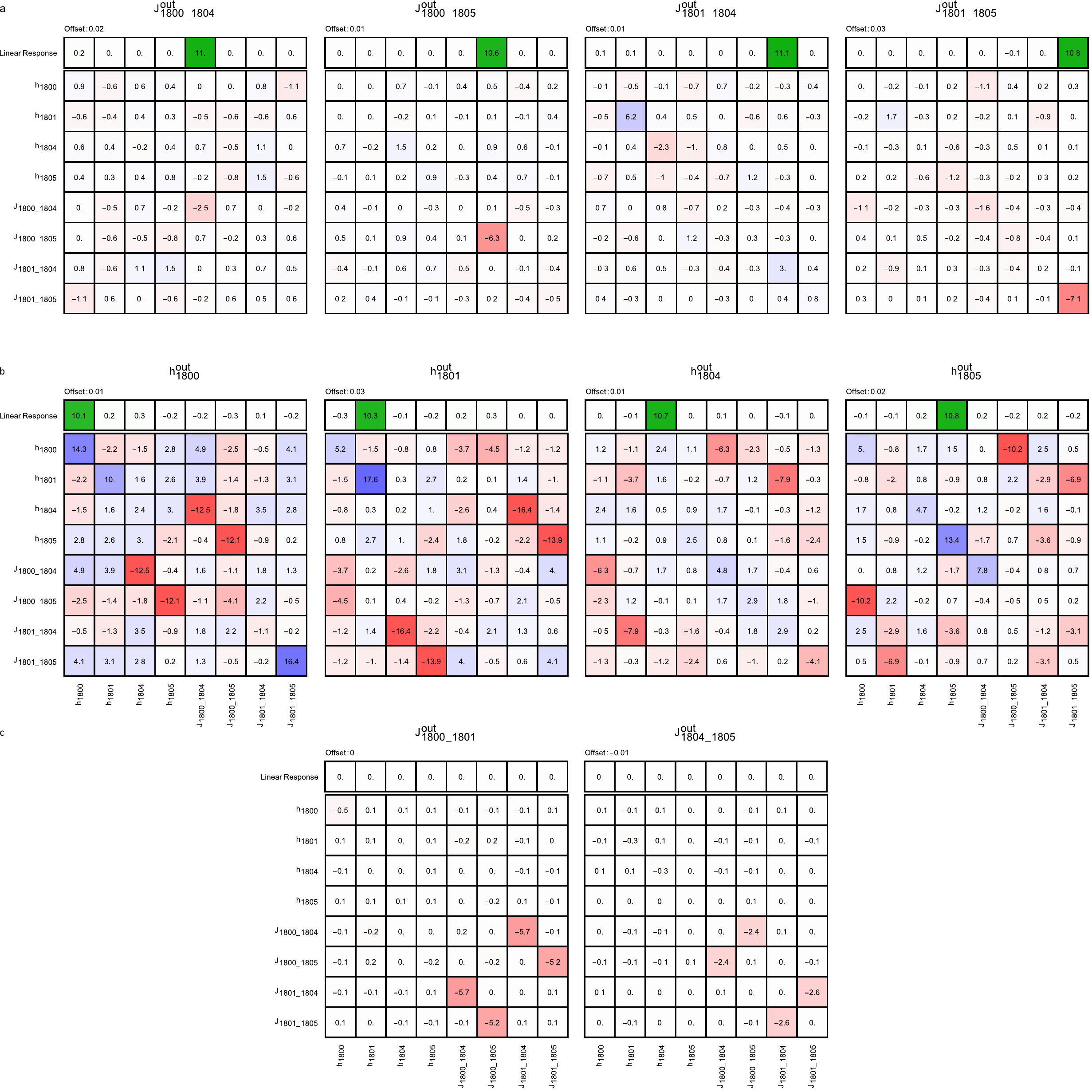}
\caption{Quadratic response experiment results for the \texttt{DW\_2000Q\_3} chip: (a) Chimera couplings; (b) Local Fields; (c) Spurious couplings.}
\label{fig:quad_response_2k3}
\end{figure*}

\begin{figure*}
\centering
\includegraphics[width=\linewidth]{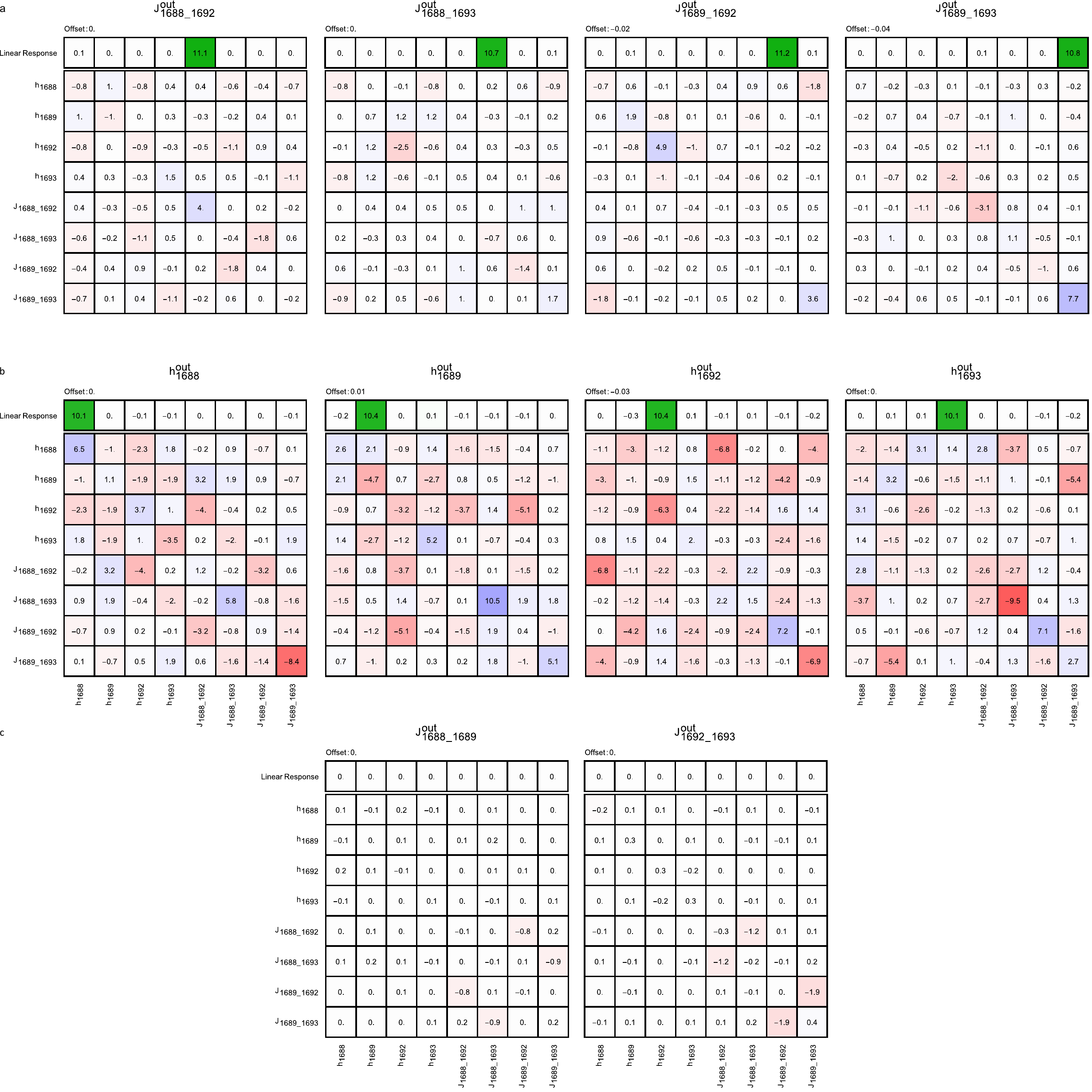}
\caption{Quadratic response experiment results for the lower-noise \texttt{DW\_2000Q\_5} chip: (a) Chimera couplings; (b) Local Fields; (c) Spurious couplings.}
\label{fig:quad_response_2k5}
\end{figure*}

\begin{figure*}
\centering
\includegraphics[width=\linewidth]{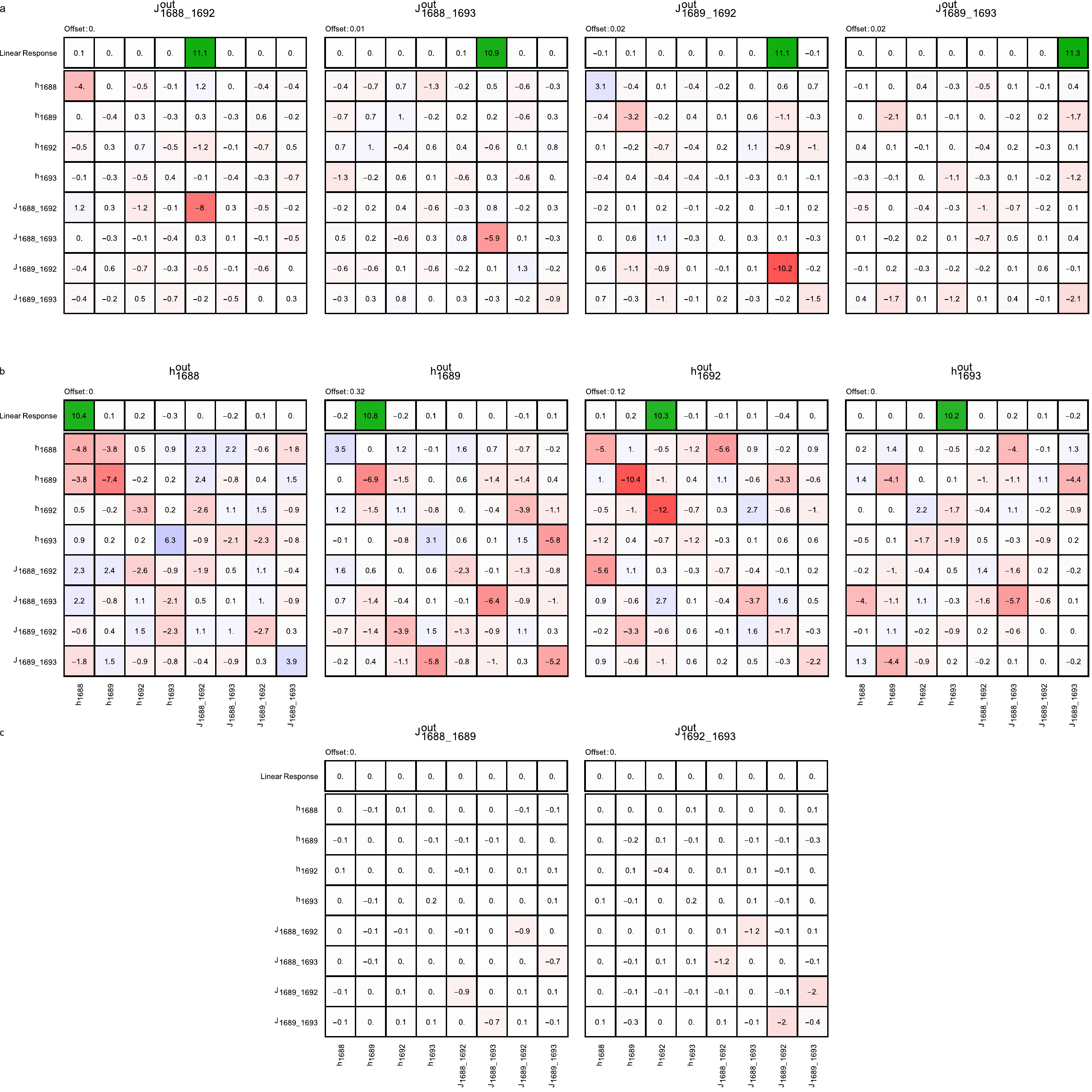}
\caption{Quadratic response experiment results for the lower-noise \texttt{DW\_2000Q\_6} chip: (a) Chimera couplings; (b) Local Fields; (c) Spurious couplings.}
\label{fig:quad_response_2k6}
\end{figure*}

Figs.~\ref{fig:quad_response_2x}-\ref{fig:quad_response_2k6} present the quadratic response motifs from all of the QPUs that have been tested. Overall, the results are remarkably similar, which suggests the universality of the quadratic response characterization that is capturing fundamental properties of D-Wave's quantum annealing implementation, like effective temperature, persistent biases, and instantaneous noise in the local fields. With the exception of the new {\em lower-noise} QPU, the existence and strength of the spurious links is consistent across hardware realizations. We have observed that some QPU implementations feature asymmetric spurious links while others are symmetric. Identifying the root-cause of this distinction is an ongoing point of investigation.

\begin{figure*}
\centering
\includegraphics[width=0.88\linewidth]{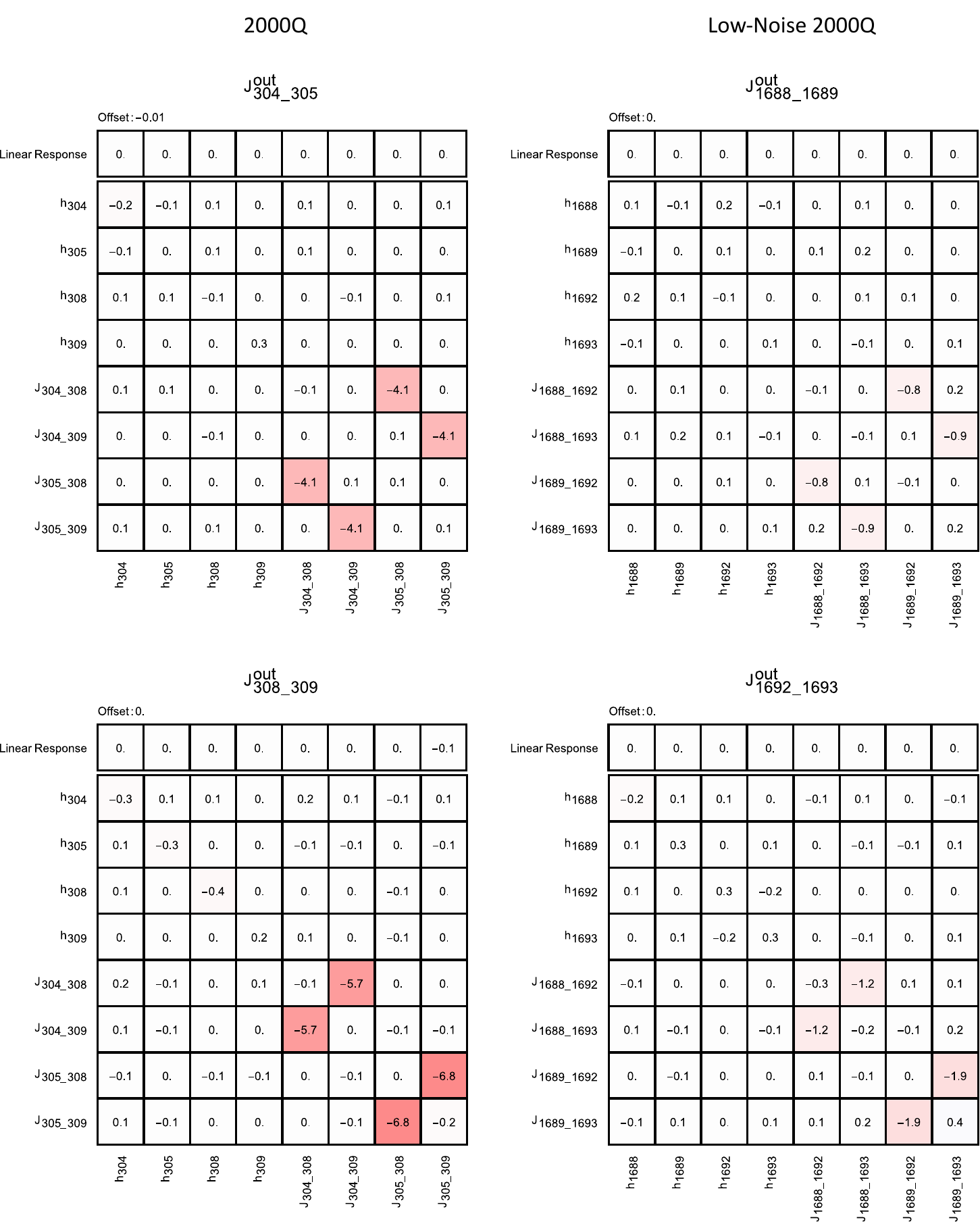}
\caption{Heat maps representing the quadratic terms of the quadratic response function for the spurious link output parameters on a regular (left) and lower-noise (right) 2000Q quantum annealers.  The significant reduction in the link strength in the lower-noise response (from -4.1 to -0.9 and -6.8 to -1.9) confirms the theoretical model of high-frequency qubit noise.}
\label{fig:quad_spurious_terms_low_noise}
\end{figure*}

A significant difference in the response function has been obtained for the \emph{lower-noise} version of the D-Wave 2000Q annealer \cite{D-Wave_low_noise}. We observe a drastic reduction of the susceptibility responsible for the strength of the spurious couplings, while the linear scale terms remain on par with other 2000Q implementations, see Fig.~\ref{fig:quad_spurious_terms_low_noise}. This result therefore provides strong evidence in support of the noise-based model introduced in this work. We anticipate that the quantitative measurement of the susceptibility associated with the spurious couplings using methods developed in this work will provide a valuable characterization of the qubit noise in the next generations \cite{D-Wave_pegasus} of quantum annealers and other analog machines with binary output statistics.

\section{Open-sourced Tools}
\label{sec:Algortihmic_Tools}

The experiments conducted in this work require the collection of billions of samples from D-Wave's quantum annealer and reconstruction of graphical models with multi-body interactions.
However, neither of these tasks is readily supported by established software and the following software was developed and released as open-source to support this work.
The first software is the D-Wave Ising Sample Collector (DWISC, \url{github.com/lanl-ansi/dwisc}), which enables the collection of millions runs on D-Wave hardware by orchestrating a series of jobs that conform to D-Wave's single-job run time limit of three seconds.
The second software is GraphicalModelLearning (GML, \url{github.com/lanl-ansi/GraphicalModelLearning.jl}), which takes empirical state distributions and reconstructs effective multi-body graphical models in a factor graph representation, leveraging the Interaction Screening method described in Section~\ref{sec:learning_general_distributions} and state-of-the-art second-order nonlinear optimization algorithms to provide model reconstructions that require the least amount of data. The notable improvement of reconstruction accuracy of interaction screening framework over  established approaches, such as mean-field, is discussed at length in \cite{lokhov2018optimal}. DWISC and GML form the foundation of the experiments in this work by providing the data and algorithms required to reconstruct high-accuracy multi-body models of the output from D-Wave's quantum annealer.

\end{document}